\documentclass[preprint,12pt]{elsarticle}
\graphicspath{{figures/}}

\usepackage{amssymb}
\usepackage{subfigure}
\usepackage{amsmath}
\usepackage{color}
\usepackage{breqn}
\usepackage{hyperref}
\usepackage{gensymb}

\usepackage{booktabs}

\usepackage{siunitx}

\def\rc{\ensuremath{\Gamma_t} }

\def\vec#1{\boldsymbol{#1}}
\def\sp{\ensuremath{\boldsymbol{x}} }
\def\disp{\ensuremath{\vec u} }

\DeclareMathAlphabet\mathbfcal{OMS}{cmsy}{b}{n}

\usepackage{marginnote}

\makeatletter
\def\ps@pprintTitle{%
	\def\@oddfoot{}%
	\let\@oddhead\@oddfoot
	\let\@evenhead\@oddfoot
	\let\@evenfoot\@oddfoot
}
\makeatother

\begin{document}

\begin{frontmatter}

 \title{The divergence-conforming immersed boundary method: Application to vesicle and capsule dynamics}

% \title{The divergence-conforming immersed boundary method: Application to polymeric capsules and lipid bilayers}

\author[label1]{Hugo Casquero\corref{cor1}}
\ead{hugocp@andrew.cmu.edu}
\author[label2]{Carles Bona-Casas}
\author[label3]{Deepesh Toshniwal}
\author[label4]{Thomas J.R. Hughes}
\author[label5]{Hector Gomez}
\author[label1]{Yongjie Jessica Zhang}
\address[label1]{Department of Mechanical Engineering, Carnegie Mellon University, Pittsburgh, PA 15213, United States}
\address[label2]{Departament de F\'isica $\&$ IAC$^{ \, 3}$, Universitat de les Illes Balears, Palma de Mallorca, 07122, Spain}
\address[label3]{Delft Institute of Applied Mathematics, Delft University of Technology, Van Mourik Broekmanweg 6, XE Delft 2628, The Netherlands}
\address[label4]{Oden Institute for Computational Engineering and Sciences, 201 East 24th Street, C0200, Austin, TX 78712-1229, United States}
\address[label5]{School of Mechanical Engineering, Purdue University, West Lafayette, IN 47907, United States}

\cortext[cor1]{Corresponding author.}

% \address[1]{Affiliation 1, Address, City and Postal Code, Country}

%\received{1 May 2013}
%\finalform{10 May 2013}
%\accepted{13 May 2013}
%\availableonline{15 May 2013}
%\communicated{S. Sarkar}

\begin{abstract}
%%%
We extend the recently introduced divergence-conforming immersed boundary (DCIB) method \cite{casquero2018non} to fluid-structure interaction (FSI) problems involving closed co-dimension one solids. We focus on capsules and vesicles, whose discretization is particularly challenging due to the higher-order derivatives that appear in their formulations. In two-dimensional settings, we employ cubic B-splines with periodic knot vectors to obtain discretizations of closed curves with $C^2$ inter-element continuity. In three-dimensional settings, we use analysis-suitable bi-cubic T-splines to obtain discretizations of closed surfaces with at least $C^1$ inter-element continuity. Large spurious changes of the fluid volume inside closed co-dimension one solids is a well-known issue for IB methods. The DCIB method results in volume changes orders of magnitude lower than conventional IB methods. This is a byproduct of discretizing the velocity-pressure pair with divergence-conforming B-splines, which lead to negligible incompressibility errors at the Eulerian level. The higher inter-element continuity of divergence-conforming B-splines is also crucial to avoid the quadrature/interpolation errors of IB methods becoming the dominant discretization error. Benchmark and application problems of vesicle and capsule dynamics are solved, including mesh-independence studies and comparisons with other numerical methods.

% In the context of immersed boundary (IB) methods in variational form, direct discretizations of extensible shear-resistant capsules and inextensible bending-resistant vesicles require $C^1$ and $C^2$ inter-element continuity, respectively.
\end{abstract}

\begin{keyword}
Fluid-structure interaction \sep Immersed boundary method \sep Volume conservation  \sep Isogeometric analysis \sep Vesicles \sep Capsules
 % \sep Variational methods \sep Convergence rates \sep  Fully-implicit time integration \sep  Benchmarking Red blood cell  \sep Divergence-conforming B-splines
\end{keyword}

%\begin{keyword}
%% Keywords
%\KWD Keyword1\sep Keyword2\sep Keyword3
%\end{keyword}

\end{frontmatter}

%\linenumbers


\renewcommand{\thefootnote}{\fnsymbol{footnote}}

% \newpage

\section{Introduction}

Immersed approaches for fluid-structure interaction (FSI) have been used to tackle a wide range of open questions involving the interaction of incompressible viscous fluids with incompressible elastic solids. However, the reliability of immersed approaches, such as the immersed boundary (IB) method \cite{Peskin1972, Peskin1977, Peskin2002, boffi2008hyper, boffi2003finite, liu2006immersed, saadat2018immersed, mittal2005immersed} and the ficticious domain (FD) method \cite{glowinski1999distributed,glowinski2001fictitious, baaijens2001fictitious, van2004combined,yu2005dlm}, is often jeopardized by large errors in imposing the incompressibility constraint at the Eulerian and/or Lagrangian levels \cite{peskin1993improved, griffith2012volume, strychalski2016intracellular, boilevin2019numerical, boilevin2019loosely}. Although this issue was first mentioned more than two decades ago \cite{peskin1993improved}, solving this problem at its root and without side effects that limit the applicability of the resulting immersed method has proven to be extremely challenging. In the case of discretizations based on finite differences or finite volumes, the dominant error is due to the fact that the Lagrangian velocity obtained from the Eulerian velocity using discretized delta functions is far from being a solenoidal field even if the Eulerian velocity is properly enforced to be divergence-free with respect to the finite-difference/finite-volume approximation of the divergence operator \cite{peskin1993improved, bao2017immersed, mendez2014unstructured}. In the case of discretizations based on finite elements, the dominant error is due to the fact that the weakly divergence-free Eulerian velocity is far from being a solenoidal field \cite{galvin2012stabilizing, john2017divergence, boilevin2019numerical}. Most of the proposed solutions to reduce the incompressibility errors do not tackle the aforementioned root causes and try to mitigate their effects instead. To name a few, penalty terms are often added to try to decrease the spurious change of fluid volume inside closed co-dimension one solids \cite{peng2010multiscale, yazdani2012three, shen2017interaction}, a post-processing correction using a Lagrange multiplier is used after each time step to slightly change the nodal coordinates of closed co-dimension one solids to preserve their inner volume \cite{li2012volume,mendez2014unstructured, saadat2018immersed}, and extremely large grad-div stabilization is added near the fluid-solid interface to obtain a velocity field that is closer to a solenoidal field in this region \cite{galvin2012stabilizing, boilevin2019numerical}. There are some recent solutions that effectively tackle the aforementioned root causes, but compromise the applicability of the resultant numerical method. In \cite{bao2017immersed}, a finite-difference discretization is proposed that results in Lagrangian velocity fields that are solenoidal, but the method is limited to periodic domains. In \cite{alauzet2016nitsche, boilevin2019numerical}, an extended finite-element discretization is proposed that leads to weakly divergence-free Eulerian velocities that are good approximations of a solenoidal field by capturing the pressure discontinuity at the fluid-solid interface, but the method has only been developed for two-dimensional settings thus far.

Since the advent of isogeometric analysis (IGA) \cite{1003.000, 1002.000}, spline-based discretizations of immersed approaches for FSI problems have proliferated \cite{ruberg2014fixed, Casquero2015, Kamensky2015, Casquero2016a, Hsu2015, kamensky2017projection, bazilevs2017new, kadapa2016fictitious, kadapa2018stabilised, heltai2017natural, maestre20173d, moutsanidis2018hyperbolic}. This includes NURBS-based and T-spline-based generalizations of the IB method \cite{Casquero2015, Casquero2016a} and the FD method \cite{Kamensky2015, Hsu2015, kadapa2016fictitious}, which are two of the most widespread immersed approaches for challenging FSI applications. Unfortunately, the issue found in classical finite-element discretizations of weakly divergence-free Eulerian velocities not being a good approximation to a solenoidal field persists in spline-based discretizations \cite{Kamensky2015, casquero2017nurbs}. A definitive solution to this problem is to use Eulerian discretizations that result in pointwise satisfaction of the incompressibility constraint. However, such discretizations are scarce and a price is often paid in an other aspect of the numerical method in comparison with a standard discretization. Scott-Vogelius elements \cite{scott1985norm} were initially developed for two-dimensional settings. Although there have been attempts to generalize Scott-Vogelius elements to three-dimensional settings \cite{zhang2011divergence,neilan2015discrete}, a full generalization is still out of reach \cite{neilan2015discrete}. In addition, the proof of \textit{inf-sup} stability of Scott-Vogelius elements is not trivial and different proofs are available in the literature under different assumptions \cite{scott1985norm, falk2013stokes, guzman2017scott}. Raviart-Thomas elements \cite{raviart1977mixed} are not $H^1$-conforming and therefore the Galerkin method cannot be used to discretize the Navier-Stokes equations in primal form. In \cite{buffa2011isogeometricA, buffa2011isogeometricB}, the higher inter-element continuity of splines was leveraged to define a smooth generalization of Raviart-Thomas elements, the so-called divergence-conforming B-splines. Divergence-conforming B-splines are \textit{pointwise} divergence-free, \textit{inf-sup} stable, $H^1$-conforming, non-negative, and pressure-robust \cite{evans2013isogeometric, evans2013isogeometric2, john}. Moreover, their higher inter-element continuity is highly beneficial in IB and FD methods avoiding the necessity of performing numerical integration of integrands that include Eulerian functions with jumps in the interior of Lagrangian elements \cite{boffi2008hyper, boffi2011finite, hesch2012continuum, casquero2018non}. As a result, divergence-conforming B-splines are an ideal candidate for the Eulerian discretization of immersed approaches for FSI\footnote{In \cite{guzman2013conforming}, \textit{inf-sup} stable, pointwise divergence-free, $H^1$-conforming, and pressure-robust tetrahedral elements on simplicial triangulations are constructed. The pressure space is simply the space of piecewise constants and the velocity space consists of piecewise cubic polynomials enriched with rational functions. Although these tetrahedral elements do not have the higher inter-element continuity of divergence-conforming B-splines, their use in immersed approaches for FSI is also worth consideration.}. Divergence-conforming B-splines have already been used in a generalization of the FD method that considers open co-dimension one solids, the so-called immersogeometric method \cite{kamensky2017immersogeometric}, and in a generalization of the IB method that considers co-dimension zero solids, the so-called divergence-conforming immersed boundary (DCIB) method \cite{casquero2018non}. In \cite{kamensky2017immersogeometric, casquero2018non}, negligible incompressibility errors were obtained at the Eulerian level. In \cite{casquero2018non}, it was also shown that the spurious change of solid volume was orders of magnitude lower than in other IB methods.

The capability of divergence-conforming B-splines to preserve the fluid volume inside a closed co-dimension one solid has not been studied yet. This is needed in a variety of FSI applications, such as studying the behavior of capsules and vesicles under flow. Both capsules and vesicles contain a viscous fluid in their interiors, but their walls are very different. A capsule wall is a thin polymeric sheet, has shear resistance, and is extensible. A vesicle wall is a phospholipid bilayer, has bending resistance, and is inextensible. Both capsules and vesicles can be engineered in laboratories and are often used in bioengineering applications, such as drug delivery \cite{lim2019biomedical}, encapsulation of hemoglobin for artificial blood \cite{li2005liposome}, and construction of cell-like bioreactors \cite{noireaux2004vesicle}. Vesicles are also present at the boundary of biological cells, playing a key role in intercellular communication and in pathological processes such as cancer and autoimmune diseases \cite{andaloussi2013extracellular, yanez2015biological}. Capsule and vesicle formulations are often combined to be used as numerical proxies for red blood cells \cite{freund2014numerical, lanotte2016red, mauer2018flow}. Developing discretizations of capsule and vesicle formulations as well as coupling these formulations with a fluid solver are active fields of research \cite{barthes2016motion, abreu2014fluid}. The main benchmark problem consists in replicating the motions of vesicles and capsules in Couette flow. The two principal motions are tank treading (TT) and tumbling (TU). In the TT motion, the wall and the inner fluid rotate resembling the tread of a tank. In the TU motion, the wall and the inner fluid undergo a flipping motion resembling a rigid body. A significant challenge of capsule and vesicle formulations is to accurately compute the high-order derivatives present in their formulations. Instead of using directly the formulae of differential geometry to evaluate the normal vector, the curvature, or the second derivatives of the curvature, various workarounds have been used such as spring-like discretizations for two-dimensional vesicles \cite{tsubota2006particle, tsubota2010effect}, $C^0$-continuous triangular meshes using trigonometric formulae for three-dimensional vesicles \cite{boedec20113d, biben2011three}, $C^0$-continuous triangular meshes using quadratic interpolation for three-dimensional vesicles \cite{yazdani2012three, zhao2011dynamics}, and  $C^0$-continuous triangular meshes for three-dimensional capsules \cite{charrier1989free,shrivastava1993large, eggleton1998large}. However, the higher inter-element continuity of splines opens the door to direct evaluation of the aforementioned quantities using the formulae of differential geometry.

% and at the boundary of cell-internal organelles

%Instead of using directly the formulae of differential geometry to find the curvature, the normal vector or the second derivatives of the curvature, various workarounds have been used to compute these quantities on triangular meshes such as trigonometrical formulae \cite{boedec20113d, biben2011three}, direct variation of the energy \cite{zhao2011dynamics}, and quadratic interpolation \cite{yazdani2012three, zhao2011dynamics}. However, the higher inter-element continuity of splines opens the door of computing the aforementioned quantities by directly evaluating the formulae of differential geometry. 

%As explained in \cite{}, Thus, a priori, the proposed algorithm is not sufficiently precise to calculate the curvature force for a given surface. However, thanks to the stiffness of the problem, the reverse problem of reconstruction of the shape by given values of the force can be solved with good precision by the present method.

%These models can be used to
%perform an inverse analysis of experimental data on flowing capsules, which leads to an evaluation
%of the membrane’smechanical properties. Another advantage of such models is that they compute
%quantities that cannot be measured, such as the stress level in the membrane, and thus allow the
%evaluation of the risk of damage.

In this work, the DCIB method is applied to FSI problems involving closed co-dimension one solids. The accuracy of the DCIB method preserving the fluid volume within closed co-dimension one solids is compared with respect to conventional IB methods \cite{bhalla2013unified, griffith2012volume} and IB methods tailored to have improved volume conservation \cite{peskin1993improved, bao2017immersed}. In order to discretize closed curves with up to $C^{p-1}$ inter-element continuity, instead of working with open knot vectors as it is done in conventional IGA, we work with B-splines of degree $p$ defined on periodic knot vectors. In order to discretize closed surfaces with at least $C^1$ inter-element continuity, we use analysis-suitable T-splines \cite{Li2012, wei2017truncated, casquero2017arbitrary, toshniwal2017smooth, casquero2019}. These spline-based discretizations of closed co-dimension one solids in conjunction with performing integration by parts enables the computation of the forces exerted by capsules and vesicles on the fluid by directly evaluating formulae of differential geometry. We consider a Heaviside function and its associated equation is coupled to the Navier-Stokes equations in order to consider inner and outer fluids with different viscosities, which is the most straightforward manner to trigger the transition from TT motion to TU motion in Couette flow. Several quantitative comparisons are included with respect to the boundary integral method (BIM) \cite{thiebaud2013rheology}, an IB method based on the lattice-Boltzmann method (LBM) \cite{shen2017interaction}, and IB methods based on finite differences \cite{doddi2008lateral, bao2017immersed}.

% , DaVeiga2013

The paper is outlined as follows. Section 2 defines the kinematic concepts that are needed to handle closed co-dimension one solids in the DCIB method. Section 3 sets forth the governing equations for closed co-dimension one solids that have inner and outer fluids with different viscosities. Section 4 describes the variational form of the FSI problem, which is the starting point to perform the discretization process. Section 5.1 presents the spline-based spatial discretization of the DCIB method for closed co-dimension one solids. Section 5.2 describes the fully-implicit time discretization based on the generalized-$\alpha$ method. Section 5.3 describes the block-iterative solution strategy used to solve the final system of nonlinear algebraic equations. Section 6 includes five numerical examples. The first example is a two-dimensional problem involving a closed curve with active behavior. This is a benchmark problem for studying how accurately IB methods preserve the area of fluid inside the closed curve. The performance of the DCIB method is compared with other IB methods. The second and third examples consider common benchmark problems for formulations of vesicles and capsules under flow, respectively. Quantities of interest, such as the TT inclination angle, the TU period, the Taylor deformation parameter, and shear stresses, are compared with other numerical methods. Taking advantage of the geometrical flexibility of the DCIB method, the fourth and fifth examples study vesicle dynamics in Taylor-Couette flow and capsule segregation in Hagen-Poiseuille flow, respectively. Conclusions are drawn in Section 7.

\section{Kinematics}

The IB method solves the (Eulerian) Navier-Stokes equations in both the fluid and solid domains with added (Lagrangian) source terms that depend on the position of the solid. In order to have a closed system of equations, the Navier-Stokes equations are coupled with the kinematic equation that relates the Lagrangian displacement of the solid with the Eulerian velocity of the Navier-Stokes equations.

Let $d = \{ 2, 3 \}$ and $(0,T)$ be the number of spatial dimensions and the time interval of interest, respectively. In the rest of this section, we define Lagrangian and Eulerian quantities that are needed to state the governing equations of the IB method for closed co-dimension one solids, with emphasis on two-dimensional vesicles and three-dimensional capsules.

\subsection{Lagrangian description}

%We consider solids whose kinematics are described by closed curves and closed surfaces in $d = 2, 3$, respectively. The solid occupies, at time $t$, the region $ \Gamma_t \subset \mathbb{R}^{d}$, which can be obtained as the image of a parametric domain $ \vec{\xi} \subset \mathbb{R}^{d-1}$ through a mapping $\vec{\varphi} : \vec{\xi}\times(0,T)\mapsto\Gamma_t$. A reference configuration is defined as $\vec X :\vec{\xi} \mapsto \mathbb{R}^{d}$.  The Lagrangian variable of the mathematical model is the Lagrangian displacement $\vec u : \vec{X} \times (0,T) \mapsto \mathbb{R}^d$, verifying $\vec{\varphi} =\vec X + \vec u$.

We consider solids whose kinematics are described by closed curves and closed surfaces in $d = 2, 3$, respectively. The solid occupies, at time $t$, the region $ \Gamma_t \subset \mathbb{R}^{d}$, which can be obtained as the image of a reference configuration $\Gamma_R \subset \mathbb{R}^{d}$ through the mapping $\vec{\varphi} : \Gamma_R \times(0,T)\mapsto\Gamma_t$. Let $\vec X \in \Gamma_R$ be a material particle. The Lagrangian unknown of the mathematical model is the Lagrangian displacement $\vec u : \Gamma_R \times (0,T) \mapsto \mathbb{R}^d$, verifying $\vec{\varphi} =\vec X + \vec u$. The deformed configuration, the reference configuration, and the Lagrangian displacement can be parametrized in space using $d-1$ parametric coordinates.

% We now introduce some basic concepts of differential geometry.

\subsubsection{Closed curve}

Let $\xi^L$ be the parametric coordinate. When working with two-dimensional vesicles, it is common to reparametrize the closed curve in terms of its arc length $s$, which is done taking into account that
\begin{equation}
ds =  \Biggr\vert \Biggr\vert \frac{{\rm d} \vec{\varphi}}{{\rm d} \xi^L} \Biggr\vert \Biggr\vert d \xi^L \text{,}
\end{equation}
\noindent where $||\cdot||$ denotes the length of a vector. Using the arc length as the parametric coordinate, the unit tangent vector to the closed curve is obtained by
\begin{equation}
\vec{t} =  \frac{{\rm d} \vec{\varphi}}{{\rm d} s} \text{.}
\end{equation}
The curvature of the closed curve is obtained by
% The positive sign in the above equation is a unit normal vector pointing toward the center of curvature
\begin{equation} \label{curvature}
	\mathcal{C} =  \text{det} \displaystyle{
	\begin{pmatrix}
		\displaystyle{ \frac{{\rm d} {\varphi}_x}{{\rm d} s} } & \displaystyle{ \frac{{\rm d} {\varphi}_y}{{\rm d} s} } \\
		\displaystyle{ \frac{{\rm d}^2 {\varphi}_x}{{\rm d} s^2} } & \displaystyle{ \frac{{\rm d}^2 {\varphi}_y}{{\rm d} s^2} } \\
	\end{pmatrix} }  \text{.}
\end{equation}
\noindent The sign of $\mathcal{C}$ depends on the orientation of the curve provided by the parametrization. We always use closed curves whose parametric coordinate moves counterclockwise. As a result, a circle has positive curvature. The unit outward normal vector to the closed curve is obtained by
\begin{equation} \label{nvector}
\vec{n} =  \displaystyle{ \left\lbrace \frac{{\rm d} {\varphi}_y}{{\rm d} s} , - \frac{{\rm d} {\varphi}_x}{{\rm d} s} \right\rbrace } \text{.}
\end{equation}
%

% | \mathcal{C} | = ||\frac{{\rm d}^2 \vec{\varphi}}{{\rm d} s^2}||

%\begin{equation} \label{nvector}
%\vec{n} =  \begin{cases}
%- \displaystyle{ \frac{{\rm d}^2 \vec{\varphi}}{{\rm d} s^2} \Biggr/ \Biggr\vert \Biggr\vert \frac{{\rm d}^2 \vec{\varphi}}{{\rm d} s^2} \Biggr\vert \Biggr\vert }  \quad \text{if} \quad \mathcal{C} > 0  \text{,} \\
%+ \displaystyle{ \frac{{\rm d}^2 \vec{\varphi}}{{\rm d} s^2} \Biggr/ \Biggr\vert \Biggr\vert \frac{{\rm d}^2 \vec{\varphi}}{{\rm d} s^2} \Biggr\vert \Biggr\vert } \quad \text{if} \quad \mathcal{C} < 0   \text{,}
%\end{cases}
%\end{equation}

\subsubsection{Closed surface}

Let $ \vec{\xi}^L = \{ \xi_1^L, \xi_2^L \}$ be the parametric coordinates. In the following, indices in Greek letters take the values ${1,2}$ and summation over repeated Greek indices is implied. Non-unit tangent vectors to the closed surface are obtained by
\begin{equation} \label{firsteq}
\vec{a}_{\alpha}=  \frac{\partial \vec{\varphi} }{\partial \xi_{\alpha}^L} \text{.}
\end{equation}
The unit outward normal vector to the closed surface is obtained by
\begin{equation} 
\vec{n} = \frac{\vec{a}_1\times\vec{a}_2}{||\vec{a}_1\times\vec{a}_2||} \text{,}
\end{equation}
where the parametric coordinates $\xi_1^L$ and $\xi_2^L$ have been chosen in such a way that $\vec{a}_1\times\vec{a}_2$ points in the outward direction. The covariant metric coefficients of the closed surface are defined as
\begin{equation} 
a_{\alpha\beta}=\vec{a}_{\alpha}\cdot\vec{a}_{\beta} \text{.}
\end{equation}
The contravariant metric coefficients can be computed as the inverse matrix of the covariant coefficients, i.e., $\left[ a^{\alpha\beta} \right] = \left[ a_{\alpha\beta} \right]^{-1}$. The contravariant metric coefficients are used to obtain the contravariant base vectors from the covariant base vectors, namely, $\vec{a}^{\alpha} = a^{\alpha\beta} \vec{a}_{\beta}$. The covariant curvature coefficients of the closed surface are defined as
\begin{equation} 
b_{\alpha\beta} = \frac{\partial \vec{a}_{\alpha}}{\partial \xi_{\beta}^L} \cdot \vec{n} \text{.}
\end{equation}
The Christoffel symbols are defined as
\begin{equation} \label{lasteq}
\Gamma_{\alpha \beta}^{\gamma} =  \frac{ \partial \vec{a}_{\alpha}}{\partial \xi_{\beta}^L} \cdot  \vec{a}^{\gamma} \text{.}
\end{equation}
The quantities in Eqs. \eqref{firsteq}-\eqref{lasteq} are defined with respect to the deformed configuration, but analogous quantities are defined with respect to the reference configuration and denoted by $\mathring{\vec{a}}_{\alpha}$, $\mathring{\vec{n}}$, $\mathring{a}_{\alpha\beta}$, $\mathring{a}^{\alpha\beta}$, $\mathring{\vec{a}}^{\alpha}$, $\mathring{b}_{\alpha\beta}$, and $\mathring{\Gamma}_{\alpha \beta}^{\gamma}$.

\subsection{Eulerian description}

Let $\Omega \subset \mathbb{R}^d$ be the time-independent region containing both the fluid and the solid and let $\vec x \in \Omega$ be a spatial position. For the sake of brevity and since it holds true in all the examples of this paper, we assume that the solid is fully immersed in the fluid, i.e., $ \Gamma_t \cap \partial \Omega = \emptyset  \; \forall t \in (0,T)$. The Eulerian variables of the mathematical model are the Eulerian velocity $\vec v:\Omega\times(0,T)\mapsto\mathbb{R}^d$, the pressure $p:\Omega\times(0,T)\mapsto\mathbb{R}$, and the Heaviside function $H:\Omega\times(0,T)\mapsto\mathbb{R}$ (the Heaviside function is only needed when fluids with different inner and outer viscosities are considered). The physical domain $\Omega$ and the Eulerian variables can be parametrized in space using $d$ parametric coordinates.

\section{Governing equations}

At the continuous level, the mathematical model of the IB method is equivalent to the boundary-fitted FSI formulation \cite{boffi2008hyper, boffi2011finite}. The strong form of the IB method for closed co-dimension one solids with different inner and outer viscosities is stated as follows: Given $\rho \in \mathbb{R}^+$, $\mu_{o} \in \mathbb{R}^+$, $\mu_{i} \in \mathbb{R}^+$, $\vec g_V \in \Omega\times(0,T)\mapsto\mathbb{R}^d$, $\vec v_0:\Omega\mapsto\mathbb{R}^d$, $\vec u_0: \Gamma_R \mapsto\mathbb{R}^d$, and $\vec v_B:\partial \Omega \times(0,T) \mapsto\mathbb{R}^d$, find $\vec v:\Omega\times(0,T)\mapsto\mathbb{R}^d$, $p:\Omega\times(0,T)\mapsto\mathbb{R}$, $H:\Omega\times(0,T)\mapsto\mathbb{R}$, and $\vec u: \Gamma_R \times(0,T) \mapsto \mathbb{R}^d$, such that,
\begin{align}
 & \rho \left( \frac{\partial \vec v}{\partial t} \bigg |_{\vec x} + \nabla_{\vec x} \cdot (\vec v  \otimes  \vec v) \right) = \nabla_{\vec x}\cdot (2 \mu \nabla_{\vec x}^{\rm sym}\vec v) & & \\ 
 & - \nabla_{\vec x} p + \vec g_V  + \displaystyle { \int_{\Gamma_t} \vec f   \delta (\vec x - \vec\varphi(\vec X,t)) \: {\rm d} \Gamma_t }  \quad & \text{in}      &  \quad\Omega\times(0,T)    \text{ ,} \label{b1}\\ 
 & \nabla_{\sp}\cdot \vec v               = 0 \quad & \text{in} & \quad\Omega\times(0,T)    \text{,}  \label{b2} \\
 & \frac{ \partial \vec u}{\partial t} \bigg |_{\vec X} = \displaystyle { \int_{\Omega} \vec v   \delta (\vec x - \vec\varphi(\vec X,t)) \: {\rm d} \Omega }  \quad & \text{in}      &  \quad \Gamma_t \times(0,T)       \text{,} \label{b3} \\
 & \Delta_{\vec x} H               = - \nabla_{\sp} \cdot \displaystyle { \int_{\Gamma_t}  \vec n  \delta (\vec x - \vec\varphi(\vec X,t)) \: {\rm d} \Gamma_t } \quad & \text{in} & \quad\Omega\times(0,T)    \text{,} \label{b4} \\
 & \mu = \mu_i H + \mu_o (1 - H)  \quad & \text{in} & \quad\Omega\times(0,T) \text{,} \label{b5} \\
 & \vec v =  \vec v_0 \quad  & \text{on} & \quad\Omega\times\{0\} \text{,} \label{b6} \\
 & \disp =  \vec u_0 \quad  & \text{on} & \quad \Gamma_0 \times \{0\} \text{,} \label{b7} \\
 & \vec v = \vec v_{B}& \text{on} & \quad\partial \Omega\times(0,T) \text{,} \label{b8} \\
 & H = 0 & \text{on} & \quad\partial \Omega\times(0,T) \text{,} \label{b9} 
\end{align}
where $\rho$ is the density, $\mu_{i}$ and $\mu_{o}$ are the inner and outer fluid viscosities, respectively, $\vec g_V$ is an external force per unit of volume acting on the system, $\nabla_{\sp}^{\rm sym} ( \cdot )$ is the symmetric gradient operator defined by $\nabla_{\sp}^{\rm sym}\vec v=(\nabla_{\sp} \vec v+\nabla_{\sp}\vec v^T)/2$, $\vec v_0$ is the initial velocity, $\vec u_0$ is the initial displacement, $\vec v_{B}$ is the boundary velocity, $\delta (\vec x - \vec\varphi(\vec X,t)) = \prod_{i=1}^{d} \delta (x_i - \varphi_i(\vec X,t))$ is the $d$-dimensional Dirac delta function, and $\vec f:\Gamma_R \times(0,T) \mapsto \mathbb{R}^d$ is the force exerted by the solid on the fluid. The notation $\displaystyle { \frac{\partial \vec v}{\partial t} \bigg |_{\vec x} }$ and $\displaystyle { \frac{ \partial \vec u}{\partial t} \bigg |_{\vec X} }$ indicates that the time derivative is taken holding $\vec x$ and $\vec X$ fixed, respectively.

Eqs. \eqref{b1}-\eqref{b3} represent the linear momemtum balance equation, the mass conservation equation, and the kinematic equation that relates the Lagrangian displacement with the Eulerian velocity, respectively. Eqs. \eqref{b4}-\eqref{b5} are added to the standard IB formulation when $\mu_{i} \neq \mu_{o}$. The derivation of Eqs. \eqref{b4}-\eqref{b5} can be found in \cite{tryggvason2001front}. For brevity and since it holds true in all the examples of this paper, we assume that the inner fluid, the outer fluid, and the solid have the same density $\rho$. Inner and outer fluids with different densities can be considered using the Heaviside function $H$ akin to the case $\mu_{i} \neq \mu_{o}$. A closed co-dimension one solid with different density than the fluid can be considered in an analogous manner as we considered co-dimension zero solids with different density than the fluid in \cite{casquero2018non} provided that a solid formulation with a consistent time evolution of its thickness is used. Eqs. \eqref{b6}-\eqref{b7} define the initial condition for the velocity and the displacement, respectively. Eqs. \eqref{b8}-\eqref{b9} define the boundary condition for the velocity and the Heaviside function, respectively. For brevity Dirichlet boundary conditions for the velocity are applied on the whole boundary in Sections 3, 4, and 5, but Neumann and periodic boundary conditions can be applied in the DCIB method following the standard procedures of variational methods as done in Section 6.

The expressions of $\vec{f}$ for two-dimensional vesicles and three-dimensional capsules are detailed below.

\subsection{Two-dimensional vesicles}

As derived in \cite{kaoui2008lateral}, the Lagrangian force per unit length exerted by the vesicle on the fluid is
\begin{equation} \label{fvesicle}
\vec f =  \left( \kappa \frac{{\rm d}^2 \mathcal{C}}{{\rm d} s^2} + \kappa \frac{\mathcal{C}^3}{2} - \mathcal{C} \zeta \right) \vec n + \frac{{\rm d} \zeta}{{\rm d} s} \vec t \text{.}
\end{equation}
\noindent where $\kappa$ is the bending rigidity of the vesicle, $\zeta$ is a Lagrange multiplier that enforces the vesicle deformations to be locally inextensible, and a null spontaneous curvature is considered. Note that Eq. \eqref{fvesicle} depends on the deformed configuration, but not the reference configuration. In order to circumvent numerical instabilities, we replace $\zeta$ by the following expression
\begin{equation} \label{zeta}
\zeta =  4 C_I \lambda \left(\lambda^2 - 1 \right) \text{,}
\end{equation}
\noindent where $C_I$ is the dilatation modulus and $\lambda = \vert \vert {\rm d} \vec{\varphi} (\xi^L, t) / {\rm d} \xi^L \vert \vert / \vert \vert {\rm d} \vec{\varphi} (\xi^L, 0) / {\rm d} \xi^L \vert \vert$ is the stretch ratio. Eq. \eqref{zeta} is equivalent to considering a capsule with perimeter strain-energy function or Helmholtz free energy given by
\begin{equation} \label{perimeterw}
W_I =  C_I \left(\lambda^2 - 1 \right)^2  \text{.}
\end{equation}
Eq. \eqref{perimeterw} is a two-dimensional analog to the dilatational component of the surface strain-energy function proposed in \cite{skalak1973strain}.

% dimensionless numbers

\subsection{Three-dimensional capsules}

As derived in \cite{lac2004spherical}, the Lagrangian force per unit area exerted by the capsule on the fluid is
\begin{equation} \label{capfirst}
\vec f = \left( \frac{\partial T^{\alpha \beta}}{\partial \xi^L_\alpha} +  \Gamma^{ \alpha}_{\alpha \lambda} T^{\lambda \beta} +  \Gamma^{ \beta}_{\alpha \lambda} T^{\alpha \lambda} \right) \vec a_{\beta} + T^{\alpha \beta} b_{\alpha \beta} \vec n \text{,}
\end{equation}
\noindent with
\begin{align}
T^{\alpha \beta}  & = \frac{2}{J_s} \frac{\partial W_s}{\partial I_1} \mathring{a}^{\alpha \beta} + 2 J_s \frac{\partial W_s}{\partial I_2} a^{\alpha \beta}  \text{,} \\
I_1  & =   \mathring{a}^{\alpha \beta} a_{\alpha \beta} - 2 \text{,} \\
I_2  & =   \text{det} (\mathring{a}^{\alpha \beta}) \text{det} (a_{\alpha \beta}) - 1  \text{,} \\
J_s  & =   \sqrt{I_2 + 1}  \text{,} \label{caplast}
\end{align}
\noindent where $T^{\alpha \beta}$ are the contravariant coefficients of the membrane forces in the deformed configuration, $W_s$ is the surface strain-energy function, $I_1$ and $I_2$ are the invariants of the deformation, and  $J_s$ is the ratio of deformed local surface area to reference local surface area. Note that Eq. \eqref{capfirst} depends on both the deformed and reference configurations. Eqs. \eqref{capfirst}-\eqref{caplast} define a membrane formulation, i.e., transverse shear and bending are neglected. This membrane formulation takes into account both geometric and material nonlinearities. Examples of constitutive equations are the neo-Hookean law and the Skalak law \cite{skalak1973strain} defined as
\begin{align}
W_s  & =   \frac{G_s}{2} \left( I_1 - 1 + \frac{1}{I_2 + 1} \right)  \text{,} \\
W_s  & =   \frac{G_s}{4} (I^2_1 + 2I_1 - 2I_2) + \frac{C_I}{4} (J_s^2 - 1)^2   \text{,}
\end{align}
\noindent respectively, where $G_s$ is the surface shear modulus. When $C_I \gg G_s$, the Skalak law leads to locally inextensible membranes, but it can be used to model other types of membranes when $C_I \simeq G_s$. The neo-Hookean and Skalak laws result in significantly different material responses under large deformations, e.g., under uniaxial extension, the neo-Hookean law is strain-softening whereas the Skalak law is strain-hardening \cite{barthes2002effect}.

\section{Variational formulation}

Given suitable trial solution spaces ($\mathcal{S}_{v}$, $\mathcal{S}_{p}$, $\mathcal{S}_{u}$, and $\mathcal{S}_{H}$) and weighting function spaces ($\mathcal{V}_{v}$, $\mathcal{V}_{p}$, $\mathcal{V}_{u}$, and $\mathcal{V}_{H}$), the variational form of the IB method is stated as follows: Find $\vec v \in \mathcal{S}_{v}$, $p \in \mathcal{S}_{p}$, $\vec u \in \mathcal{S}_{u}$, and $H \in \mathcal{S}_{H}$, such that,
\begin{equation}\label{wf}
B \left( (   \vec w, q, \vec s, M ), (   \vec v, p, \vec u, H )  \right)   -   L \left(  \vec w \right)  = 0  \quad \quad   \forall (\vec w, q, \vec s, M)  \in     \mathcal{V}_{v} \times  \mathcal{V}_{p}  \times \mathcal{V}_{u} \times \mathcal{V}_{H} \text{,}
\end{equation}

\noindent with

\begin{align} \label{bbb}
B \left( (   \vec w, q, \vec s, M ), (   \vec v, p, \vec u, H )  \right)  & = \left(\vec w, \rho \frac{\partial\vec v}{\partial t} \bigg |_{\vec x} \right)_{\Omega}  - \left(\nabla_{\sp} \vec w, \rho \vec v \otimes \vec v \right)_{\Omega} \nonumber \\[.2cm]
& - \left(\nabla_{\sp}\cdot\vec w, p\right)_{\Omega} + \left(\nabla_{\sp}\vec w,2\mu_o\nabla_{\sp}^{\rm sym}\vec v \right)_{\Omega}  \nonumber \\
& + \left(\nabla_{\sp}\vec w,2H (\mu_i - \mu_o)\nabla_{\sp}^{\rm sym}\vec v \right)_{\Omega} + \left( q ,\nabla_{\sp}\cdot\vec v \right)_{\Omega}   \nonumber \\
& - \left( \vec w \circ  \vec{\varphi} ,  \vec f   \right)_{\rc} + \left(  \vec s , \frac{ \partial \vec u}{\partial t} \bigg |_{\vec X} - \vec v \circ  \vec{\varphi} \right)_{\rc} \nonumber \\
& + \left(\nabla_{\sp} M, \nabla_{\sp} H \right)_{\Omega} + \left( \nabla_{\sp} M \circ  \vec{\varphi} ,  \vec n \right)_{\rc} \text{,} \\
L  \left(  \vec w \right) & = \left(  \vec w,  \vec g_V  \right)_{\Omega} \label{lll} \text{,}
\end{align} 

where $(\cdot,\cdot)_\Omega$ and $(\cdot,\cdot)_{\rc}$ denote the $L^2$ inner product over the domains $\Omega$ and $\rc$, respectively. As mathematically shown in \cite{boffi2008hyper, boffi2003finite}, the variational formulation of the IB method enables the elimination of the Dirac delta functions. Note that Eq. \eqref{b4} is not considered in \cite{boffi2008hyper, boffi2003finite}, but its Dirac delta function is eliminated in the same way.

\subsection{Two-dimensional vesicles}

In this case, $\vec f$ contains a term involving one factor with fourth-order derivatives (${\rm d}^2 \mathcal{C} / {\rm d} s^2$) and another factor with second-order derivatives ($\vec n$). Therefore, integration by parts in the variational form can be applied once to remove the fourth-order derivatives from the variational form, namely, 
\begin{equation} 
\displaystyle { \int_{\Gamma_t} \kappa \frac{{\rm d}^2 \mathcal{C}}{{\rm d} s^2} \vec n \cdot ( \vec w \circ  \vec{\varphi} )  \: {\rm d} s} =  \displaystyle { - \int_{\Gamma_t} \kappa \frac{{\rm d} \mathcal{C}}{{\rm d} s} \frac{{\rm d} \vec n}{{\rm d} s} \cdot ( \vec w \circ  \vec{\varphi} )   \: {\rm d} s - \int_{\Gamma_t} \kappa \frac{{\rm d} \mathcal{C}}{{\rm d} s} \vec n \cdot \frac{{\rm d} ( \vec w \circ  \vec{\varphi} )}{{\rm d} s}    \: {\rm d} s} \text{,}
\end{equation}
where we have assumed that $\kappa$ is a constant. As a result, a discretization of the closed curve with $C^{2}$ inter-element continuity is needed to directly compute the term $\left( \vec w \circ  \vec{\varphi} ,  \vec f   \right)_{\rc}$.

\subsection{Three-dimensional capsules}

In this case, $\vec f$ contains a term involving one factor with second-order derivatives ($b_{\alpha\beta}$) and another factor with first-order derivatives ($T^{\alpha \beta}$). Therefore, integration by parts cannot be used to decrease the order of these derivatives and a discretization of the closed surface with $C^{1}$ inter-element continuity is needed to directly compute the term $\left( \vec w \circ  \vec{\varphi} ,  \vec f   \right)_{\rc}$.

%The continuum-based model that has been proven to capture more accurately the dynamics of red blood cells consists in representing the cytoskeleton by the Skalak capsule and the lipid bilayer by a vesicle. 
%
%The Lagrangian force per unit area that the vesicle exerts on the fluid is the following:
%%
%\begin{equation}
%\vec f^{VE} = - E_b  \left(  2 \Delta_{LB}\kappa + (2\kappa + c_0) (2 \kappa^2 - 2\kappa_g - c_0 \kappa)   \right)  \vec n \text{,}
%\end{equation}
%%
%where $E_b$ is the bending modulus, $\kappa$ is the mean curvature, $\kappa_g$ is the Gaussian curvature, $c_0$ is the spontaneous curvature, and $\Delta_{LB}$ is the Laplace-Beltrami operator.

\section{Discretization}

This section describes the DCIB method for closed co-dimension one solids. The DCIB method discretizes the variational formulation defined in Eqs. \eqref{wf}-\eqref{lll}. Thus, devising smeared delta functions is not needed in the DCIB method. By contrast, IB methods that discretize the strong form defined in Eqs. \eqref{b1}-\eqref{b9} require smeared delta functions.

\subsection{Spatial discretization}

In addition to the difficulty of accurately imposing the incompressibility constraint in IB methods, an additional challenge is to accurately compute integrals in which both Eulerian and Lagrangian functions are involved. The spatial discretization of the DCIB method has been purposefully designed to tackle these two challenges.

\subsubsection{Closed curves}

We discretize closed curves using B-splines of degree $p$ with periodic knot vectors and global $C^{p-1}$ continuity. Let us start defining the mapping $\vec F^{L} : [0,1] \mapsto \Gamma_R$ using B-splines as follows
\begin{equation}\label{gm}
\vec F^{L} \left( \xi^L \right) = \sum_{C=1}^{n^{L}} \vec{P}^{L}_C \widehat{N}^{L}_C \left( \xi^L \right)\text{,} \quad \xi^L \in [0,1]  \text{,}
\end{equation}
where the superscript $L$ stands for ``Lagrangian", $\{\vec{P}^{L}_C\}_{C=1}^{n^L}$ are the control points that define the reference configuration of the solid, and $\{\widehat{N}^{L}_C\}_{C=1}^{n^L}$ are uni-variate B-spline basis functions.

Uni-variate B-spline basis functions are computed from a knot vector $\bold{\Xi}$ using the Cox-de Boor recursion formula \cite{piegl2012nurbs, 1003.000}. A knot vector is a finite non-decreasing sequence of real numbers $\bold{\Xi} = \{  {\xi}_{1} ,  {\xi}_{2}, ...,  {\xi}_{n^{L} + p + 1} \}$, where ${\xi}_{i}$ is the $i$-th knot, $p$ is the polynomial degree, and $n^{L}$ is the number of uni-variate B-spline basis functions. The parametric domain of the curve is defined by the interval $[{\xi}_{p+1}, {\xi}_{n^{L}+1}]$ (we impose ${\xi}_{p+1} = 0$ and ${\xi}_{n^{L}+1} = 1$). A knot span is the difference between two consecutive knots (${\Delta\xi}_{i} = {\xi}_{i+1} - {\xi}_{i}$). Inside nonzero knot spans, B-spline basis functions are polynomials of degree $p$. Since the knots can be repeated, we define a sequence of knots without repetitions (called breakpoints) $\boldsymbol{\eta} = \{  {\eta}_{1} ,  {\eta}_{2}, ...,  {\eta}_{m} \}$ and another sequence with the knot multiplicities $\boldsymbol{r} = \{  {r}_{1} ,  {r}_{2}, ...,  {r}_{m} \}$. Knot multiplicity controls the continuity of B-spline basis functions at breakpoints, namely, B-spline basis functions have $\alpha_{i} = p - {r}_{i}$ continuous derivatives at ${\eta}_i$.

\begin{figure}[h!]
\centering
\subfigure[B-spline basis]{\includegraphics[scale=0.52]{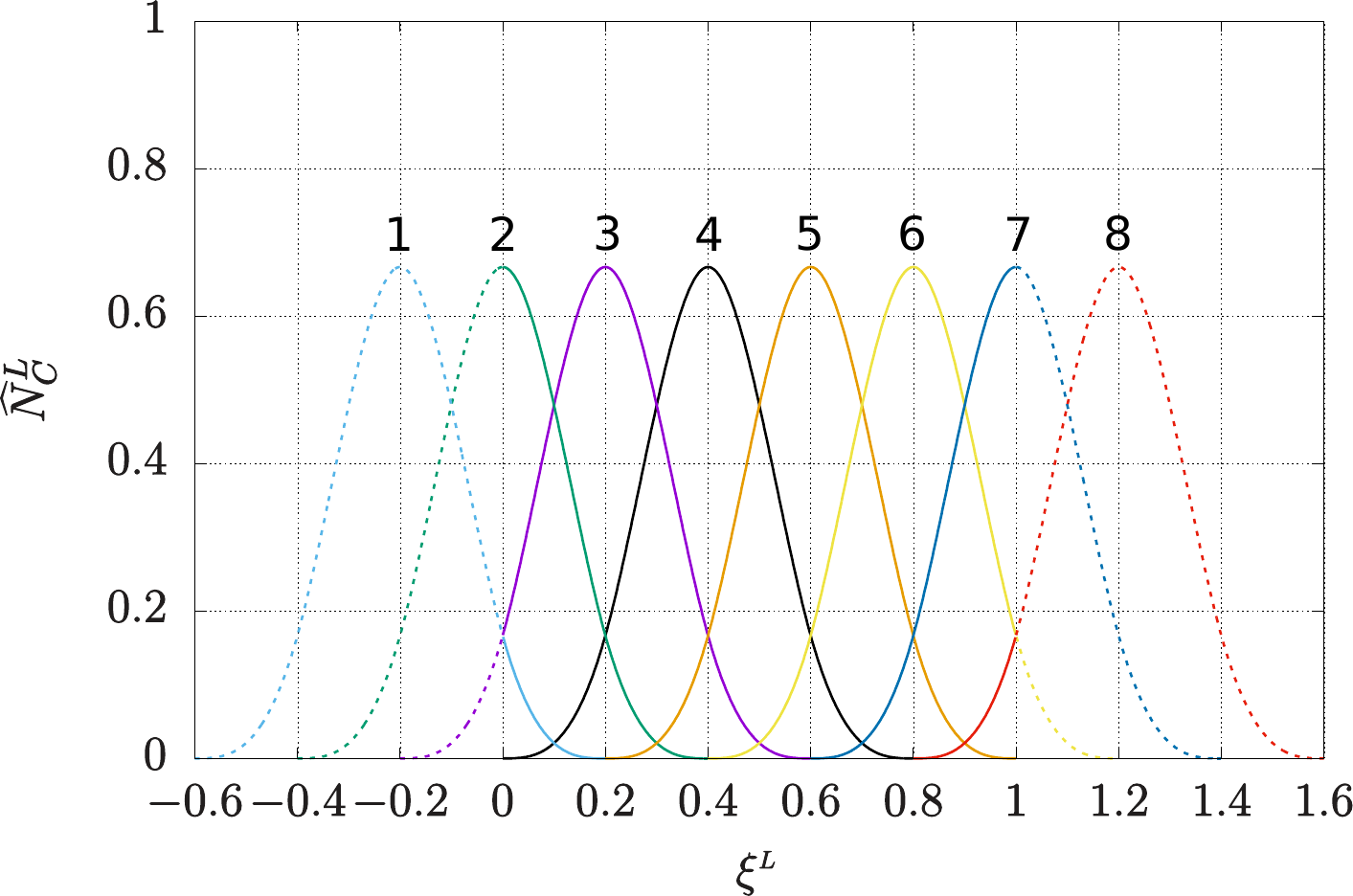}}
\caption{Given the periodic knot vector $\bold{\Xi} = \{  -0.6 ,  -0.4, -0.2,  0.0, 0.2, 0.4, 0.6, 0.8, 1.0, 1.2, 1.4, 1.6 \}$ and $p=3$, the B-spline basis functions $\{\widehat{N}^{L}_C\}_{C=1}^{8}$ are plotted.} \label{periodiccurve1}
\end{figure}

\begin{figure}[h!]
\centering
\subfigure[Open B-spline curve]{\includegraphics[scale=0.52]{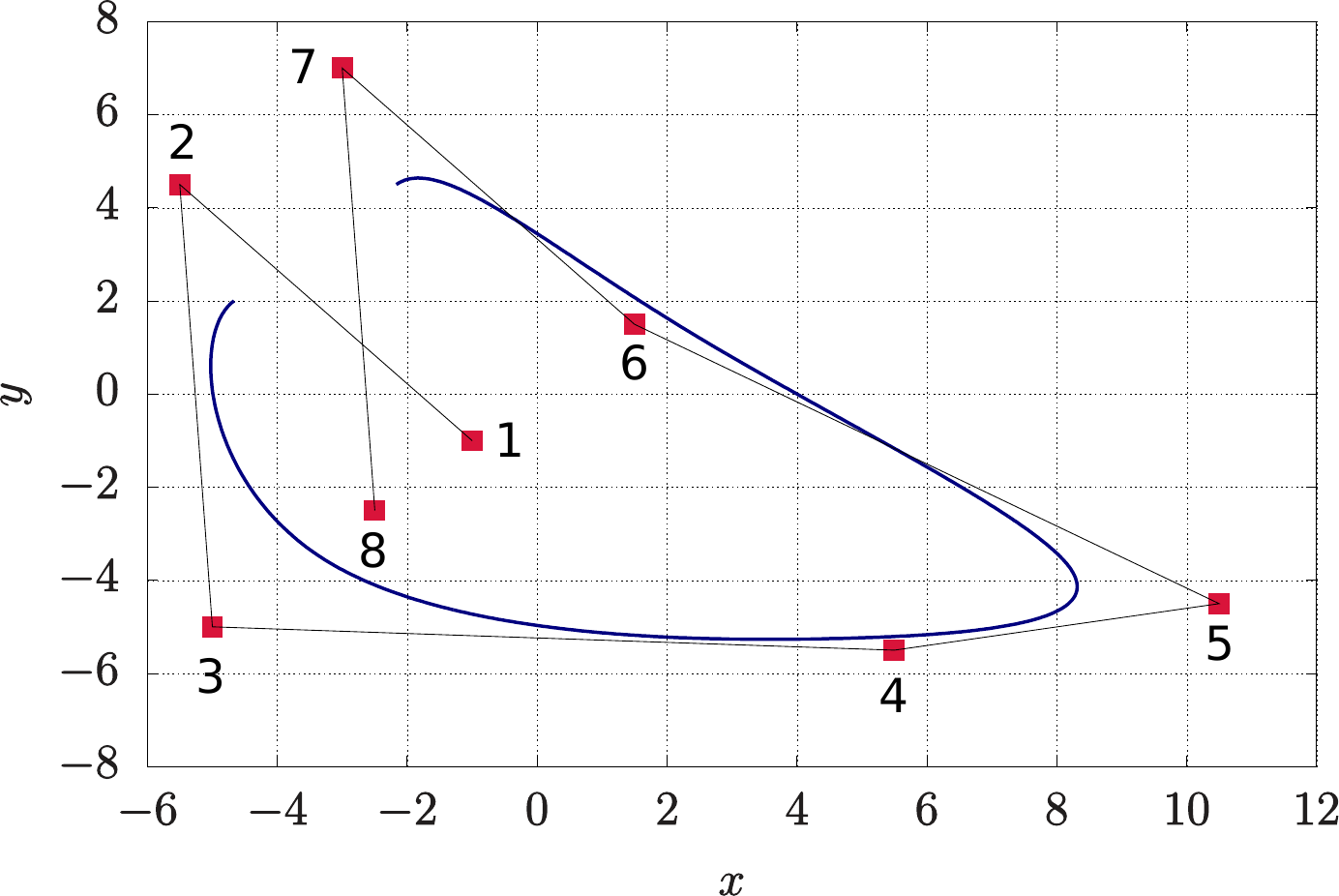}} \,
\subfigure[Closed B-spline curve]{\includegraphics[scale=0.52]{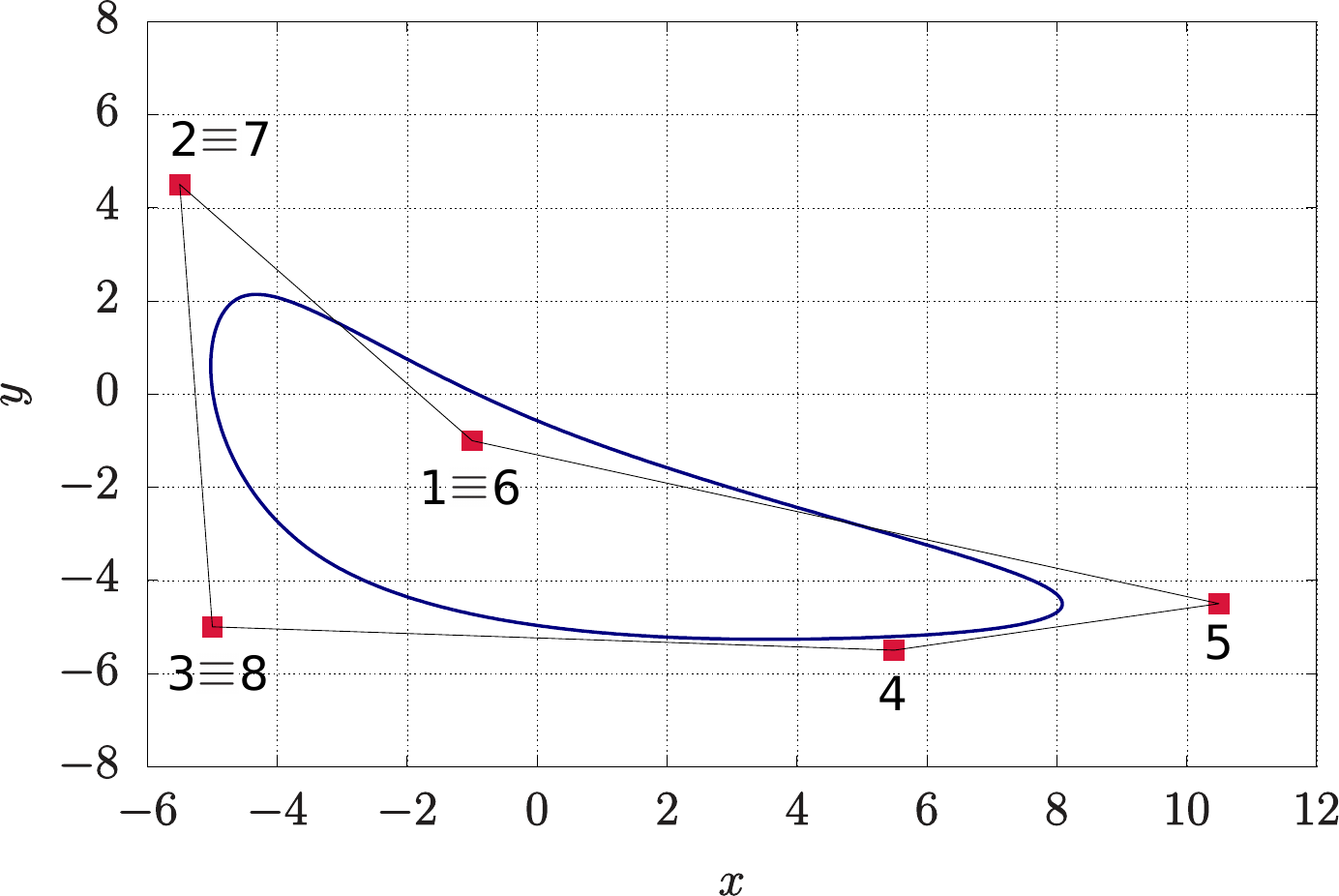}} \\
\caption{ Two B-spline curves are plotted using $\bold{\Xi} = \{  -0.6 ,  -0.4, -0.2,  0.0, 0.2, 0.4, 0.6, 0.8, 1.0, 1.2, 1.4, 1.6 \}$ and $p=3$. (a) An open curve in which all the control points have different coordinates. (b) A $C^{2}$-continuous closed curve obtained by wrapping the last 3 control points with the first 3 control points, i.e., $\vec{P}^{L}_{6} = \vec{P}^{L}_1$, $\vec{P}^{L}_{7} = \vec{P}^{L}_2$, and $\vec{P}^{L}_{8} = \vec{P}^{L}_3$. The control points are denoted by red squares and the B-spline curves are colored in blue.} \label{periodiccurve2}
\end{figure}

An open (or clamped) knot vector is obtained when $r_1 = r_m = p + 1$. Open knot vectors facilitate the strong imposition of Dirichlet boundary conditions, which has made open knot vectors the standard choice in IGA. However, when using open knot vectors, imposing periodic boundary conditions requires building a system of constraints. These constraints get more complicated as the continuity at the periodic boundary is increased (see \cite{liu2013isogeometric} for a description of how to impose $C^1$ periodicity with open knot vectors). An alternative is to build the periodicity into the B-spline space. This is accomplished using periodic (or uniform) knot vectors instead of open knot vectors. Periodic knot vectors have no repeated knots and the knots are evenly distributed. Reference \cite{Dalcin2016} builds the geometry using open knot vectors. After that, the knot vectors are unclamped, i.e., transformed into periodic knot vectors, which requires modifying the control points to preserve the geometry. This strategy is appropriate to impose $C^{p-1}$ periodicity in geometries that are not closed, e.g., imposing $C^{p-1}$ periodicity to the velocity and the pressure in a straight channel. However, when the geometry is closed and $C^{p-1}$ periodicity must be imposed to both the geometry and the unknowns of the problem, periodic knot vectors must be used from the beginning. This is the strategy that we follow here for the closed curve and the Lagrangian displacement. To obtain $C^{p-1}$ periodicity, the last $p$ control points must be wrapped with the first $p$ control points, namely, $\vec{P}^{L}_{n^{L}-p+1} = \vec{P}^{L}_1$, $\vec{P}^{L}_{n^{L}-p+2} = \vec{P}^{L}_2$, ..., $\vec{P}^{L}_{n^{L}} = \vec{P}^{L}_p$. An example of a B-spline closed curve is shown in Figs. \ref{periodiccurve1} and \ref{periodiccurve2}.

Invoking the isoparametric concept, the Lagrangian displacement is parameterized as follows
\begin{equation}\label{gm2}
\widehat{\vec u}^{ \, h^L} \left( \xi^L, t \right) = \sum_{C=1}^{n^{L}} \vec{u}_C (t) \widehat{N}^{L}_C \left( \xi^L \right)\text{,} \quad \xi^L \in [0,1]  \text{,}
\end{equation}
where $\vec u_C(t)$ are the control variables of the Lagrangian displacement. As a result, the deformed configuration is parameterized as follows
\begin{equation}\label{gm3}
\widehat{\vec{\varphi}}^{ \, h^L} \left( \xi^L, t \right) = \sum_{C=1}^{n^{L}} \vec{P}^{L}_C \widehat{N}^{L}_C \left( \xi^L \right) + \sum_{C=1}^{n^{L}} \vec{u}_C (t) \widehat{N}^{L}_C \left( \xi^L \right) \text{,} \quad \xi^L \in [0,1]  \text{.}
\end{equation}
The nonzero knot spans define the elements of a mesh $\mathcal{M}^{L}$ in $[0,1]$, this mesh is called the Lagrangian mesh from now on. The Lagrangian mesh can be pushed forward to the reference configuration and the deformed configuration using Eq. \eqref{gm} and Eq. \eqref{gm3}, respectively. All the integrals posed on the domain ${\Gamma}^{h^L}_t = \widehat{\vec{\varphi}}^{ \, h^L} \left( [0,1], t \right)$ are computed using Gauss quadrature in the elements of the Lagrangian mesh, namely, $p+1$ quadrature points are used per element. Using the Bubnov-Galerkin method, we obtain
\begin{equation}\label{gm4}
\widehat{\vec s}^{ \, h^L} \left( \xi^L \right) = \sum_{C=1}^{n^{L}} \vec{s}_C \widehat{N}^{L}_C \left( \xi^L \right)\text{,} \quad \xi^L \in [0,1]  \text{.}
\end{equation}

\subsubsection{Closed surfaces}

We discretize closed surfaces using bi-cubic analysis-suitable T-splines (AST-splines) with at least $C^1$ inter-element continuity. AST-splines are unstructured and eight extraordinary points with valence three are enough to build simple closed surfaces such as a sphere in the $``$soccer ball$"$ parameterization. A detailed explanation of how to construct the AST-spline surfaces that we use in this work can be found in \cite{toshniwal2017smooth}.

\subsubsection{Navier-Stokes equations}

The spatial discretization of the Eulerian velocity, the pressure, and their weighting functions is performed using divergence-conforming B-splines, which are smooth generalizations of Raviart-Thomas mixed finite elements.

Let us start by defining the mapping $\vec F^{E} : \widehat{\Omega} \mapsto \Omega^{h^E}$ using non-uniform rational B-splines (NURBS) as follows
\begin{equation}\label{gm5}
\vec F^{E} \left( \vec {\xi}^{E} \right) = \sum_{i=1}^{n^{E}} \vec{P}^{E}_i \frac{ w_i \widehat{N}^{E}_i \left( \vec {\xi}^{E} \right)}{ \sum_{j=1}^{n^{E}} w_j \widehat{N}^{E}_j \left( \vec {\xi}^{E} \right)}\text{,} \quad\vec\xi^{E} \in  \widehat{\Omega}  \text{,}
\end{equation}
where the superscript $E$ stands for ``Eulerian", $\widehat{\Omega} = [0,1]^d$, $\{\vec{P}^{E}_i\}_{i=1}^{n^E}$ are the control points that define the geometry of the physical domain, $\{w_i\}_{i=1}^{n^E}$ are the weights associated with the control points, and $\{\widehat{N}^{E}_i\}_{i=1}^{n^E}$ are a set of $d$-variate basis functions. The $d$-variate B-spline basis functions are obtained as the tensor product of uni-variate B-spline basis functions \cite{piegl2012nurbs, 1003.000}. NURBS basis functions reduce to B-spline basis functions when $w_i = 1 \; \forall i \in 1,2,...,n^{E}$ since B-spline basis functions form a partition of unity. The main reason to consider weights $w_i \neq 1$ is to exactly represent certain geometries such as a circular cylinder.

The nonzero knot spans of the $d$ knot vectors used to obtain the basis functions $\{\widehat{N}^{E}_i\}_{i=1}^{n^E}$ define the elements of a mesh $\mathcal{M}^{E}$ in $\widehat{\Omega}$; this mesh is called the Eulerian mesh from now on. The Eulerian mesh can be pushed forward to the physical domain $\Omega^{h^E}$ using Eq. \eqref{gm5}. All the integrals posed on the domain $\Omega^{h^E}$ are computed using Gauss quadrature in the elements of the Eulerian mesh.

The discrete trial and weighting function spaces for the velocity and the pressure are defined using the Piola and integral-preserving transformations, respectively, as follows
\begin{align} \label{divfirst}
\mathcal{S}^{h^E}_{v}  =  & \bigg \{ {\vec v}^{h^E} \: \: | \:  {\vec v}^{h^E}  \circ  \vec F^{E}  =  \frac{D \vec F^{E}  \widehat{ \vec v }^{ \, h^E}}{\text{det}(D \vec F^{E})} , \quad \widehat{ \vec v }^{ \, h^E} (\cdot,t)  \in  \widehat{VEL}^{h^E} , \nonumber \\
&   \widehat{\vec v}^{ \, h^E} \cdot \widehat{ \vec n }^{ \, h^{E}} =  \widehat{\vec v_{B}} \cdot \widehat{ \vec n }^{ \, h^{E}} \quad \text{on} \quad \partial \widehat{\Omega} \times(0,T) \bigg \} \text{,}  \\
\mathcal{S}^{h^E}_{p}  = & \bigg \{ p^{  h^E} \: \: | \: p^{  h^E} \circ  \vec F^{E}  =  \frac{\widehat{ p }^{ \, h^E}}{\text{det}(D \vec F^{E})} , \quad \widehat{ p }^{ \, h^E} (\cdot,t)  \in  \widehat{PRE}^{h^E} , \quad \int_{\widehat{\Omega}} \widehat{ p }^{ \, h^E} \: {\rm d} \widehat{\Omega}  = 0 \bigg \} \text{,}  \\
\mathcal{V}^{h^E}_{v}  = & \bigg \{  \vec w^{h^E} \: \: | \: {\vec w}^{h^E}  \circ  \vec F^{E}  =  \frac{D \vec F^{E}  \widehat{ \vec w }^{ h^E}}{\text{det}(D \vec F^{E})} , \quad \widehat{\vec w}^{h^E}(\cdot)  \in \widehat{VEL}^{h^E} , \nonumber \\
&   \widehat{\vec w}^{h^E} \cdot \widehat{ \vec n }^{ \, h^{E}} =  0 \quad \text{on} \quad \partial \widehat{\Omega} \times(0,T) \bigg \} \text{,} \\
\mathcal{V}^{h^E}_{p} = & \bigg \{  q^{h^E} \: \: | \: q^{ \, h^E} \circ  \vec F^{E}  =  \frac{\widehat{ q }^{ \, h^E}}{\text{det}(D \vec F^{E})} , \quad \widehat{q}^{ \, h^E}(\cdot)  \in  \widehat{PRE}^{h^E}  \bigg \} \text{,}
\end{align}
with
\begin{align}
\widehat{VEL}^{h^E} & = \begin{cases}
\mathcal{S}_{k, k-1}^{k + 1, k} (\mathcal{M}^{E}) \times \mathcal{S}_{k-1,k}^{k, k + 1} (\mathcal{M}^{E}) \quad \quad \text{if} \: d = 2  \text{,} \\
\mathcal{S}_{k, k-1, k-1}^{k + 1, k, k} (\mathcal{M}^{E}) \times \mathcal{S}_{k-1,k,k-1}^{k, k + 1, k} (\mathcal{M}^{E}) \times  \mathcal{S}_{k-1,k-1,k}^{k, k, k + 1} (\mathcal{M}^{E}) \quad \quad \text{if} \: d = 3  \text{,} \end{cases} \\
\widehat{PRE}^{h^E} & = \begin{cases}
\mathcal{S}_{k-1,k-1}^{k, k} (\mathcal{M}^{E}) \quad \quad \text{if} \: d = 2  \text{,} \\
\mathcal{S}_{k-1,k-1,k-1}^{k, k, k} (\mathcal{M}^{E}) \quad \quad \text{if} \: d = 3  \text{,}
\end{cases} \label{divlast}
\end{align} 
where $D \vec F^{E}$ is the gradient of the mapping $\vec F^{E}$, $\widehat{ \vec n }^{h^{E}}$ is the unit outward normal to $\widehat{\Omega}$, and $\mathcal{S}_{\alpha_1,\alpha_2, ..., ,\alpha_d}^{p_1, p_2, ..., p_d} (\mathcal{M}^{E})$ is the $d$-variate B-spline space of basis functions defined on $\mathcal{M}^{E}$ that, in direction $i$, have degree $p_i$ and $C^{\alpha_i}$ continuity at all interior knots. In Eqs. \eqref{divfirst}-\eqref{divlast}, we defined divergence-conforming B-spline spaces of maximal continuity since these will be the spaces used throughout the examples of this paper. Note that only the normal Dirichlet boundary condition has been imposed on the velocity discrete space, the tangential Dirichlet boundary conditions will be imposed weakly using Nitsche's method in the semi-discrete form. Let us define the basis functions $\{\widehat{N}^E_{v_l,A_l}\}_{A_l=1}^{n^{v_l}}$ and $\{\widehat{N}^E_{p,B}\}_{B=1}^{n^{p}}$ such that $\text{span} \{\widehat{N}^{E}_{v_1,A_1}(\vec \xi^{E})\}_{A_1=1}^{n^{v_1}} \times ... \times \text{span} \{\widehat{N}^{E}_{v_d,A_d}(\vec \xi^{E})\}_{A_d=1}^{n^{v_d}}  = \widehat{VEL}^{h^E}$ and $\text{span} \{\widehat{N}^{E}_{p,B}(\vec \xi^{E})\}_{B=1}^{n^{p}}  = \widehat{PRE}^{h^E}$, where $n^{v_l}$ and $n^{p}$ are the total number of degrees of freedom that the $l$-th component of the velocity and the pressure have, respectively.

The above choices of spaces are inf-sup stable and the discrete function spaces for the velocity and the pressure are $H^1$-conforming and $L^2$-conforming, respectively, for $k \geq 1$. In addition, imposing the discrete Eulerian velocity to be weakly divergence-free results in a solenoidal field, i.e.,
\begin{equation}
\left( q^{h^{E}} ,\nabla_{\sp}\cdot\vec v^{h^{E}} \right)_{\Omega^{h^{E}}} =   0  \quad  \forall q^{h^{E}} \in  \mathcal{V}_{p}^{h^{E}}   \implies   \nabla_{\sp}\cdot \vec{v}^{h^{E}}               = 0    \quad  \forall \vec{x} \in {\Omega}^{h^{E}}, \forall t \in [0,T] \text{,}
\end{equation}
as mathematically shown in \cite{evans2013isogeometric, john}. That is, weak incompressibility implies strong (i.e., pointwise) incompressibility for divergence-conforming B-splines.

\subsubsection{Heaviside function}

The spatial discretization of the Heaviside function and its weighting function is performed using B-splines. The discrete trial and weighting function spaces for the Heaviside function are defined as
\begin{align} 
\mathcal{S}^{h^E}_{H} & = \left\{ H^{  h^E} \: \: | \: H^{  h^E} \circ  \vec F^{E}  =  \widehat{ H }^{ \, h^E} , \quad \widehat{ H }^{ \, h^E} (\cdot,t)  \in  \widehat{HEA}^{h^E} , \;   \widehat{H}^{ \, h^E} =  0 \quad \text{on} \quad \partial \widehat{\Omega} \times(0,T)  \right\} \text{,}  \\
\mathcal{V}^{h^E}_{H} & = \left\{ M^{  h^E} \: \: | \: M^{  h^E} \circ  \vec F^{E}  =  \widehat{ M }^{ \, h^E} , \quad \widehat{ M }^{ \, h^E} (\cdot,t)  \in  \widehat{HEA}^{h^E} , \;   \widehat{M}^{ \, h^E} =  0 \quad \text{on} \quad \partial \widehat{\Omega} \times(0,T)  \right\} \text{,}
\end{align}
with
\begin{align}
\widehat{HEA}^{h^E} & = \begin{cases}
\mathcal{S}_{k-1,k-1}^{k, k} (\mathcal{M}^{E}) \quad \quad \text{if} \: d = 2  \text{,} \\
\mathcal{S}_{k-1,k-1,k-1}^{k, k, k} (\mathcal{M}^{E}) \quad \quad \text{if} \: d = 3  \text{.}
\end{cases} 
\end{align} 
Let us define the basis functions $\{\widehat{N}^E_{H,D}\}_{D=1}^{n^{H}}$ such that $\text{span} \{\widehat{N}^{E}_{H,D}(\vec \xi^{E})\}_{D=1}^{n^{H}}  = \widehat{HEA}^{h^E}$, where $n^H$ is the number of degrees of freedom that the discrete Heaviside function has. As mentioned in \cite{tryggvason2001front}, some overshoots and undershoots may take place near the interface. When the value of $H^{h^E}$ at a quadrature point is greater than one or less than zero, this value is substituted by one and zero, respectively.

\subsubsection{Semi-discrete form}

The semi-discrete form of the DCIB method for closed co-dimension one solids is stated as follows: Find $\vec v^{h^{E}} \in \mathcal{S}_{v}^{h^{E}}$, $p^{h^{E}} \in \mathcal{S}_{p}^{h^{E}}$, $\vec u^{h^{L}} \in \mathcal{S}_{u}^{h^{L}}$, and $H^{h^{E}} \in \mathcal{S}_{H}^{h^{E}}$, such that for all $\vec w^{h^{E}} \in \mathcal{V}_{v}^{h^{E}} $, $q^{h^{E}} \in \mathcal{V}_{p}^{h^{E}}$, $ \vec s^{h^{L}}  \in \mathcal{V}_{u}^{h^{L}}$, and $M^{h^{E}} \in \mathcal{V}_{H}^{h^{E}}$
\begin{equation}\label{sdf1}
B \left( (   \vec w^{h^{E}}, q^{h^{E}}, \vec s^{h^{L}}, M^{h^{E}}  ), (   \vec v^{h^{E}}, p^{h^{E}}, \vec u^{h^{L}}, H^{h^{E}} )  \right)   - b  \left(    \vec w^{h^{E}},    \vec v^{h^{E}}   \right) -   L \left(  \vec w^{h^{E}} \right)  + l \left(  \vec w^{h^{E}} \right)= 0   \text{,}
\end{equation}

\noindent with  

\begin{align} 
b  \left(    \vec w^{h^{E}},    \vec v^{h^{E}}   \right)  & =  \sum_{F \in \partial {\Omega}^{h^{E}}} \int_F 2 \mu_o (\vec w^{h^{E}})_{||} \cdot  (\nabla_{\sp}^{\rm sym} \vec v^{h^{E}} \vec n^{h^{E}}) d\partial {\Omega} \nonumber \\
& - \sum_{F \in \partial {\Omega}^{h^{E}}} \int_F  \rho (\vec w^{h^{E}})_{||} \cdot  \vec v^{h^{E}}  (\vec v_{B} \cdot \vec n^{h^{E}})_{+} d\partial {\Omega} \nonumber \\
&  +  \sum_{F \in \partial {\Omega}^{h^{E}}} \int_F 2 \mu_o  (\nabla_{\sp}^{\rm sym} \vec w^{h^{E}} \vec n^{h^{E}}) \cdot (\vec v^{h^{E}})_{||} d\partial {\Omega} \nonumber \\
& - \sum_{F \in \partial {\Omega}^{h^{E}}} \int_F 2 \mu_o \frac{C_{pen}}{h_F} (\vec w^{h^{E}})_{||} \cdot (\vec v^{h^{E}})_{||} d\partial {\Omega}  \text{,} \\ 
l \left(  \vec w^{h^{E}} \right)  & = \sum_{F \in \partial {\Omega}^{h^{E}}} \int_F  \rho (\vec w^{h^{E}})_{||} \cdot  \vec v_{B}  (\vec v_{B} \cdot \vec n^{h^{E}})_{-} d\partial {\Omega} \nonumber \\
& + \sum_{F \in \partial {\Omega}^{h^{E}}} \int_F 2 \mu_o  (\nabla_{\sp}^{\rm sym} \vec w^{h^{E}} \vec n^{h^{E}}) \cdot  (\vec v_{B})_{||}  d\partial {\Omega} \nonumber \\
& - \sum_{F \in \partial {\Omega}^{h^{E}}} \int_F 2 \mu_o \frac{C_{pen}}{h_F} (\vec w^{h^{E}})_{||}  \cdot (\vec v_{B})_{||}  d\partial {\Omega} \label{sdf2} \text{,}
\end{align} 

\noindent where the terms $b  \left(    \vec w^{h^{E}},    \vec v^{h^{E}}   \right)$ and $l \left(  \vec w^{h^{E}} \right)$ have been added to the semi-discrete form to weakly impose the tangential Dirichlet boundary conditions using Nitsche's method, $\vec n^{h^{E}}$ is the outward unit normal vector to $\Omega^{h^{E}}$, $ (\cdot)_{||} = (\cdot) - ((\cdot) \cdot \vec n^{h^{E}}) \vec n^{h^{E}}$ is the vector tangential component, $(\vec v_{B} \cdot \vec n^{h^{E}})_{+} = \vec v_{B} \cdot \vec n^{h^{E}}$ if $\vec v_{B} \cdot \vec n^{h^{E}} > 0$ and $0$ otherwise, $(\vec v_{B} \cdot \vec n^{h^{E}})_{-} = \vec v_{B} \cdot \vec n^{h^{E}}$ if $\vec v_{B} \cdot \vec n^{h^{E}} \leq 0$ and $0$ otherwise, $h_F$ is the mesh size in the direction normal to the face $F$, and $C_{pen}$ is the Nitsche's penalization parameter. All the numerical results of this paper use the value $C_{pen} = 5(k + 1)$ as proposed in \cite{evans2013isogeometric}.

Note that the integrals of IB methods posed on the solid domain include Eulerian functions such as $\vec{v}^{h^{E}}$, $\vec{w}^{h^{E}}$, $\vec{M}^{h^{E}}$, and their first derivatives. In order to evaluate these Eulerian functions at a quadrature point with given parametric coordinates  $\vec{\xi}_G^L$ in the Lagrangian mesh, we follow two steps. First, we compute the physical coordinates $\vec{x}_G \in \Omega^{h^{E}}$ associated with the parametric coordinates $\vec{\xi}_G^L$ as $\vec{x}_G = \widehat{\vec{\varphi}}^{h^L}(\vec{\xi}_G^L,t)$. Then, we compute the parametric coordinates $\vec{\xi}_G^E$ in the Eulerian mesh associated with $\vec{x}_G$ as $\vec{\xi}_G^E = (\vec F^{E})^{-1}(\vec{x}_G)$. The NURBS mapping $\vec F^{E}$ can be inverted analytically in many cases of practical interest and when that is not the case, the mapping is inverted solving a $d \times d$ system of nonlinear algebraic equations. Once $\vec{\xi}_G^E$ is known, Eulerian functions can be evaluated using standard procedures. Thus, no interpolation or approximation has been used to evaluate the Eulerian functions at the quadrature points of the Lagrangian mesh. However, the quadrature errors of the integrals posed on ${\Gamma}^{h^L}_t$ are larger than in standard variational methods since the element boundaries of the Eulerian mesh (where the Eulerian functions have reduced regularity) often intersect the interior of the integration regions being used (the elements of the Lagrangian mesh). As opposed to standard finite elements, where the inter-element continuity of $\vec{v}^{h^{E}}$, $\vec{w}^{h^{E}}$, $\vec{M}^{h^{E}}$ is $C^0$ (and its first derivatives have jumps)\footnote{In \cite{hesch2012continuum}, to avoid performing numerical integration of integrals that include Eulerian functions with jumps inside the integration regions, $\vec{v}^{h^{E}}$ and $\vec{w}^{h^{E}}$ are projected into the Lagrangian mesh by evaluating the Eulerian functions at the nodes of the Lagrangian mesh to use those values as nodal values of their Lagrangian counterparts. Then, the Lagrangian counterparts are used for computing the integrals. In this case, the quadrature error has been changed to an interpolation error arising from the computation of the Lagrangian counterparts of $\vec{v}^{h^{E}}$ and $\vec{w}^{h^{E}}$.}, divergence-conforming B-splines enable us to raise the inter-element continuity by increasing $k$ until the quadrature error is not the dominant discretization error. Throughout the examples of this paper, the elements of the Lagrangian mesh are chosen  to be slightly smaller than the elements of the Eulerian mesh.

\subsection{Time discretization}

\allowdisplaybreaks

The time integration chosen to compute the Lagrangian displacement from the Eulerian velocity has a significant impact on how accurately the fluid volume inside a closed co-dimension one solid is preserved. The DCIB method performs fully-implicit time discretization of Eqs. \eqref{sdf1}-\eqref{sdf2} based on the generalized-$\alpha$ method. Since the mathematical model of the IB method has only first-order time derivatives, we follow the generalized-$\alpha$ method for first-order problems developed in \cite{Jansen2000}, which results in second-order accuracy, optimal high frequency damping, and unconditional stability for linear problems.

Let us start dividing $[0,T]$ into a sequence of subintervals $(t_n,t_{n+1})$ with fixed time-step size $\Delta t=t_{n+1}-t_n$. We define the residual vectors

\begin{align}
& \vec R^{M} = \left\{ \vec R^{M}_l \right\}\text{,} \quad  \vec R^{M}_l = \left\{ R^{M}_{l,A_l} \right\}\text{,} \quad \vec R^I = \left\{ R^{I}_{B} \right\}\text{,} \nonumber \\
& \vec R^K = \left\{ \vec R^{K}_{l} \right\}\text{,} \quad \vec R^K_l = \left\{ R^{K}_{l,C} \right\} \text{,} \quad \vec R^H = \left\{ R^{H}_{D} \right\} \text{,}
\end{align}
with
\begin{align} 
 R^{M}_{l,A_l} & =  B \left( (   \vec w_{l,A_l}, 0, \vec 0, 0  ), (   \vec v^{h^{E}}, p^{h^{E}}, \vec u^{h^{L}}, H^{h^{E}} )  \right)   - b  \left(    \vec w_{l,A_l},    \vec v^{h^{E}}   \right) -   L \left( \vec w_{l,A_l} \right) + l \left(  \vec w_{l,A_l} \right) \text{,} \\
 R^I_B       & =  B \left( (   \vec 0, q_B, \vec 0, 0  ), (   \vec v^{h^{E}}, p^{h^{E}}, \vec u^{h^{L}}, H^{h^{E}} )  \right) - b  \left(    \vec 0,    \vec v^{h^{E}}   \right) -   L \left( \vec 0 \right) + l \left(  \vec 0 \right) \text{,} \\
 R^K_{l,C}   & =  B \left( (   \vec 0, 0, \vec s_{l,C}, 0  ), (   \vec v^{h^{E}}, p^{h^{E}}, \vec u^{h^{L}}, H^{h^{E}} )  \right) - b  \left(    \vec 0,    \vec v^{h^{E}}   \right) -   L \left( \vec 0 \right) + l \left(  \vec 0 \right) \text{,} \\
  R^H_D       & =  B \left( (   \vec 0, 0, \vec 0, M_D  ), (   \vec v^{h^{E}}, p^{h^{E}}, \vec u^{h^{L}}, H^{h^{E}} )  \right) - b  \left(    \vec 0,    \vec v^{h^{E}}   \right) -   L \left( \vec 0 \right) + l \left(  \vec 0 \right) \text{,} 
\end{align}
where $l$ is a dimension index which runs from 1 to $d$, $A_l\in\{1,\dots,n^{v_l}\}$, $B\in\{1,\dots,n^{p}\}$, $C\in\{1,\dots,n^L\}$, $D\in\{1,\dots,n^H\}$, $\vec w_{l,A_l} \circ  \vec F^{E}  =   D \vec F^{E}  \widehat{N}^E_{v_l,A_l} \vec e_l / \text{det}(D \vec F^{E}) $, $q_B \circ  \vec F^{E}  =   \widehat{N}^E_{p,B} / \text{det}(D \vec F^{E}) $, $\vec s_{l,C} \circ  \vec F^{L} = \widehat{N}^L_C \vec e_l$, $M_D \circ  \vec F^{E}  =   \widehat{N}^E_{H,D} $, and $\vec{e}_l$ is the $l$-th versor of the coordinate system.

Let us now define $\vec V_n$, $\vec P_n$, $\vec A_n$, $\overline{\vec U}_n$, $\overline{\vec V}_n$, and $\vec H_n$ as the global vectors of control variables of $\vec v^{h^{E}}(\cdot,t_n)$, $p^{h^{E}}(\cdot,t_n)$, $\frac{\partial\vec v^{h^{E}}}{\partial t}(\cdot,t_n)$, $\disp^{h^{L}} (\cdot,t_n)$, $\frac{\partial\disp^{h^{L}}}{\partial t}(\cdot,t_n)$, and $H^{h^{E}}(\cdot,t_n)$, respectively. Using this notation, the fully-discrete form of the DCIB method for closed co-dimension one solids is stated as follows: Given $\vec V_n$, $\vec A_n$, $\overline{\vec U}_n$, and $\overline{\vec V}_n$, find $\vec V_{n+1}$, $\vec A_{n+1}$, $\vec V_{n+\alpha_f}$,  $\vec A_{n+\alpha_m}$, $\vec P_{n+1}$, $\overline{\vec U}_{n+1}$, $\overline{\vec V}_{n+1}$, $\overline{\vec U}_{n+\alpha_f}$, $\overline{\vec V}_{n+\alpha_m}$, and , $\vec H_{n+1}$ such that
\begin{align}
& \vec R^M (\boldsymbol {V}_{n+\alpha_f}, \boldsymbol {A}_{n+\alpha_m}, \boldsymbol {P}_{n+1}, \overline{\vec U}_{n+\alpha_f}, \boldsymbol {H}_{n+1}) = 0 \text{,} &  \label{r1} \\
& \vec R^I (\boldsymbol {V}_{n+\alpha_f}) = 0 \text{,} &   \label{r2}\\
& \vec R^{K} ( \overline{\vec V}_{n+\alpha_m}, \boldsymbol {V}_{n+\alpha_f}) = 0 \text{,} &  \label{r3} \\
& \vec R^H (\overline{\vec U}_{n+\alpha_f}, \boldsymbol {H}_{n+1}) = 0 \text{,} &   \label{r4}\\
& \boldsymbol {V}_{n+\alpha_f}=\boldsymbol {V}_{n}+\alpha _{f} ( \boldsymbol {V}_{n+1}-\boldsymbol {V}_{n} ) \text{,} &   \label{alpha1} \\
& \boldsymbol {A}_{n+\alpha_m}=\boldsymbol {A}_{n}+\alpha _{m} ( \boldsymbol {A}_{n+1} - \boldsymbol {A}_{n} ) \text{,}  &   \\
& \overline{\vec U}_{n+\alpha_f} =\overline{\vec U}_{n}+\alpha _{f} ( \overline{\vec U}_{n+1}-\overline{\vec U}_{n} ) \text{,} &   \\ 
& \overline{\vec V}_{n+\alpha_m} =\overline{\vec V}_{n}+\alpha _{m} ( \overline{\vec V}_{n+1}-\overline{\vec V}_{n} ) \text{,} &   \label{c1} \\ 
& \boldsymbol {V}_{n+1} = \boldsymbol {V}_n + \Delta t ((1-\gamma) \boldsymbol {A}_n + \gamma \boldsymbol {A}_{n+1}) \text{,} &   \label{c2} \\ 
& \overline{\boldsymbol {U}}_{n+1} = \overline{\boldsymbol {U}}_n + \Delta t ((1-\gamma) \overline{\boldsymbol {V}}_n + \gamma \overline{\boldsymbol {V}}_{n+1}) \text{,} &   \label{c33} \\
& \alpha _{m}=\dfrac {1} {2}\left( \dfrac {3-\varrho _{\infty }} {1+\varrho _{\infty }}\right) \text{,} & \label{alm} \\
& \alpha _{f}=\gamma = \dfrac {1} {1+\varrho _{\infty }} \text{.} & \label{alf}
\end{align}
We use $\varrho_{\infty} = 1/2$ throughout the examples of this paper, which corresponds to a time discretization scheme with a balance between accuracy and robustness.

\subsection{Solution strategy}

The system of nonlinear algebraic equations defined by \eqref{r1}-\eqref{alf} is solved using the Newton-Raphson method in a block iterative approach \cite{Bazilevs2012}, namely, we build two tangent matrices: (1) a tangent matrix for $\vec R^{M}$, $\vec R^{I}$, and $\vec R^{H}$ in which the Lagrangian control variables $\overline{\vec U}_{n+\alpha_f}$ are considered to be constant and (2) a tangent matrix for $\vec R^{K}$ in which the Eulerian control variables $\boldsymbol {V}_{n+\alpha_f}$ are considered to be constant. A parallel implementation of the DCIB method has been built on top of PetIGA \cite{Dalcin2016}, PetIGA-MF \cite{sarmiento2017petiga, cortes2017scalable2, espath2016energy}, and PETSc  \cite{petsc-web-page}. As nonlinear solvers, Newton-Raphson methods with critical-point line search \cite{Brune2015} and without line search are used for the systems of equations associated with the Eulerian and Lagrangian unknows, respectively. The nonlinear convergence is checked separately for each residual vector ($\vec R^{M}$, $\vec R^{I}$, $\vec R^{K}$, and $\vec R^{H}$) and the nonlinear relative tolerance is set to $10^{-4}$. As linear solvers, flexible GMRES \cite{saad1993flexible} with the block-preconditioning strategy described in \cite{cortes2017scalable2} (we observe that using Block Jacobi instead of algebraic multigrid leads to lower computational times for the mesh resolutions used in this paper) and GMRES \cite{Saad1986} with incomplete LU preconditioner are used for the systems of equations associated with the Eulerian and Lagrangian unknows, respectively. A linear relative tolerance for $\vec R^{K}$ is set to $10^{-10}$. A linear absolute tolerance for the unpreconditionated residuals $\vec R^{M}$, $\vec R^{I}$, and $\vec R^{H}$ is set to $10^{-10}$ unless otherwise specified in the examples of this paper. 

\section{Numerical examples}

In the examples of this section, the accuracy with which the incompressibility constraint is satisfied at the discrete level is measured at the Eulerian and Lagrangian levels. The incompressibility error at the Eulerian level ($e_{DIV}$) is defined as the $L^2$-norm error of Eq. \eqref{b2}, i.e.,
\begin{equation}
e_{DIV} (t) =  \left( \displaystyle { \int_{\Omega^{h^{E}}} } (\nabla_{\sp}\cdot \vec v^{h^E})^2 \; \hbox{d}\Omega  \right)^{1/2}   \text{.}
\end{equation}
Note that using divergence-conforming B-splines implies that $e_{DIV}$ is exactly zero as long as the final system of algebraic equations is solved exactly. However, this is not feasible for large systems of equations which must be solved iteratively up to a certain tolerance. In practice, this is not a shortcoming. In Fig. 2 (c) of \cite{casquero2018non}, we showed that, even for moderate tolerances, divergence-conforming B-splines were able to decrease the value of $e_{DIV}$ several orders of magnitude in comparison with standard weakly divergence-free discretizations when applied to a benchmark problem for IB methods involving co-dimension zero solids.

In a three-dimensional setting, the fluid volume inside a closed co-dimension one solid is obtained applying Gauss's theorem as $V (t) =  \int_{\Gamma^{h^{L}}_t}  \frac{1}{3}  \vec{\varphi}^{h^{L}}(\xi_1^L, \xi_2^L,t) \cdot \vec{n}^{h^{L}}(\xi_1^L, \xi_2^L,t)  \sqrt{\text{det}(a_{\alpha\beta})}\hbox{d}\xi_1^L\hbox{d}\xi_2^L $ and the incompressibility error at the Lagrangian level ($e_{VC}$) is defined as the relative change of the fluid volume inside a closed co-dimension one solid, i.e., $e_{VC} (t) = \left| V (t) - V(0)\right| / V(0)$.

In a two-dimensional setting, the fluid area inside a closed co-dimension one solid is obtained applying Green's theorem as $A (t) =  \int_{\Gamma^{h^{L}}_t}  \frac{1}{2} ( \varphi_x^{h^{L}}(s,t) \frac{d\varphi_y^{h^{L}}}{ds} (s,t) -\varphi_y^{h^{L}}(s,t) \frac{d\varphi_x^{h^{L}}}{ds} (s,t) ) \hbox{d}s $ and the incompressibility error at the Lagrangian level ($e_{VC}$) is defined as the relative change of the fluid area inside a closed co-dimension one solid, i.e., $e_{VC} (t) = \left| A (t) - A(0)\right| / A(0)$.

\begin{figure}[h!]
\centering
\includegraphics[width=11cm]{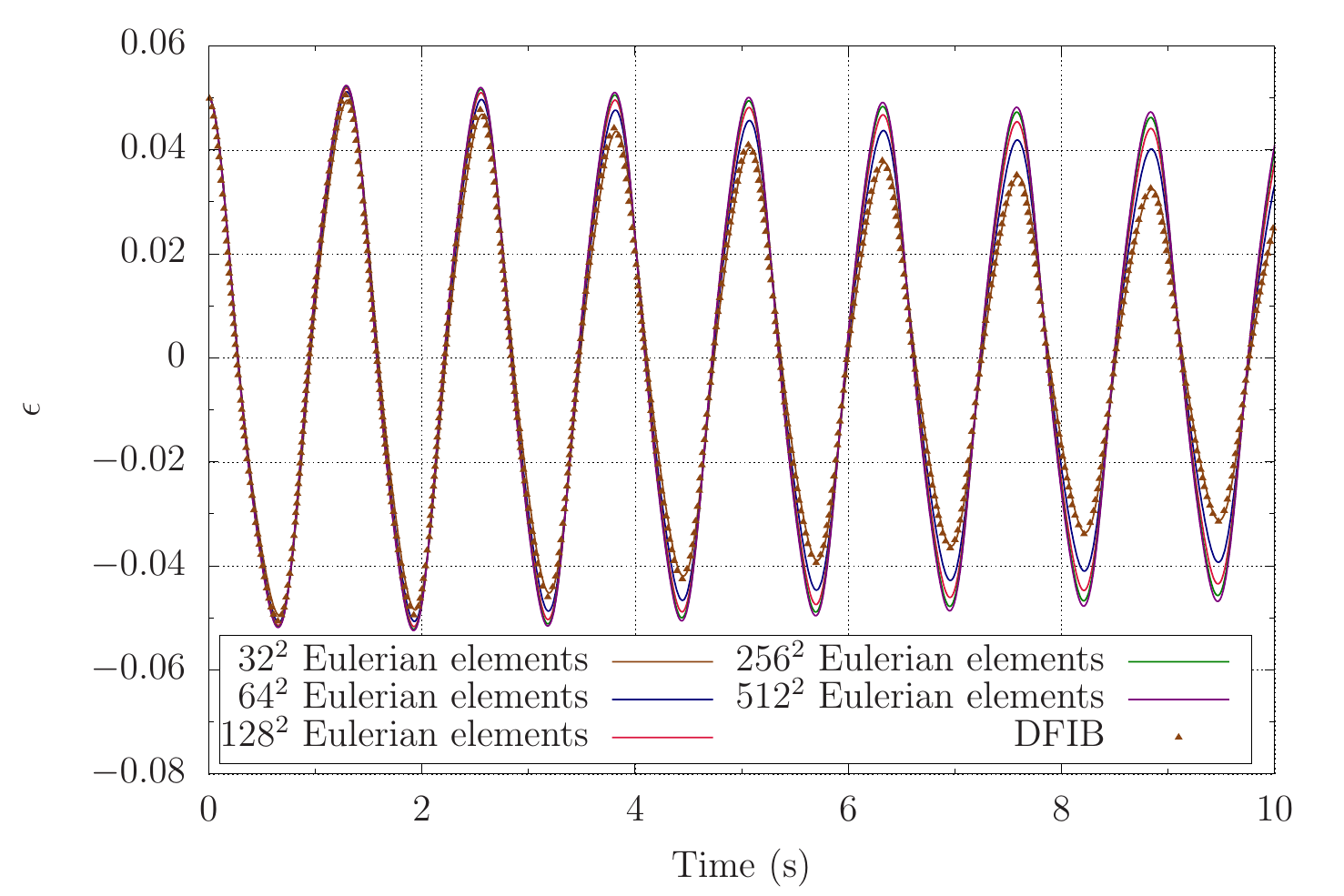}
\caption{Amplitude of the closed curve with active behavior. Mesh-independence study using the DCIB method. The results from the literature \cite{bao2017immersed} using $128^2$ Eulerian elements are also included.} 
\label{amplitude}
\end{figure}

The main dimensionless numbers in the study of vesicle and capsule dynamics under flow are the following:

\begin{itemize}

\item The viscosity contrast ($\Lambda$). It is the ratio of inner to outer viscosities, i.e., $\Lambda = \mu_i / \mu_o$.

\item The swelling degree ($\Delta$). In a three-dimensional setting, it is defined as the ratio of the vesicle/capsule volume ($V$) to the volume of a sphere with the vesicle/capsule external area ($A_e$), i.e., $\Delta = 6 \sqrt{\pi} V / \sqrt{A_e^3}$. In a two-dimensional setting, it is defined as the ratio of the vesicle/capsule area ($A$) to the area of a circle with the vesicle/capsule perimeter ($P$), i.e., $\Delta = 4 \pi A / P^2$. Note that the swelling degree is supposed to stay constant along the simulation for vesicles, but not for capsules. As a result, for capsules, it is more common to just mention its reference shape (e.g., spherical, elliptical, etc.) instead of using the swelling degree.

\item The confinement degree ($\chi$). It is the ratio of the effective radius ($R_0$) to the channel half-width ($W$), i.e., $\chi = R_0 / W$. In a three-dimensional setting, $R_0$ is defined as the radius of a sphere with the vesicle/capsule volume, i.e., $ R_0 = ( 3 V / 4 \pi)^{1/3}$. In a two-dimensional setting, $R_0$ is the radius of a circle with the vesicle/capsule area, i.e., $ R_0 = \sqrt{A/\pi}$.

\item The capillary number ($C_a$). It quantifies the relative strength of viscous forces to elastic forces. $C_a$ is defined as $C_a = \mu_o \dot{\gamma} R_0^3 / \kappa$ and $C_a = \mu_o \dot{\gamma} R_0 / G_s$ for vesicles and capsules, respectively, where $\dot{\gamma}$ is a representative shear rate of the flow.

\item The Reynolds number ($R_e$). It quantifies the relative strength of convective forces to viscous forces. $R_e$ is defined as $R_e = \rho \dot{\gamma} R_0^2 /  \mu_o $.

\end{itemize}

\subsection{Closed curve with active behavior}

We first consider a two-dimensional benchmark problem proposed in \cite{bao2017immersed} to study the accuracy with which the incompressibility constraint is imposed in immersed FSI methods involving closed co-dimension one solids. It consists of a closed curve that drives the motion of the surrounding fluid by exerting the following force on it
\begin{equation}
\vec f = \kappa(t)\frac{\partial^2  \widehat{\vec{\varphi}} }{\partial \theta^2}  \text{,}
\end{equation}
where
\begin{equation} \label{kappa}
\kappa(t) = 10 (1+0.8 \text{sin}(10 t))  \text{,}
\end{equation}
\begin{equation} \label{initialgeo}
\widehat{\vec{\varphi}}(\theta,0) = (1+ 0.05 \text{cos}(2 \theta)) (\text{cos} \theta, \text{sin} \theta) \text{,} \quad \theta \in [0,2\pi] \text{.}
\end{equation}
Eq. \eqref{kappa} defines a periodic time-dependent stiffness coefficient and  Eq. \eqref{initialgeo} defines the initial geometry of the solid, which is a unit circle with a small-amplitude radial perturbation. Note that the solid in this example is not parameterized in terms of the arc length $s$, but rather the polar angle $\theta$.

\begin{figure}[h!]
\centering
\subfigure[$t= 4.160156 \; \text{s}$]{\includegraphics[scale=0.07]{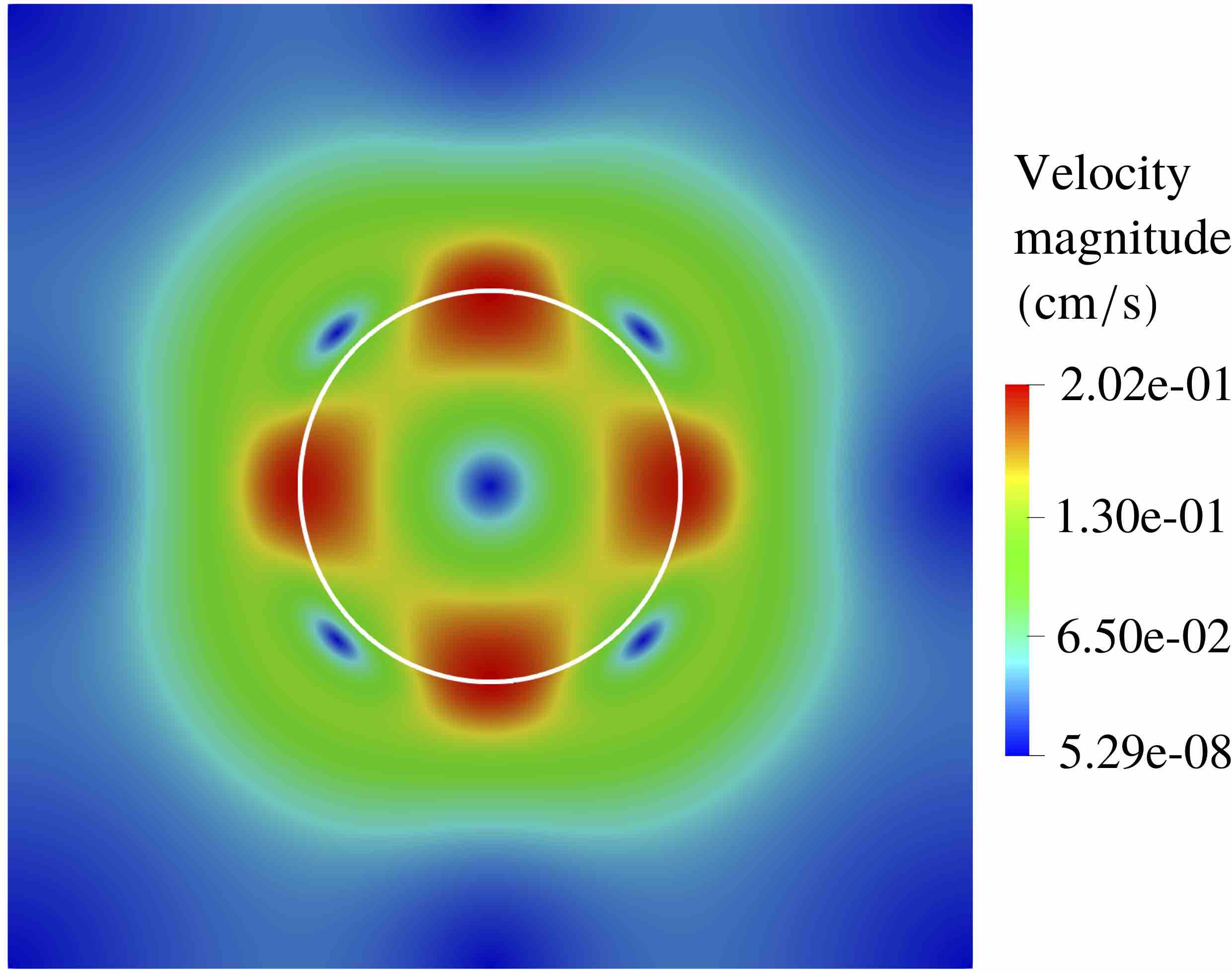}} \, \hspace*{+6mm}
\subfigure[$t= 4.386719 \; \text{s}$]{\includegraphics[scale=0.07]{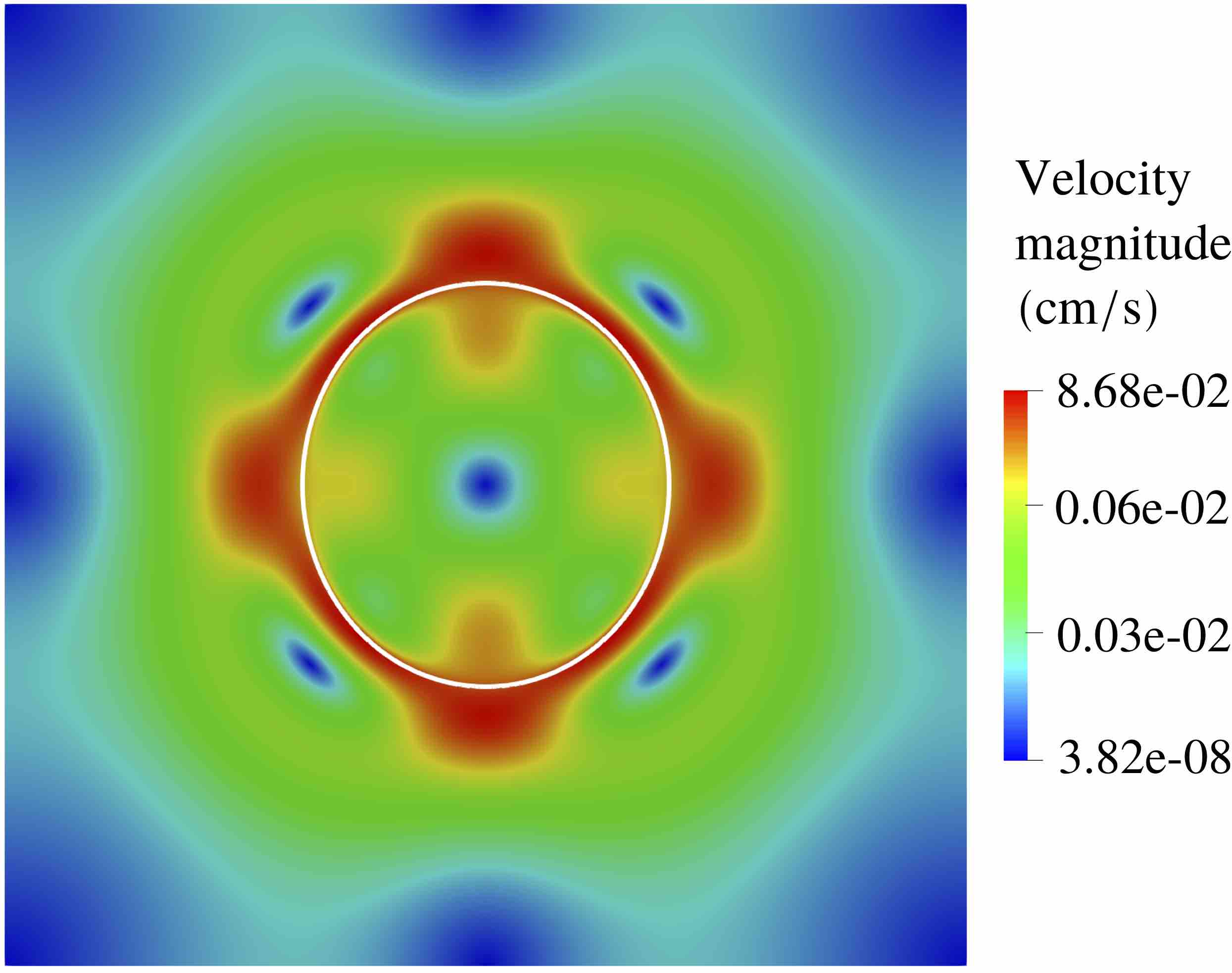}} \\
\subfigure[$t= 4.417969 \; \text{s}$]{\includegraphics[scale=0.07]{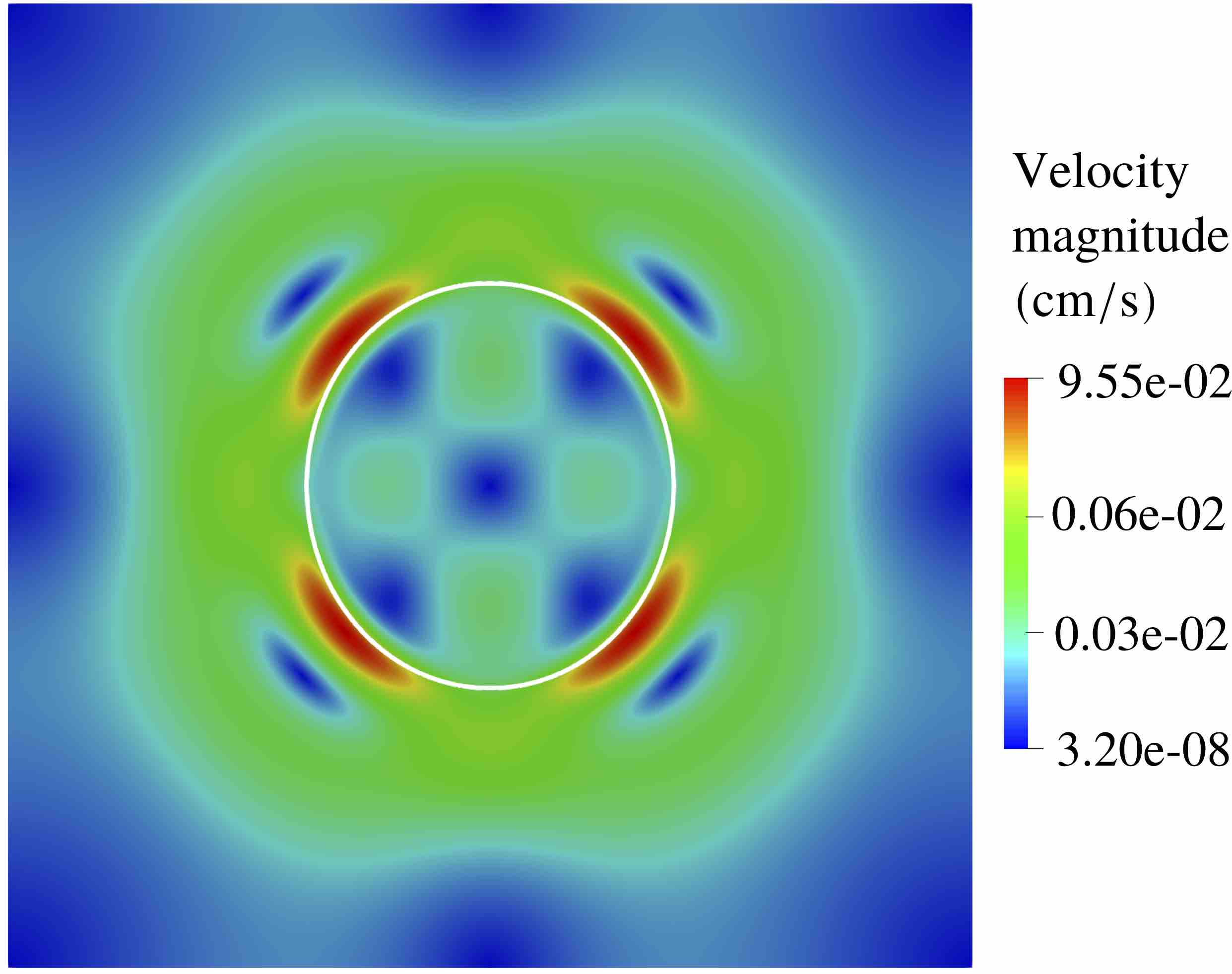}} \, \hspace*{+6mm}
\subfigure[$t= 4.460938 \; \text{s}$]{\includegraphics[scale=0.07]{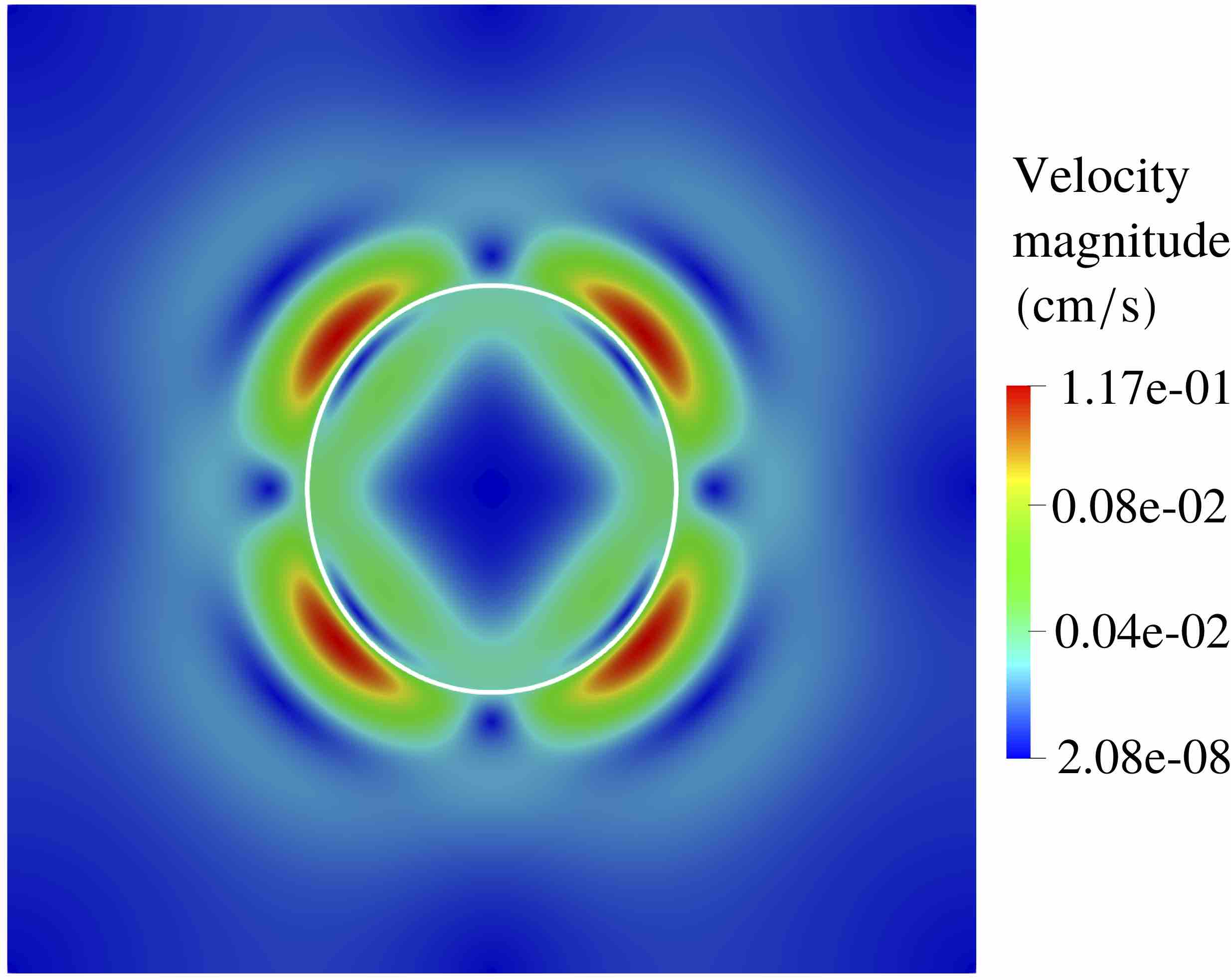}} \\
\caption{Closed curve with active behavior. (a)-(d) Velocity magnitude along with the deformed curve at four different times.} \label{closedcurve}
\end{figure}

The physical domain $\Omega$ is a square with side $L = 5 \; \textrm{cm} $. The solid is located at the center of the square. Periodic boundary conditions are imposed on all sides of the square. Both the fluid and the solid are initially at rest. The remaining physical parameters that define this problem are the following: $\rho = 1.0  \; \textrm{g} / \textrm{cm}^3$, $\mu_o = \mu_i = 0.15 \; \textrm{g} / (\textrm{cm} \cdot \textrm{s})$, and $\vec g_V = \vec 0$.

% This is a typical situation encountered in mechanobiology, where materials re-arrange its internal microstructure in response to external non-mechanical stimuli. 

\begin{figure}[h!]
\centering
\subfigure[$e_{VC}$ for different IB methods]{\includegraphics[scale=0.53]{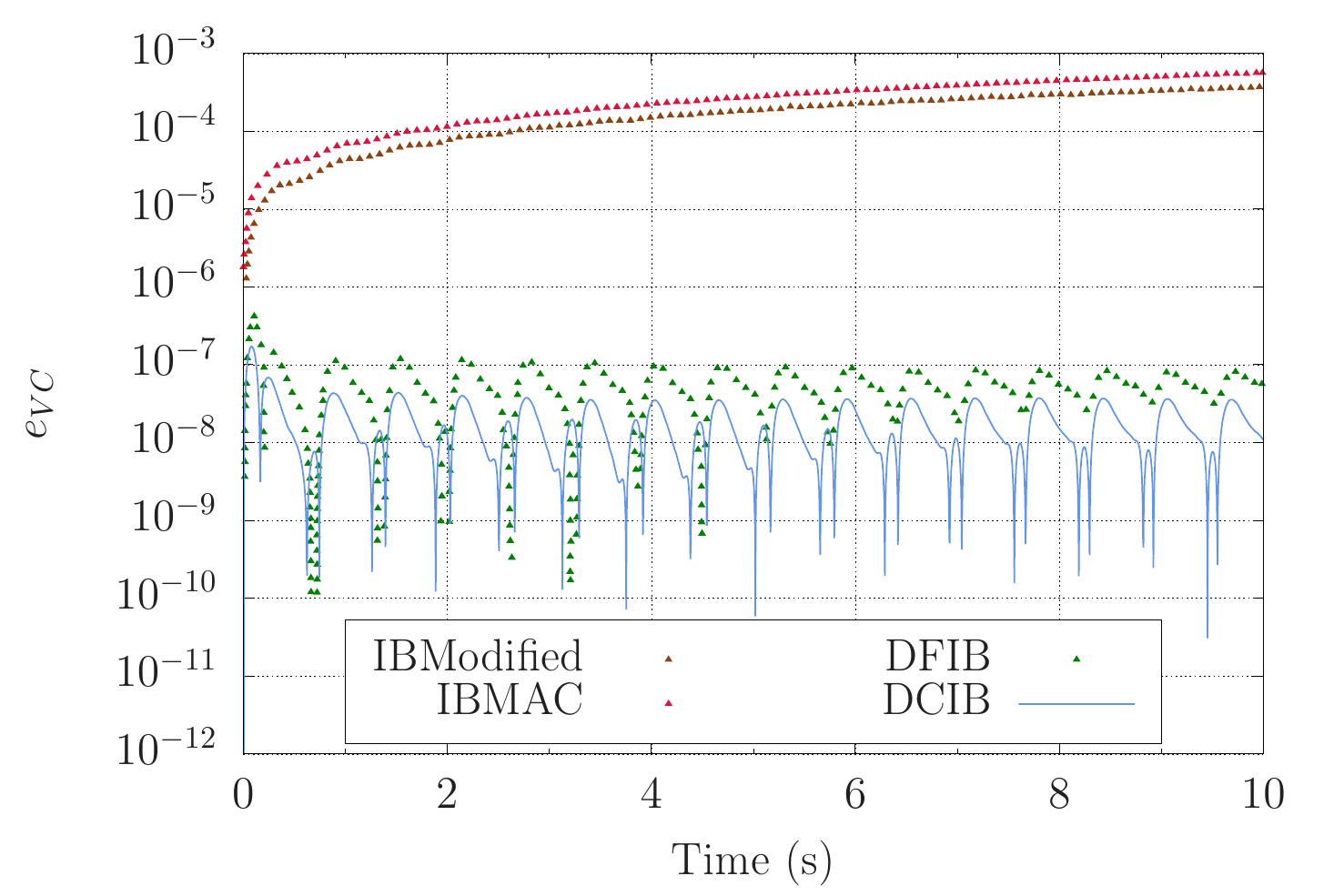}} \, \hspace*{-6mm}
\subfigure[Convergence of $e_{VC}$]{\includegraphics[scale=0.53]{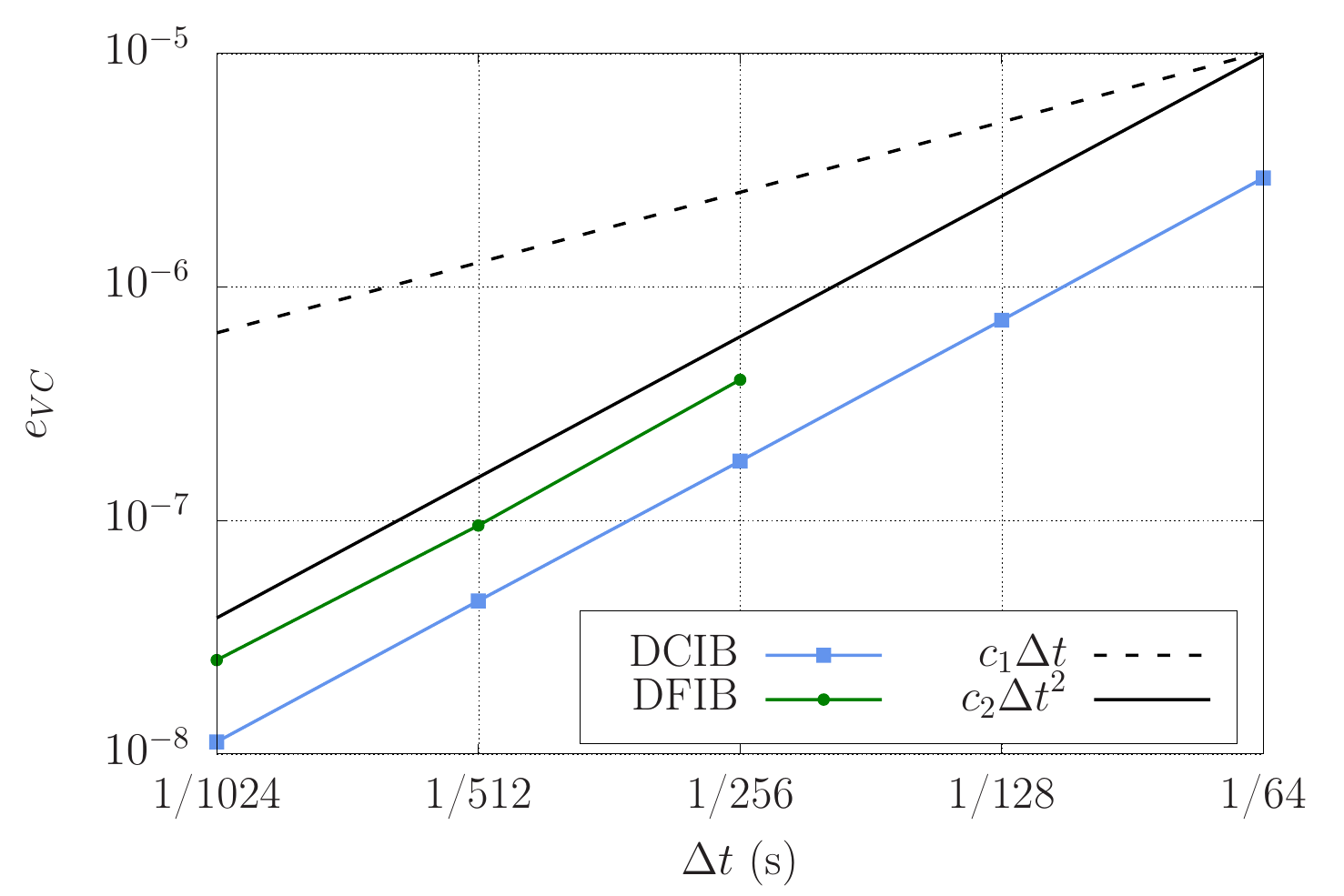}} \\
\subfigure[$e_{DIV}$ for different values of $a_{tol}$]{\includegraphics[scale=0.53]{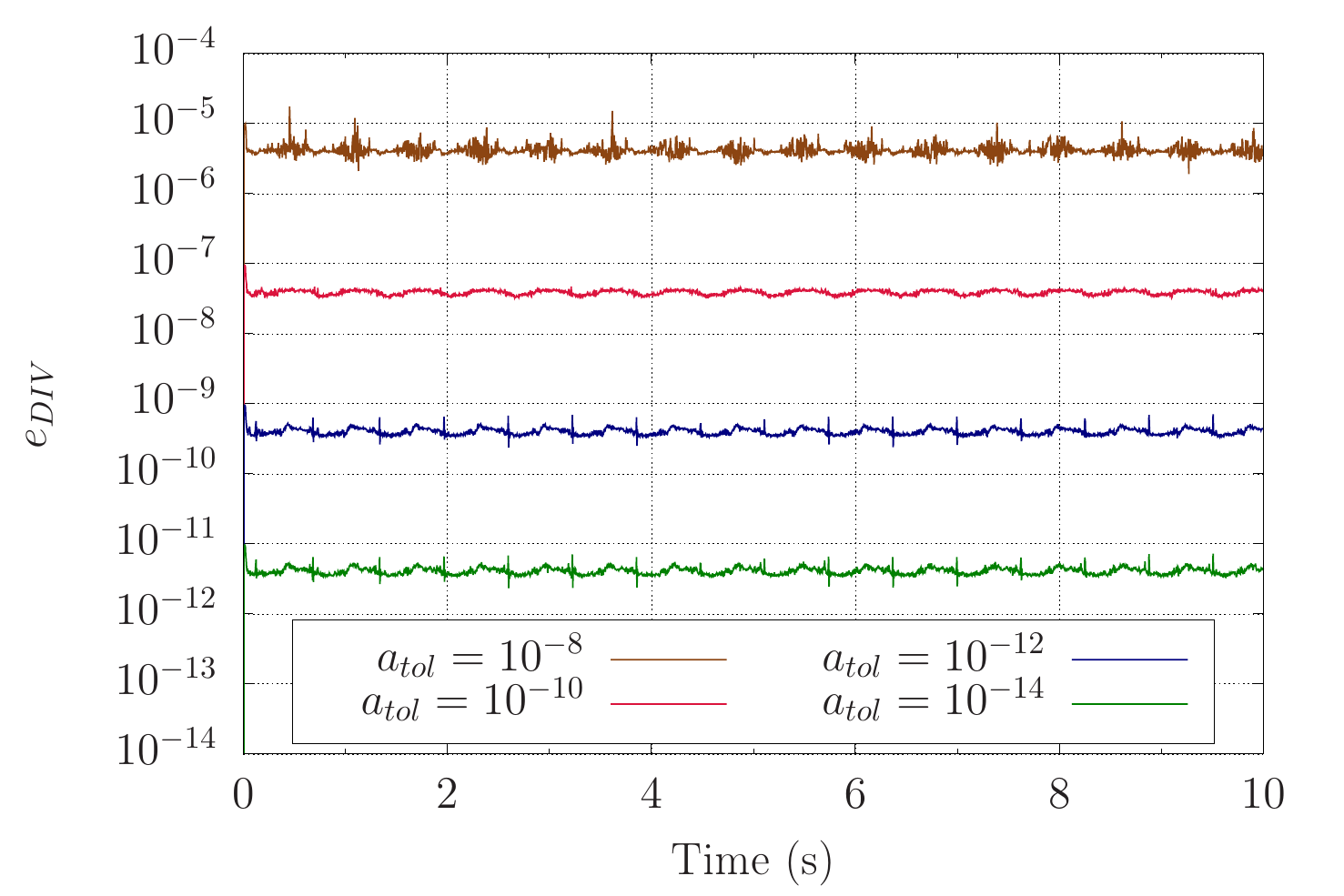}} \, \hspace*{-6mm}
\subfigure[$e_{VC}$ for different values of $a_{tol}$]{\includegraphics[scale=0.53]{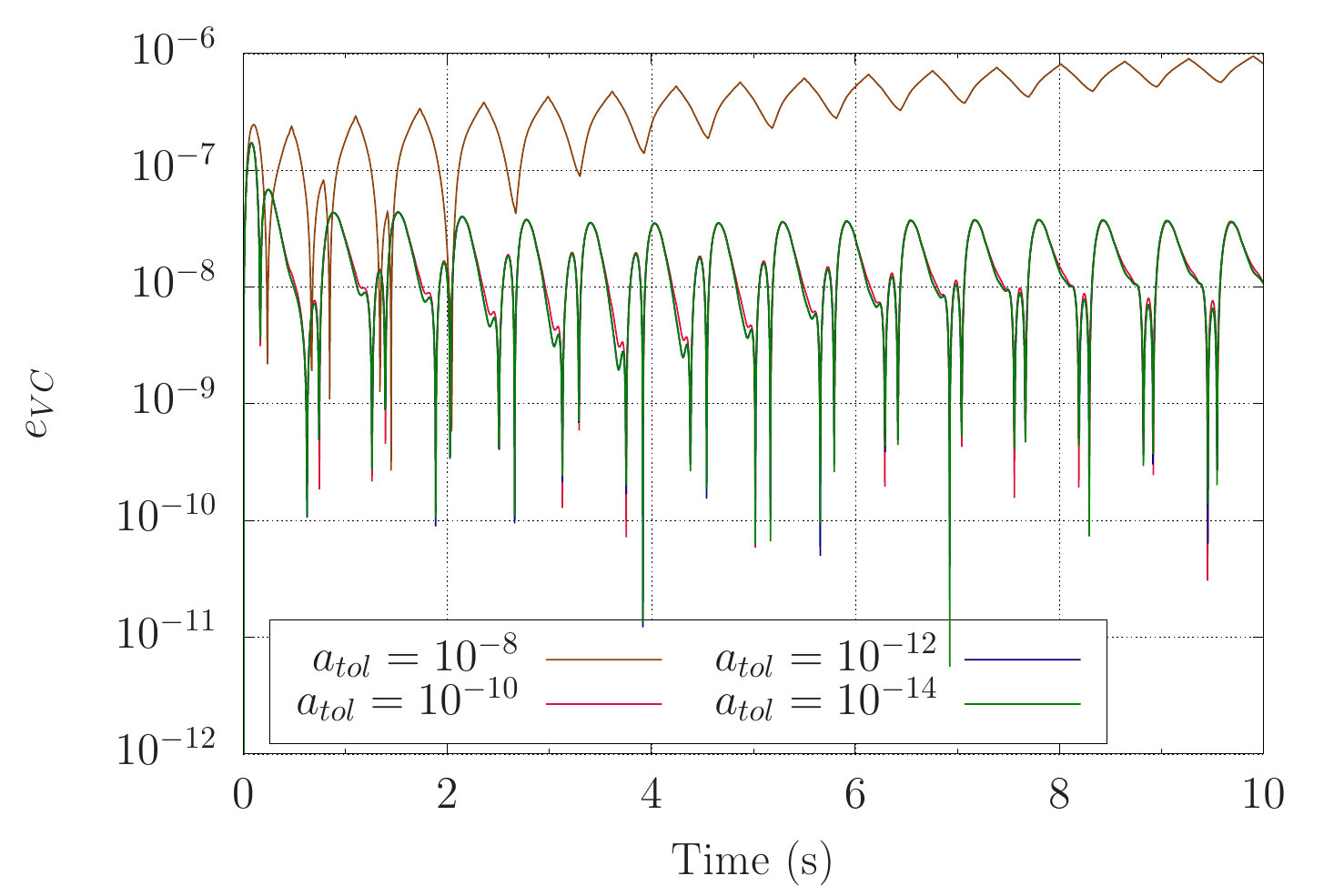}} \\
\caption{ Incompressibility errors of the closed curve with active behavior. (a) Incompressibility error at the Lagrangian level for the DCIB method and the IB methods considered in \cite{bao2017immersed} using the same element sizes and time step in all methods. (b) Convergence study as we refine the discretization of the kinematic equation while keeping the Eulerian discretization fixed. Eulerian meshes with $32 \times 32$ elements and $128 \times 128$ elements are used for the DCIB and DFIB methods, respectively. (c)-(d) Incompressibility errors at the Eulerian and Lagrangian levels for the DCIB method as the absolute tolerance for solving the Navier-Stokes equations is decreased.} \label{close}
\end{figure}

For the parameter values chosen here, the solid undergoes damped oscillations. A study of this problem for different parameter values can be found in \cite{cortez2004resonance, cortez2004resonance2}. As in \cite{bao2017immersed}, we assume the following ansatz for the position of the solid
\begin{equation} \label{geosolid}
\widehat{\vec{\varphi}}(\theta,t) = (1+ \epsilon (t) \text{cos}(2 \theta)) (\text{cos} \theta, \text{sin} \theta) 
\end{equation}
and compute the time-dependent amplitude $\epsilon (t)$ applying a fast Fourier transform. In Fig. \ref{amplitude}, we perform a mesh-independence study. We start with a coarse discretization, namely, $32 \times 32$ Eulerian elements with $k=2$, $82$ Lagrangian elements with $p=2$, and time step $\Delta t = 1.5625 \text{e--}3 \; \textrm{s}$. After that, the discretization is refined by performing uniform $h$-refinement three times on the Lagrangian and Eulerian meshes and dividing the time step by two each time a new level of refinement is introduced. A converged result is obtained as we increase the resolution. In \cite{bao2017immersed}, $\epsilon (t)$ is computed for a fixed discretization, namely, $128 \times 128$ Eulerian elements, $328$ Lagrangian elements, and time step $\Delta t = 3.90625 \text{e--} 4 \; \textrm{s}$; this resolution coincides with our resolution after two refinements. The result from \cite{bao2017immersed} is included in Fig. \ref{amplitude}. Figs. \ref{closedcurve} (a)-(d) show the rather complex velocity patterns created by the perturbed circle with periodic stiffness. 

% where $\vec{\hat{r}}(s)$ is the unit vector of the radial coordinate in a polar coordinate system with origin located at the center of the circle. 

%We first choose equidistant lagrangian points along the parametric curve and we calculate their position in physical space using radial coordinates. This process is straightforward if the parametrization is chosen so that the membrane is parametrized along the polar angle $\theta$ so that $s=\theta$. Then we calculate the 1D Fast Fourier Transform (FFT) of the difference between the measured radius of the lagrangian points and the original radius of the circle along the the parametrization angle $\theta$. Finally the amplitude is obtained by extracting the data of the FFT at the right frequency according to the choice of parameter $p$.

The incompressibility test established in \cite{bao2017immersed} consists in measuring $e_{VC}$ during the time interval $t \in [0.0 \; \textrm{s}, 10.0 \; \textrm{s}]$ using $128 \times 128$ Eulerian elements, $328$ Lagrangian elements, and time step $\Delta t = 3.90625 \text{e--}4 \; \textrm{s}$. With this discretization, $h^L \approx h^E/2$ and $\Delta t = h^E/10$. In  \cite{bao2017immersed}, the test is solved using three IB methods based on finite differences, namely, the DFIB method proposed in \cite{bao2017immersed}, the IBmodified method proposed in \cite{peskin1993improved} and the IBMAC method proposed in \cite{griffith2012volume, griffith2012immersed}. These results are included in Fig. \ref{close} (a) together with the results obtained using the DCIB method with $k=p=2$. The DCIB method is more than three orders of magnitude more accurate than the IBMAC and IBModified methods. The DCIB method is more than two times more accurate than the DFIB method. Furthermore, the DFIB method can only handle periodic boundary conditions as explained by the authors in \cite{bao2017immersed} while the DCIB method can handle Dirichlet and Neumann boundary conditions as well. Therefore, the DCIB method is as flexible as a conventional IB method, e.g., IBMAC method, and it is able to impose the incompressibility constraint accurately at the same time.

In a conventional IB method, the incompressibility error at the Lagrangian level does not decrease if the Lagrangian discretization and the time discretization are refined  for a fixed Eulerian discretization \cite{peskin1993improved}. In \cite{bao2017immersed}, it was shown that the DFIB method is able to overcome this limitation. The authors used a fixed Eulerian mesh with $128 \times 128$ elements and showed second-order convergence of $e_{VC}$ as the time discretization and the Lagrangian discretization were refined three times. The results from \cite{bao2017immersed} are plotted in Fig. \ref{close} (b). We now pick a coarse Eulerian mesh with $32 \times 32$ elements and the time and Lagrangian discretizations are refined five times. Since divergence-conforming B-splines lead to negligible incompressibility errors at the Eulerian level as long as the final linear system of equations is solved accurately (no matter how coarse the Eulerian mesh is), the DCIB method is able to decrease $e_{VC}$ with second-order convergence as shown in Fig. \ref{close} (b), thus overcoming the limitation of conventional IB methods as well.

\begin{figure}[h!]
\centering
\includegraphics[width=10cm]{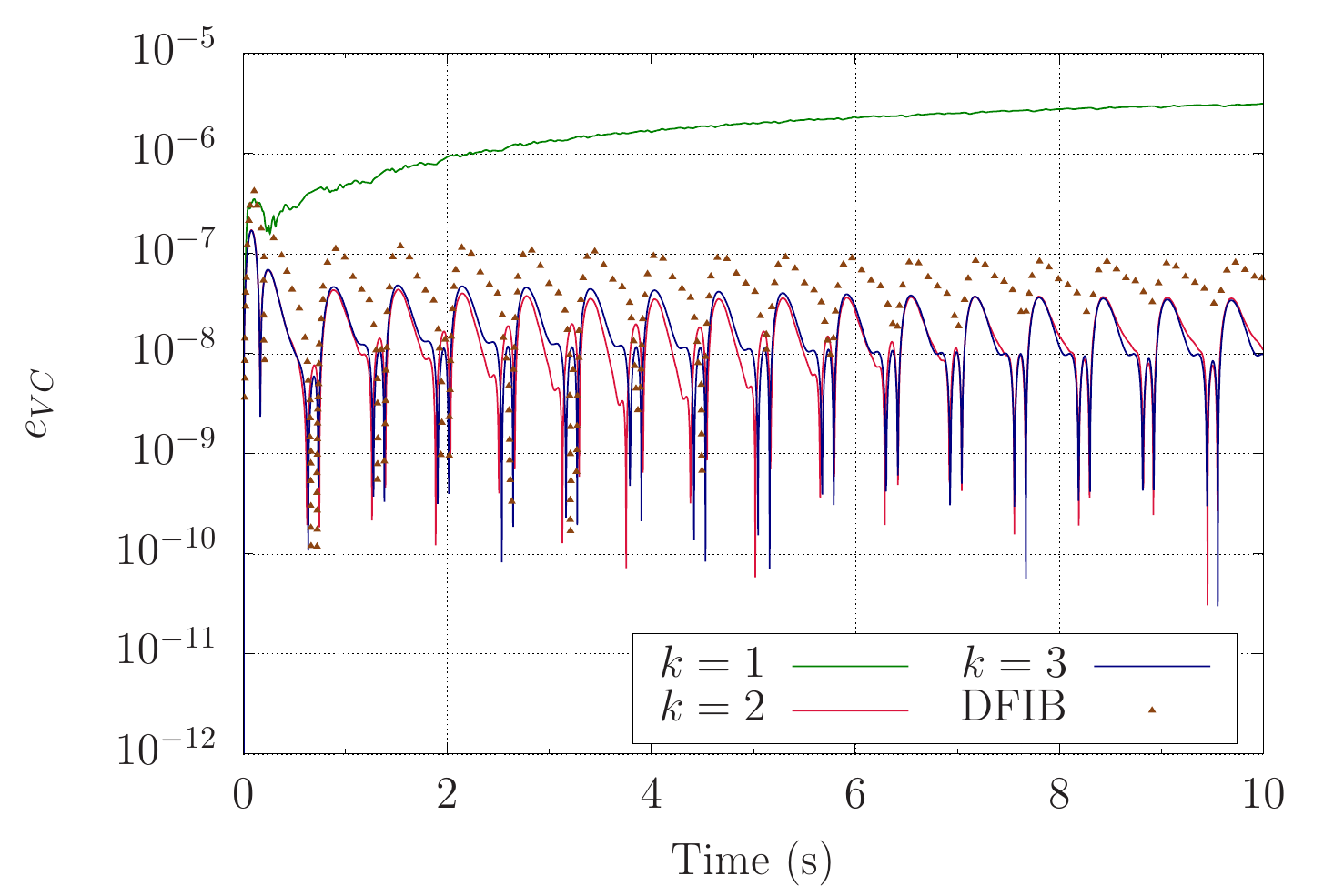}
\caption{Effect of the discretization order at the Eulerian level for the closed curve with active behavior. Incompressibility error at the Lagrangian level for different values of $k$. For $k=1$, the Eulerian velocity has kinks in the interior of the Lagrangian elements and the quadrature error made when integrating those kinks seems to be the dominant discretization error when solving the kinematic equation. For $k=2,3$, the Eulerian velocity is already smooth in the interior of the Lagrangian elements. The differences between $k=2$ and $k=3$ are negligible and the largest value of $e_{VC}$ along the considered time interval is the same for $k=2$ and $k=3$.} 
\label{orderA}
\end{figure}

% The fact that $e_{VC}$ is slightly larger for $k=3$ than $k=2$ for certain time values may be due to the fact that the positive effect of the increased continuity of $k=3$ is counterbalanced by the fact that the quadrature error of integrating a polynomial on a fixed set of quadrature points increases as the order of the polynomials being integrated increases.

In order to further show that the incompressibility error at the Lagrangian level with the DCIB method shown in Fig. \ref{close} (a) is due to discretization errors of the kinematic equation that is solved to compute the Lagrangian displacement from the Eulerian velocity (Eq. \eqref{b3}) instead of incompressibility errors coming from the Eulerian velocity, we solve the problem with an absolute tolerance for the Navier-Stokes equations ($a_{tol}$) varying from $10^{-8}$ to $10^{-14}$. The incompressibility errors at the Eulerian and Lagrangian levels are plotted in Figs. \ref{close} (c)-(d), respectively. While $e_{DIV}$ decreases uniformly as we reduce the absolute tolerance, $e_{VC}$ stops decreasing at $a_{tol} = 10^{-10}$. Therefore, $e_{VC}$ can only be decreased further by refining the time and Lagrangian discretizations, that is, refining the discretization of the kinematic equation.

\begin{figure}[h!]
\centering
\includegraphics[width=10cm]{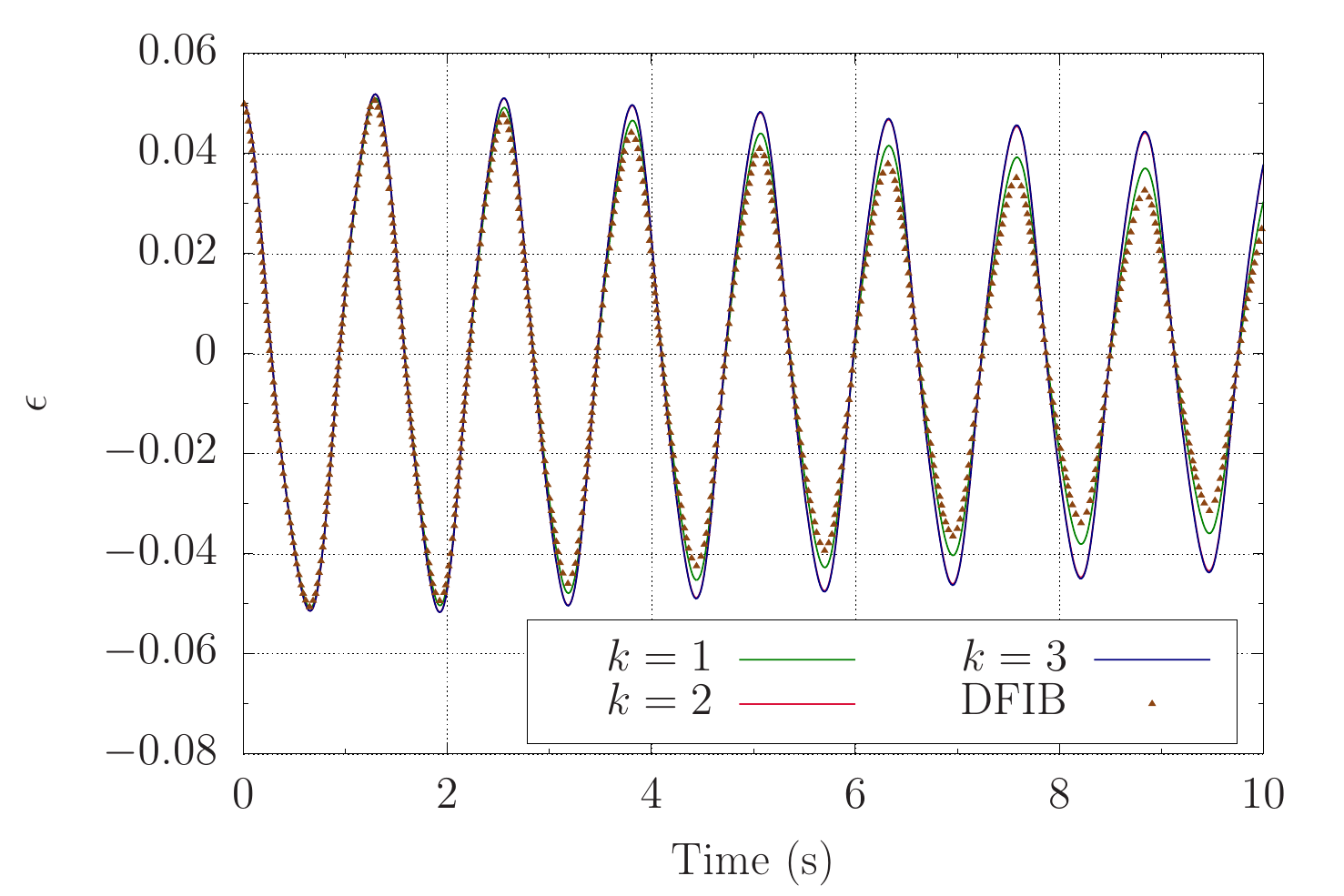}
\caption{Effect of the discretization order at the Eulerian level for the closed curve with active behavior. Amplitude of the closed curve for different values of $k$, the plots with $k=2$ and $k=3$ are overlapped.} 
\label{orderB}
\end{figure}

In addition to the standard discretization errors in space and time, conventional IB methods also have large quadrature/interpolation errors \cite{boffi2008hyper, boffi2011finite, hesch2012continuum, casquero2018non} due to the fact that the integrals that are computed in the elements of the Lagrangian mesh have functions in their integrands that are defined on the Eulerian mesh. As a result, Gauss quadrature is performed on integrands with lines of reduced continuity within the integration regions (as opposed to applying Gauss quadrature in regions where all the functions are $C^\infty$ as in standard finite-element problems). In the DCIB method, this issue is alleviated by leveraging the higher inter-element continuity of splines. In order to show this, we solved this problem using $128 \times 128$ Eulerian elements with $k=$ 1, 2, and 3, $328$ Lagrangian elements with $p=2$, and time step $\Delta t = 3.90625 \text{e--}4 \; \textrm{s}$. The time evolutions of $e_{VC}$ and $\epsilon$ for these three discretizations are plotted in Figs. \ref{orderA} and \ref{orderB}, respectively. The results suggest that the quadrature error is the dominant error for $k=1$, but not for $k=2$ since no additional accuracy is obtained with $k=3$. Note that the optimal convergence rates given by the approximation properties of the discrete spaces are not reached in IB and FD methods \cite{boffi2008hyper, casquero2018non, yu2018error} due to the reduced regularity of the exact solution, namely, the pressure and the viscous stresses are discontinuous at the fluid-solid interface \cite{lai2001remark, boffi2008hyper}. Therefore, the main motivation behind using $k=2$ is to decrease the quadrature error. 

%In Fig \ref{close}b second order convergence is shown, which is obtained with a fixed eulerian mesh of 32x32 elements and the refinement is only performed on the lagrangian mesh. The rest of the figure shows the effect of a decreasing tolerance with which the linear system of equations is solved in both the conservation error at the eulerian (Fig \ref{close}c) and lagrangian (Fig \ref{close}d) meshes. The fact that the lagrangian error is not affected despite the improvement in the eulerian error together with a second order convergence obtained with a fixed (and coarse) eulerian mesh is an indicator that, contrary to other methods with standard discretizations, the use of divergence conforming b-splines already provides a neglegible conservation error at the eulerian level, almost regardless of the resolution of the mesh or the error of the solution of the linear system. This points at the discretization error of the solution of the kinematic equation as being responsible for the solid surface conservation error in our method.

% \cite{griffith2012volume, griffith2012immersed, bhalla2013unified}

\begin{figure}[h!]
\centering
\subfigure[$\Lambda = 5.0$, $t= 0.600 \; \text{s}$]{\includegraphics[scale=0.085]{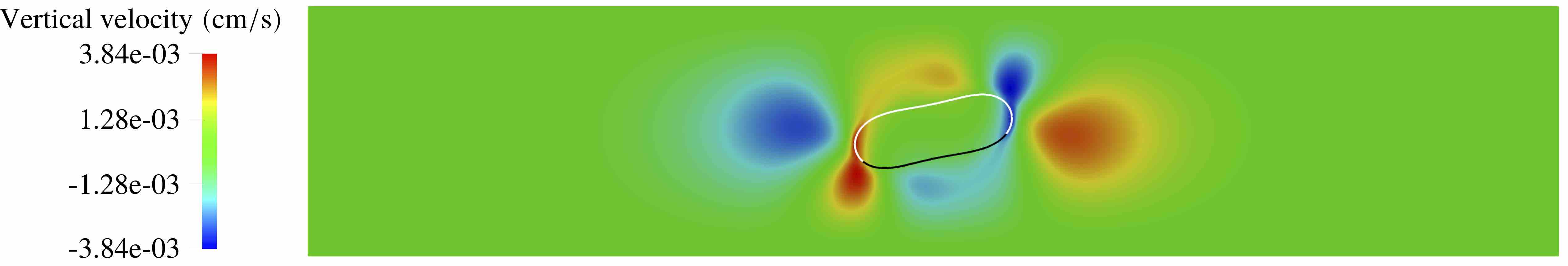}} \\
\subfigure[$\Lambda = 5.0$, $t= 1.620 \; \text{s}$]{\includegraphics[scale=0.085]{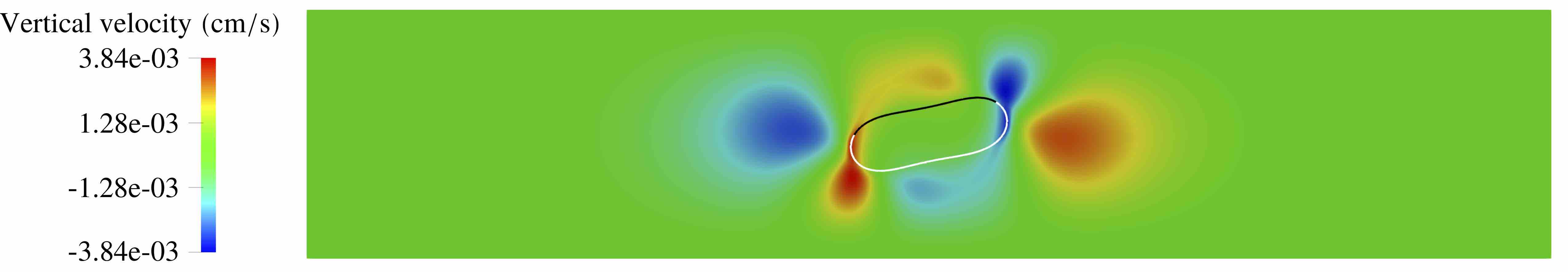}} \\
\subfigure[$\Lambda = 100$, $t=1.150 \; \text{s}$]{\includegraphics[scale=0.085]{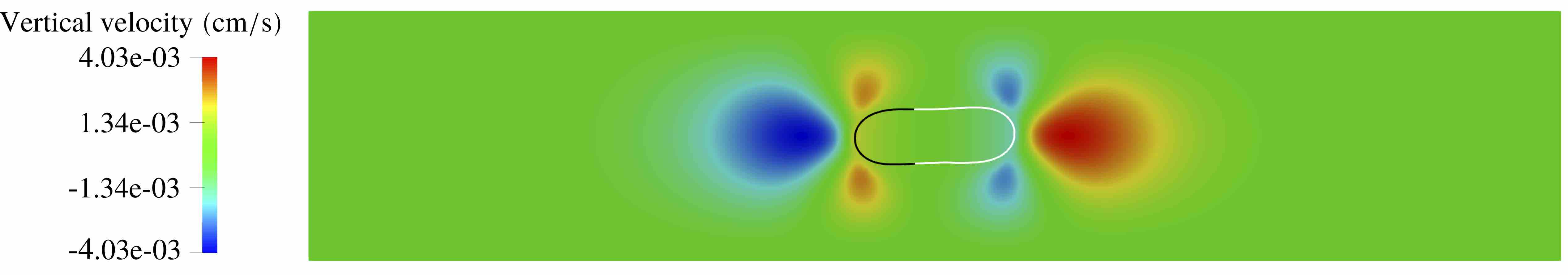}} \\
\subfigure[$\Lambda = 100$, $t=1.578 \; \text{s}$]{\includegraphics[scale=0.085]{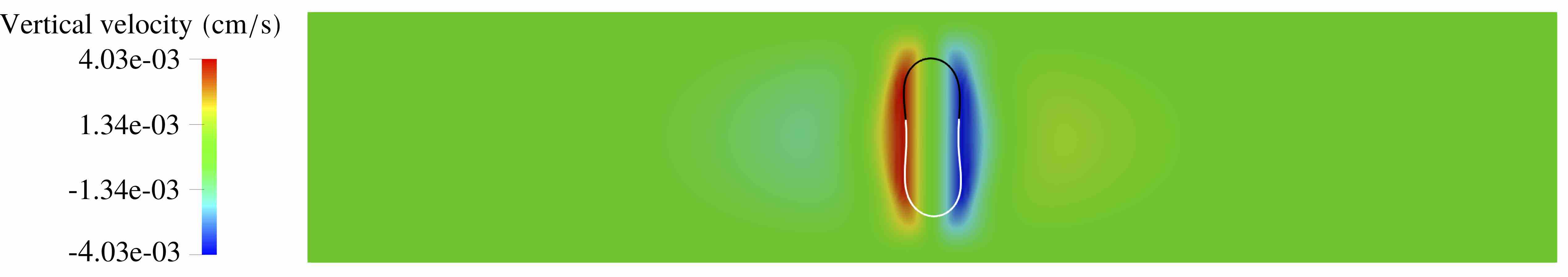}} \\
\caption{Two-dimensional vesicle dynamics in Couette flow. (a)-(b) Vesicle with viscosity ratio 5.0 undergoes tank-treading motion. (c)-(d) Vesicle with viscosity ratio 100 undergoes tumbling motion. In all figures, some Lagrangian elements are represented in black color to know whether or not tank-treading motion is taking place.} \label{shearflow}
\end{figure}

\subsection{Vesicle dynamics in Couette flow}

We next consider a two-dimensional benchmark problem \cite{thiebaud2013rheology, shen2017interaction} to verify our vesicle discretization. It consists of a vesicle in Couette flow. The viscosity contrast is varied over three orders of magnitude to study the vesicle dynamics in both TT and TU regimes.

%        \begin{table} [h]
%\caption{Discretizations considered for the vesicle in Couette flow along with the incompressibility errors and the inclination angle.} \label{table1}
%\bigskip
%\centering
%\begin{tabular}{cccccccccc}
%\toprule
% $n_{el}^{E}$ & EDOF & $k$ & $n_{el}^{L}$ & LDOF & $p$ & $\Delta t \; (\text{s})$ & $e_{DIV}$ & $e_{VC}$ & $\psi \; (\text{deg})$ \\
%\midrule
%100$\times$20 & 1,008 & 2 & 64 & 336 & 3  & 2.0e--4 & 5.56e--11 & 2.25e--6  & 12.78  \\ 
%200$\times$40 & 3,536 & 2 & 128 & 520 & 3  & 1.0e--4 & 1.02e--10 & 2.10e--7  & 12.47 \\   % 12.942885 k=3 12.2596 p=4 12.937360       
%400$\times$80 & 13,200 & 2 & 256 & 1,800 & 3  & 5.0e--5  & 1.92e--10 & 5.51e--8 & 12.29   \\    % k = 3 12.31180   12.29762
%800$\times$160 & 50,960 & 2 & 512 & 6,664 & 3  & 2.5e--5 & 2.0e--4 & 4.5708e--5  & 2.2375e--3 \\ % 11.9781   
%\bottomrule
%\end{tabular}
%\end{table}

\begin{figure}[h!]
\centering
\subfigure[Inclination angle]{\includegraphics[scale=0.52]{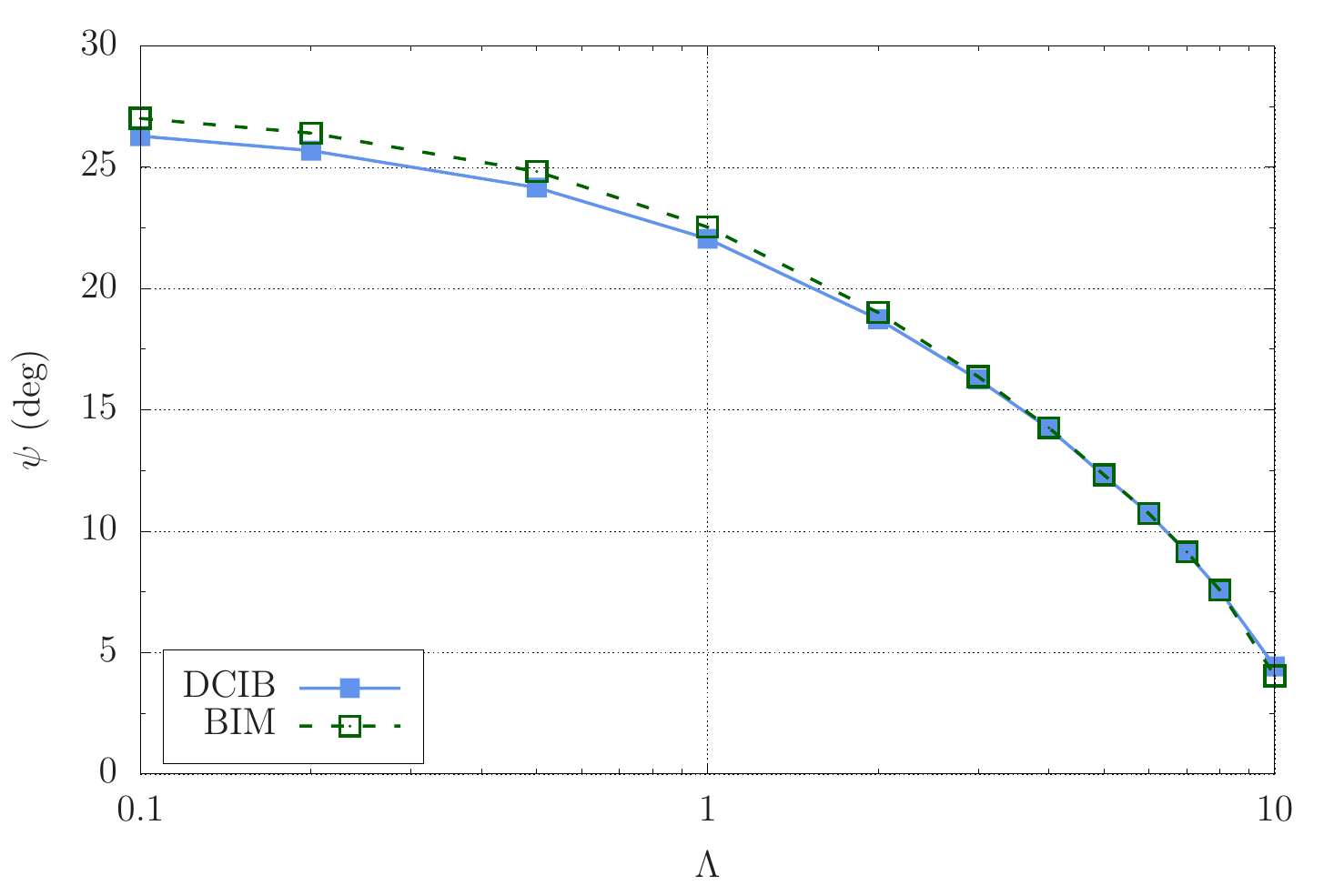}} \, \hspace*{-3mm}
\subfigure[Tumbling period]{\includegraphics[scale=0.52]{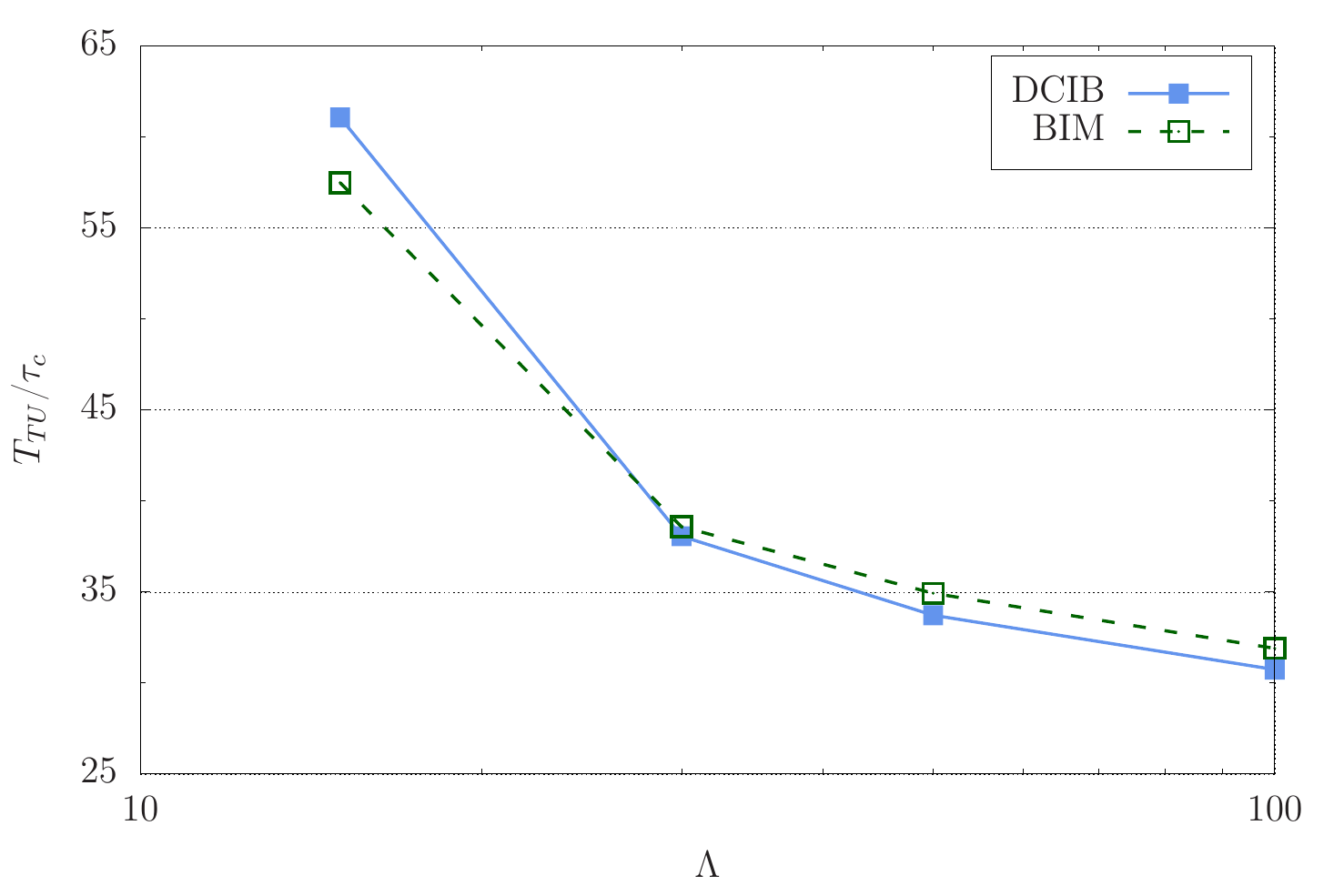}} \\
\subfigure[$\Lambda = 1.0$]{\includegraphics[scale=0.52]{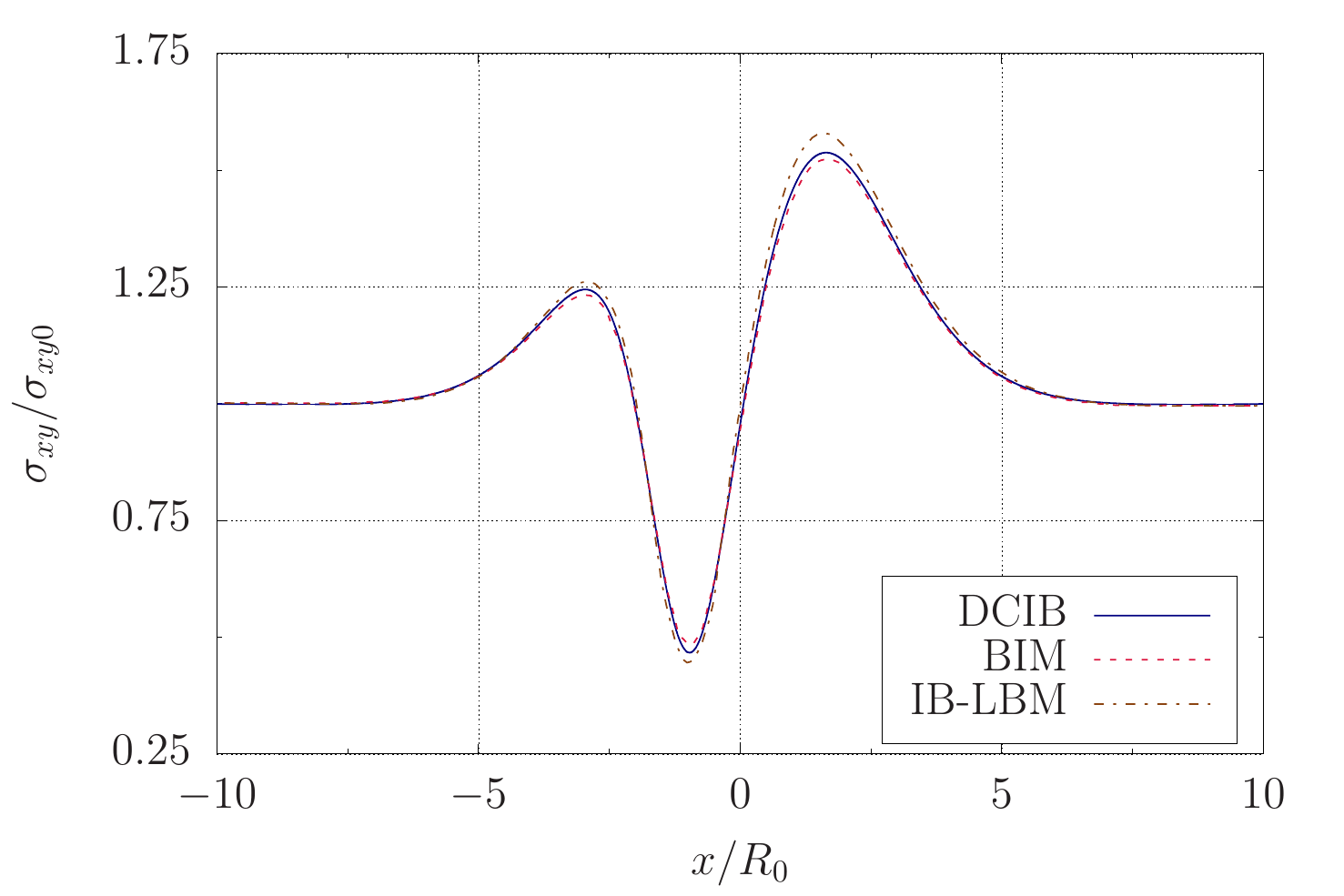}} \, \hspace*{-3mm}
\subfigure[$\Lambda = 1.0$]{\includegraphics[scale=0.52]{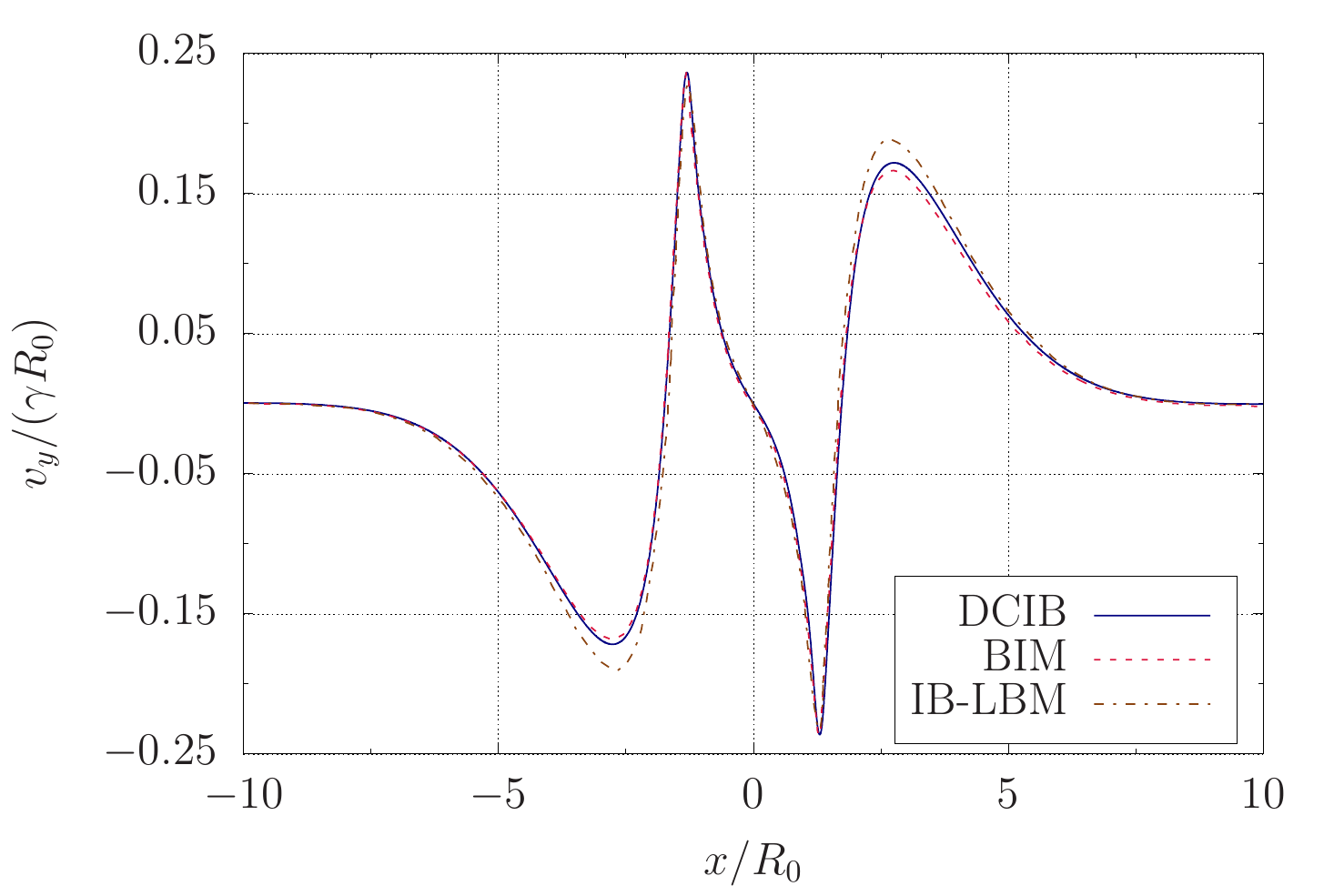}} \\
\caption{Two-dimensional vesicle dynamics in Couette flow. (a) Inclination angle in the tank-treading regime. (b) Flipping period in the tumbling regime. $\tau_c = \mu_o R_0^3 / \kappa$ is the time needed for the vesicle to recover its equilibrium shape if the flow is stopped.  (c) Shear stress at the bottom wall. $\sigma_{xy0}$ is the shear stress at the wall for a pure Couette flow with no vesicle. (d) Vertical velocity at the horizontal line that passes through the center of the channel.} \label{shearflowplots}
\end{figure}

The vesicle is an ellipse whose semiaxes are $a = 1.6625 \text{e--}3 \;  \text{cm}$ and $b = 6.015 \text{e--}4 \;  \text{cm}$. The effective radius is $R_0 = 0.001 \;  \text{cm}$. The physical domain $\Omega$ is a rectangle with sides $L_x = 25 R_0 \;  \text{cm}$ and $L_y = 5 R_0 \;  \text{cm}$. The vesicle is located at the center of the rectangle. A horizontal velocity of $0.25 \;  \text{cm} / \text{s}$ is applied at the top and bottom sides of the rectangle in opposite directions. A no-penetration boundary condition is applied at the top and bottom sides of the rectangle. Periodic boundary conditions are applied on the left and right sides of the rectangle\footnote{The presence of periodic boundary conditions means that the vesicle within our simulation domain may interact with the vesicles located to the left and to the right of the simulation domain \cite{thiebaud2013rheology}. The interaction between vesicles decreases as $L_x/R_0$ is increased. We ran a simulation with $L_x/R_0 = 50$ and $\Lambda = 1$ and the quantities measured in this benchmark stayed the same, which suggests that the interaction among vesicles for $L_x/R_0 = 25$ is already negligible. It is particularly important to keep this effect in mind when comparing simulations with experimental results with only one vesicle.}. If there was no vesicle, the applied boundary conditions would lead to a pure Couette flow with shear rate $\dot{\gamma} = 20 \; \text{s}^{-1}$. The fluid and the vesicle are initially at rest. The physical parameters defining this problem are the following: $\rho = 1.0  \; \textrm{g} / \textrm{cm}^3$, $\mu_o = 0.01 \; \textrm{g} / (\textrm{cm} \cdot \textrm{s})$, $\kappa = 2.0 \text{e--}10  \; \textrm{g} \cdot \textrm{cm}^2  /  \textrm{s}^2$, $C_I = 0.2 \; \textrm{g} / (\textrm{cm}^2 \cdot \textrm{s}^2)$, $\vec g_V = \vec 0$, and $\mu_i = 0.001, 0.002, 0.005, 0.01, 0.02, 0.03, 0.04, 0.05, 0.06, 0.07, 0.08, 0.1, 0.15, 0.3, 0.5,$ and $ 1.0 \; \textrm{g} / (\textrm{cm} \cdot \textrm{s})$. The dimensionless numbers of this benchmark are $\Delta = 0.7$, $\chi = 0.4$,  $C_a = 1$, $R_e = 0.002$, and $\Lambda = 0.1$, 0.2, 0.5, 1, 2, 3, 4, 5, 6, 7, 8, 10, 15, 30, 50, and 100.

% $\vec g_V = (0.0 \; \textrm{g} / ( \textrm{cm}^2 \cdot \textrm{s}^2 ),0.0 \; \textrm{g} / ( \textrm{cm}^2 \cdot \textrm{s}^2 ))$

The Eulerian mesh has $400 \times 80$ elements with $k=2$, the Lagrangian mesh has $256$ elements with $p=3$, and a time step $\Delta t = 1.0 \text{e--}4 \; \textrm{s}$ is used. For this discretization, $R_0/h^E = 16$. For $\Lambda = 5$, the vesicle undergoes TT motion. The inclination angle ($\psi$) is computed as the angle between the minor principal axis of inertia of the vesicle and the horizontal direction (flow direction in a Couette flow). We observed that $e_{DIV}$ remained lower than 2.0e--10 throughout the simulation. The value of $e_{VC}$ at $t = 1.0 \; \text{s}$ (once the vesicle has already reached a steady inclination angle) is 1.78e--7. For the selected dilatation modulus, the relative perimeter change of the vesicle is lower than 2.2e--4. Note that the dynamics of a vesicle are highly dependent on the swelling degree \cite{abreu2014fluid}, which encodes all the geometric information that is needed to define a vesicle. In order to perform reliable simulations of a vesicle, the relative changes of inner fluid area and perimeter of the vesicle must be negligible so that the swelling degree stays constant along the simulation. Figs. \ref{shearflow} (a)-(b) show the vertical velocity and the vesicle deformation for $\Lambda = 5$ at two different times. Some elements of the Lagragian mesh are represented in black color to show the TT motion of the vesicle. Figs. \ref{shearflow} (c)-(d) show the vertical velocity and the vesicle deformation for $\Lambda = 100$ at two different times. For this viscosity contrast, the vesicle undergoes TU motion.

In \cite{thiebaud2013rheology}, this problem is solved using the BIM. The inclination angle is measured in the TT regime and the period is measured in the TU regime. As shown in Figs. \ref{shearflowplots} (a)-(b), good agreement is found between the DCIB method and the BIM. Here, we define the TU period as the time spent by the vesicle to undergo a complete turn in its flipping motion. In \cite{thiebaud2013rheology}, the TU period is defined as the time spent by the vesicle to undergo half a turn in its flipping motion. Therefore, the results from \cite{thiebaud2013rheology} are multiplied by two in Fig. \ref{shearflowplots} (b). In \cite{shen2017interaction}, this problem is solved using the BIM and an IB method in which the Navier-Stokes equations are discretized using the LBM and the vesicle is discretized using the spring-like method developed in \cite{tsubota2006particle, tsubota2010effect}. The shear stress at the bottom wall and the vertical velocity at the horizontal line that passes through the channel center are measured in \cite{shen2017interaction}. As shown in Figs. \ref{shearflowplots} (c)-(d), good agreement is found between the DCIB method and the BIM while the IB-LBM seems to slightly overestimate the shear stress and the vertical velocity. Note that the IB-LBM used in \cite{shen2017interaction} requires to add a penalty parameter to keep $e_{VC}$ around $1\%$ while in the DCIB method no penalty parameter is added and $e_{VC}$ is five orders of magnitude lower. We have also solved this problem with $p=4$ and without integrating by parts the force exerted by the vesicle on the fluid; the differences in the quantities plotted in Fig. \ref{shearflowplots} were negligible.

\subsection{Spherical capsule dynamics in Couette flow}

We now consider a three-dimensional benchmark problem \cite{lac2004spherical, doddi2008lateral, li2008front} to verify our capsule discretization. It consists of a spherical capsule under Couette flow with $\Lambda = 1$. For this viscosity contrast, the capsule undergoes TT motion.

\begin{figure}[h!]
\centering
\includegraphics[width=12cm]{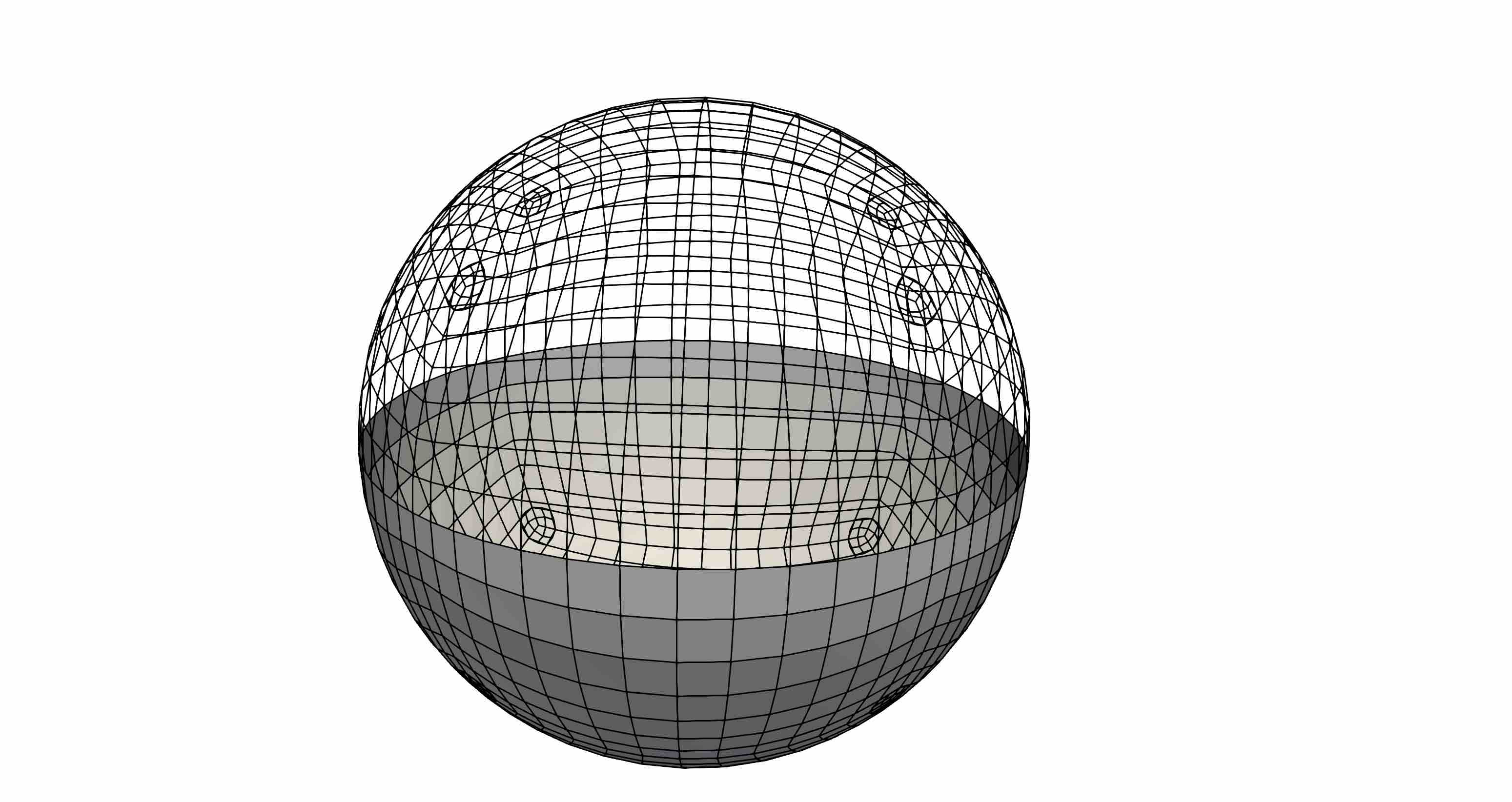}
\caption{Spherical capsule dynamics in Couette flow. The Lagrangian mesh with eight extraordinary points after one level of refinement is plotted. The element boundaries are plotted for the whole mesh and the element interiors are plotted in half of the mesh only.} 
\label{Lmesh}
\end{figure}

The capsule is a sphere with radius $R_0 = 0.001 \;  \text{cm}$. The physical domain $\Omega$ is a cube with side $L = 2 \pi R_0$. The capsule is modeled with the neo-Hookean constitutive law and is initially located at the center of the cube. Velocities of $( \pi / 10 \;  \text{cm} / \text{s}, 0 \;  \text{cm} / \text{s}, 0 \;  \text{cm} / \text{s})$ and $( - \pi / 10 \;  \text{cm} / \text{s}, 0 \;  \text{cm} / \text{s}, 0 \;  \text{cm} / \text{s})$ are applied at the top and bottom sides of the cube, respectively. Periodic boundary conditions are applied to the four lateral sides of the cube. If there was no capsule, the applied boundary conditions would lead to a pure Couette flow with shear rate $\dot{\gamma} = 100 \; \text{s}^{-1}$. Both the fluid and the capsule are initially at rest. Note that the length $L = 2 \pi R_0$ is not large enough for the effect of the periodic boundary conditions to be fully negligible, but this is the length for which more data is available in the literature \cite{doddi2008lateral}. Therefore, we keep this length in order to perform an appropriate comparison. The physical parameters defining this problem are the following: $\rho = 1.0  \; \textrm{g} / \textrm{cm}^3$, $\mu_o = \mu_i = 0.01 \; \textrm{g} / (\textrm{cm} \cdot \textrm{s})$, $G_s = 1.66667  \text{e--}3 \; \textrm{g} / (\textrm{cm}^2 \cdot \textrm{s}^2)$, and $\vec g_V = \vec 0$. The dimensionless numbers of this benchmark are $C_a = 0.6$\footnote{In \cite{doddi2008lateral}, the capillary number is defined with respect to the Young modulus ($E_s$) instead of the shear modulus $G_s$ as in the present work. Since $E_s = 3G_s$ for a Poisson ratio $\nu_s = 0.5$ \cite{barthes2002effect}, the value of $C_a$ given in \cite{doddi2008lateral} needs to be multiplied by 3 to be equivalent to the value of $C_a$ used here.}, $\chi = 1/ \pi$, $\Lambda = 1.0$, and $R_e = 0.01$.

\begin{figure}[h!]
\centering
\subfigure[$t= 0.17 \; \text{s}$]{\includegraphics[scale=0.08]{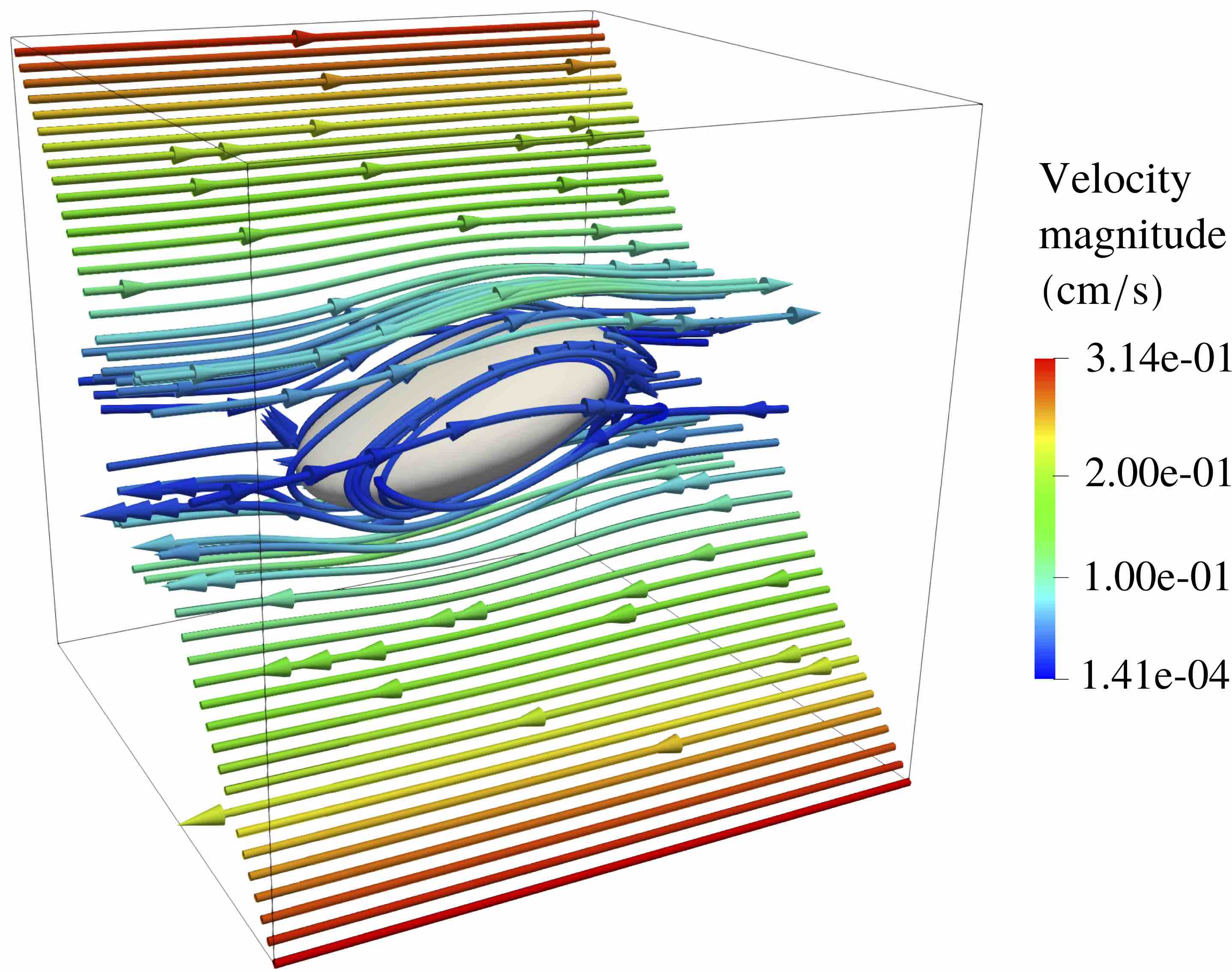}} \, \hspace*{+7mm}
\subfigure[$t= 0.17 \; \text{s}$]{\includegraphics[scale=0.08]{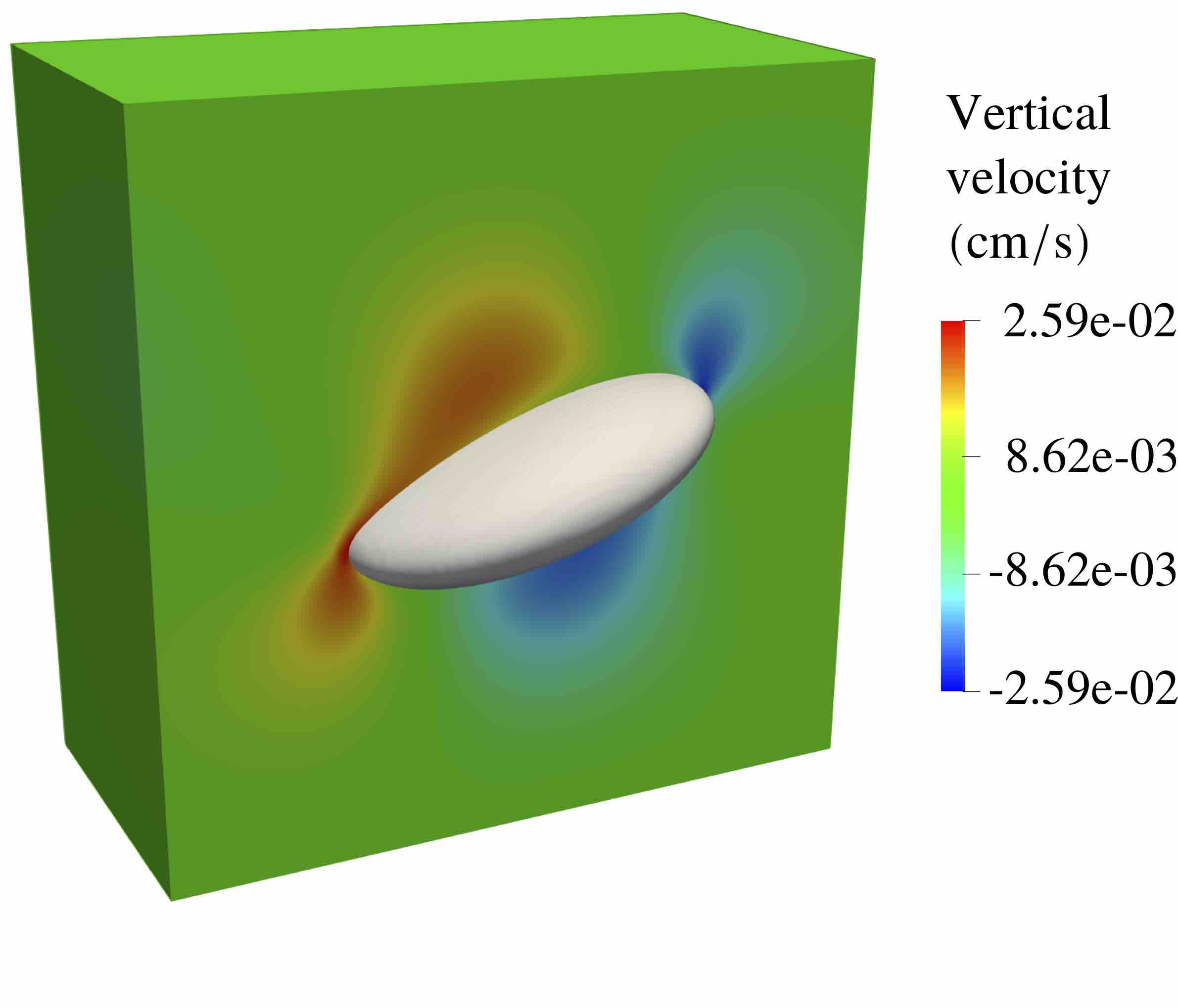}} \\
\caption{ Spherical capsule in Couette flow. (a) Streamlines colored by the velocity magnitude along with the deformed capsule. (b) Vertical velocity of half of the domain along with the deformed capsule.} \label{spherevel}
\end{figure}

\begin{figure}[h!]
\centering
\subfigure[Inclination angle]{\includegraphics[scale=0.53]{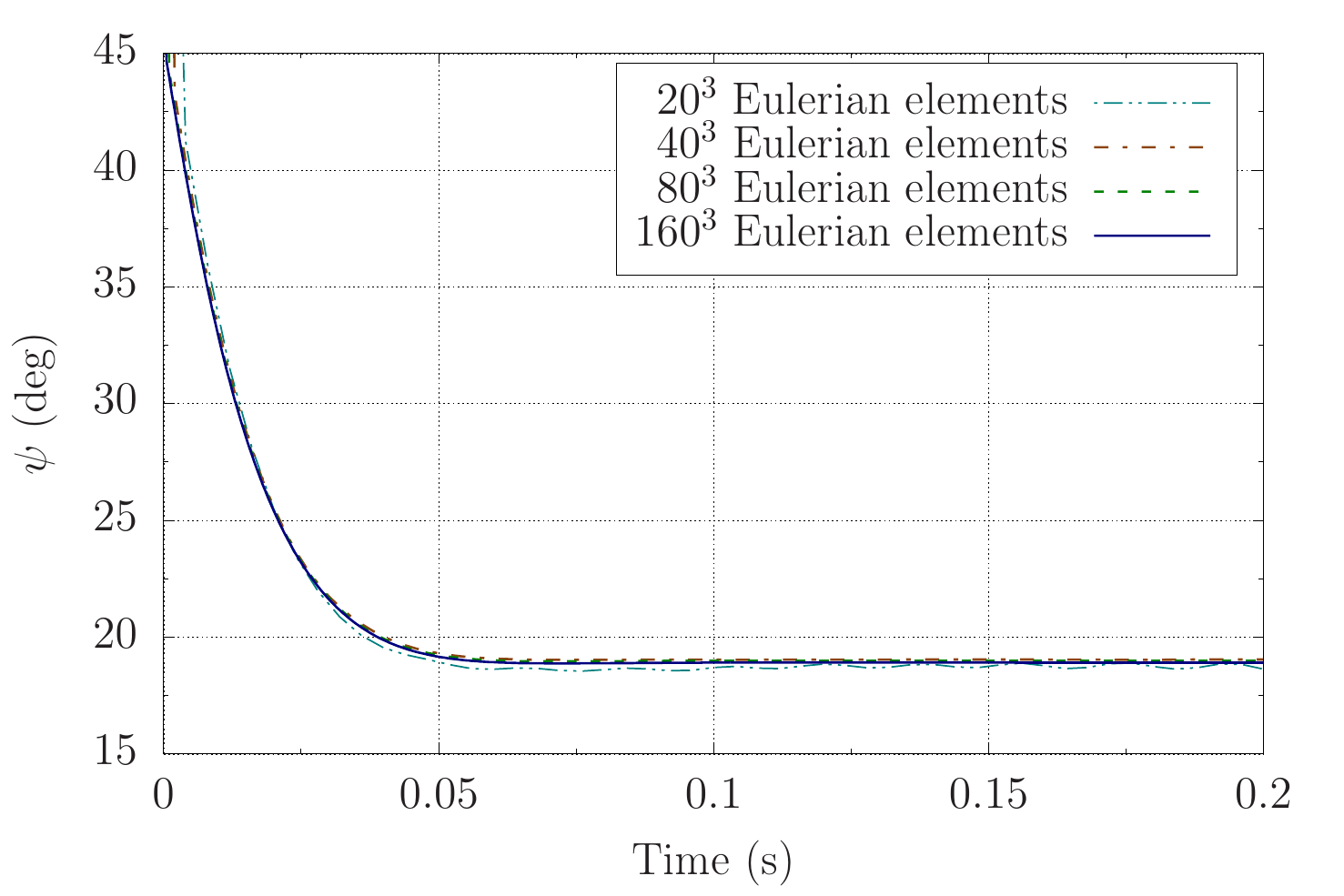}} \,\hspace*{-3mm}
\subfigure[Taylor parameter]{\includegraphics[scale=0.53]{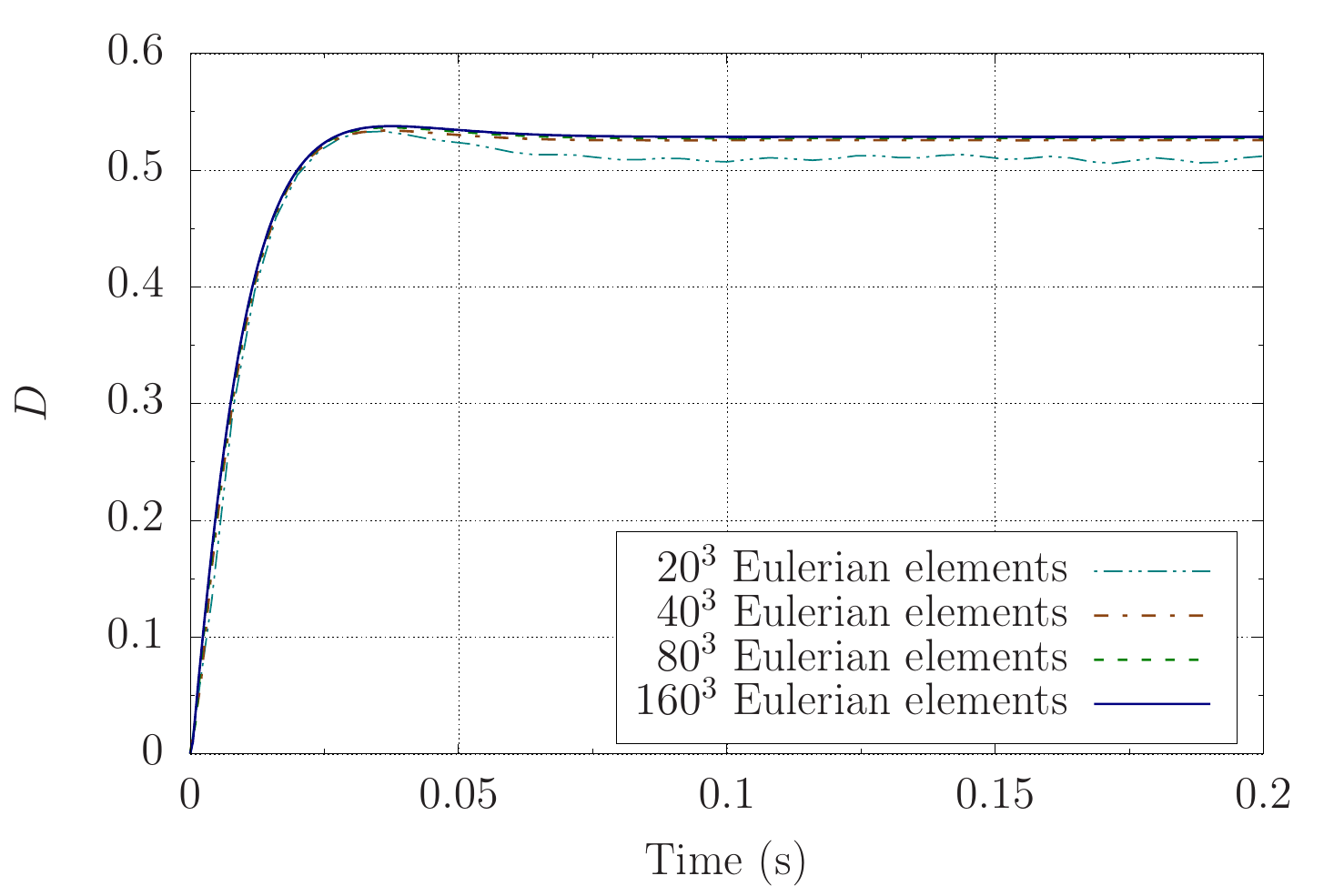}} \\
\caption{ Mesh-independence studies for the spherical capsule in Couette flow. } \label{sphere}
\end{figure}

% $\vec g_V = (0.0 \; \textrm{g} / ( \textrm{cm}^2 \cdot \textrm{s}^2 ),0.0 \; \textrm{g} / ( \textrm{cm}^2 \cdot \textrm{s}^2 ),0.0 \; \textrm{g} / ( \textrm{cm}^2 \cdot \textrm{s}^2 ))$

We start with a coarse discretization, namely, $20^3$ Eulerian elements with $k=2$, $366$ Lagrangian elements with $p=3$, and time step $\Delta t = 4.\text{e--}3 \; \textrm{s}$. After that, the discretization is refined by performing uniform $h$-refinement four times on the Lagrangian and Eulerian meshes and dividing the time step by two each time a new level of refinement is introduced. The Lagrangian mesh after one level of refinement is plotted in Fig. \ref{Lmesh}. For the coarsest discretization $R_0 / h^E = 10 / \pi$ and for the finest discretization $R_0 / h^E = 80 / \pi$. $e_{DIV}$ is smaller than 5.0e--11 for all discretizations considered. The value of $e_{VC}$ at $t = 0.2 \; \text{s}$ goes from 4.2e--3 for the coarsest discretization to 1.4e--8 for the finest discretization (second-order convergence). In order to show that in the DCIB method $e_{VC}$ is primarily produced by time-discretization errors of the kinematic equation, we now solve this problem using the coarsest Eulerian and Lagrangian meshes, but decreasing the time step an order of magnitude ($\Delta t = 4.\text{e--}4 \; \textrm{s}$). The value of $e_{VC}$ at $t = 0.2 \; \text{s}$ is 2.7e--5, that is, it decreased more than two orders of magnitude with respect to the value with $\Delta t = 4.\text{e--}3 \; \textrm{s}$ (4.2e--3) showing the second-order convergence of $e_{VC}$ with respect to the time step. In \cite{doddi2008lateral}, it is reported that $e_{VC}$ is approximately 0.001 and the time step is varied from $1.\text{e--}4 \; \textrm{s}$ to $1.\text{e--}6 \; \textrm{s}$. Figs. \ref{spherevel} (a)-(b) show the streamlines colored by the velocity magnitude and the vertical velocity once the capsule has reached a steady inclination angle, respectively.

% 1248

        \begin{table} [h]
\caption{Comparison with the literature for the spherical capsule in Couette flow.} \label{tabsc}
\bigskip
\centering
\begin{tabular}{ccc}
\toprule
 Numerical method & $\psi \; (\text{deg})$ & $D$ \\
\midrule
DCIB & 18.90 & 0.528  \\ 
Ref. \cite{doddi2008lateral} & 19.26 & 0.496 \\       
\bottomrule
\end{tabular}
\end{table}

Mesh-independence studies for the inclination angle and the Taylor parameter are shown in Figs. \ref{sphere} (a)-(b). The inclination angle is computed as the angle between the minor principal axis of inertia of the capsule in the plane of shear and the horizontal direction (flow direction in a Couette flow) \cite{ramanujan1998deformation}. The Taylor parameter ($D$) is a measure of the capsule deformation and it is defined as
\begin{equation}
D = \frac{l_1-l_2}{l_1+l_2}
\end{equation}
where $l_1$ and $l_2$ are the lengths of the minor and major principal axes of inertia in the plane of shear of an ellipsoid with the same moments of inertia as the capsule \cite{ramanujan1998deformation}. In \cite{doddi2008lateral}, this problem is solved using an IB method in which the Navier-Stokes equations are discretized using finite differences and the capsule is treated using the $C^0$ triangular discretization developed in \cite{charrier1989free,shrivastava1993large, eggleton1998large}. In \cite{doddi2008lateral}, it is reported that $D$ increases $3.1 \%$ when the Eulerian mesh is changed from $40^3$ to $80^3$ while $D$ only increases $0.3\%$ using the DCIB method, i.e., the DCIB method reaches a converged result with coarser resolutions. Table \ref{tabsc} includes the converged result for the inclination angle and the Taylor parameter obtained with the DCIB method and the IB method used in  \cite{doddi2008lateral}. In \cite{lac2004spherical}, an unbounded BIM is used to solve this problem. Although the results in \cite{doddi2008lateral} are often compared with the results in \cite{lac2004spherical}, this comparison is not appropriate since the effect of confinement (top and bottom walls) and the effect of periodic boundary conditions are not present in the results from \cite{lac2004spherical}. In an analogous way, the results from \cite{li2008front}, obtained using an IB method based on finite differences, are often compared with the results from \cite{doddi2008lateral} although a different simulation domain was used in \cite{li2008front}.

% \cite{li2008front}

% Eulerian elements 40^3 80^3 160^3 
% Lagrangian elements 1248 4776 18888
% Lagrangian cps 1266 4794 18906

% 0.3151\%

\begin{figure}[h!]
\centering
\subfigure[$\Lambda = 1.0$, $t=2.830 \; \text{s}$]{\includegraphics[scale=0.06]{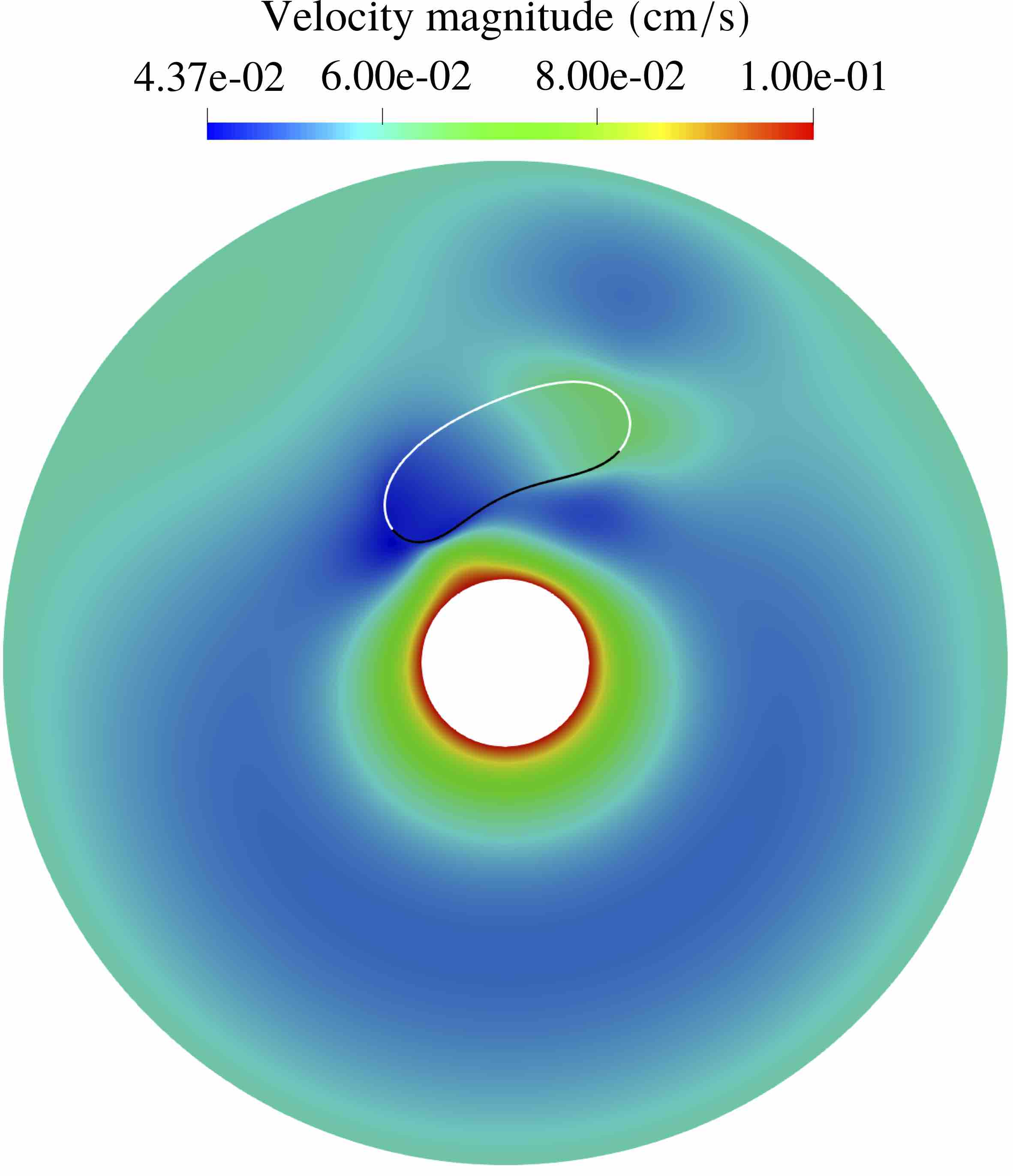}} \, \hspace*{+7mm}
\subfigure[$\Lambda = 1.0$, $t=3.125 \; \text{s}$]{\includegraphics[scale=0.06]{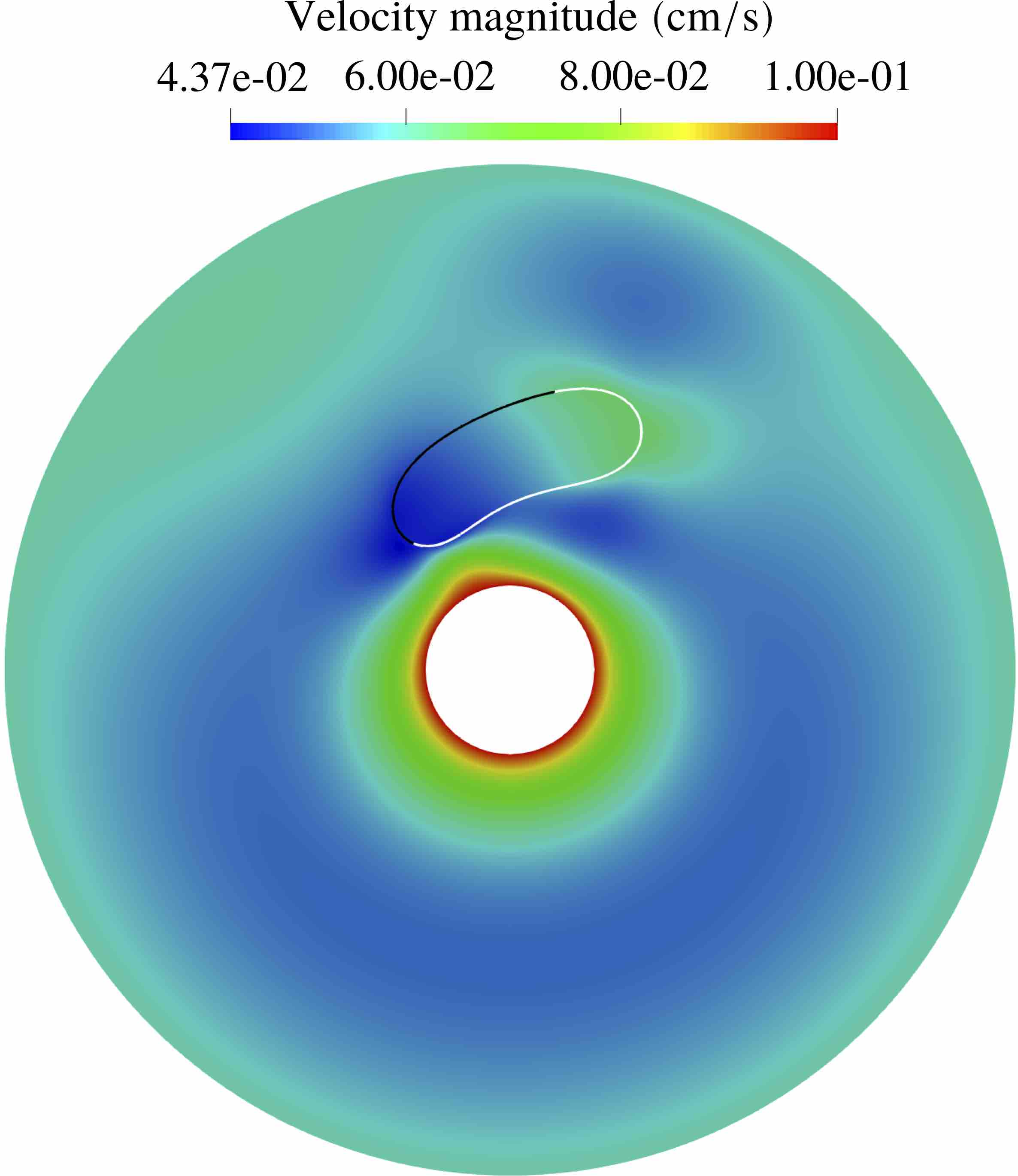}} \\
\subfigure[$\Lambda = 100$, $t=2.568 \; \text{s}$]{\includegraphics[scale=0.06]{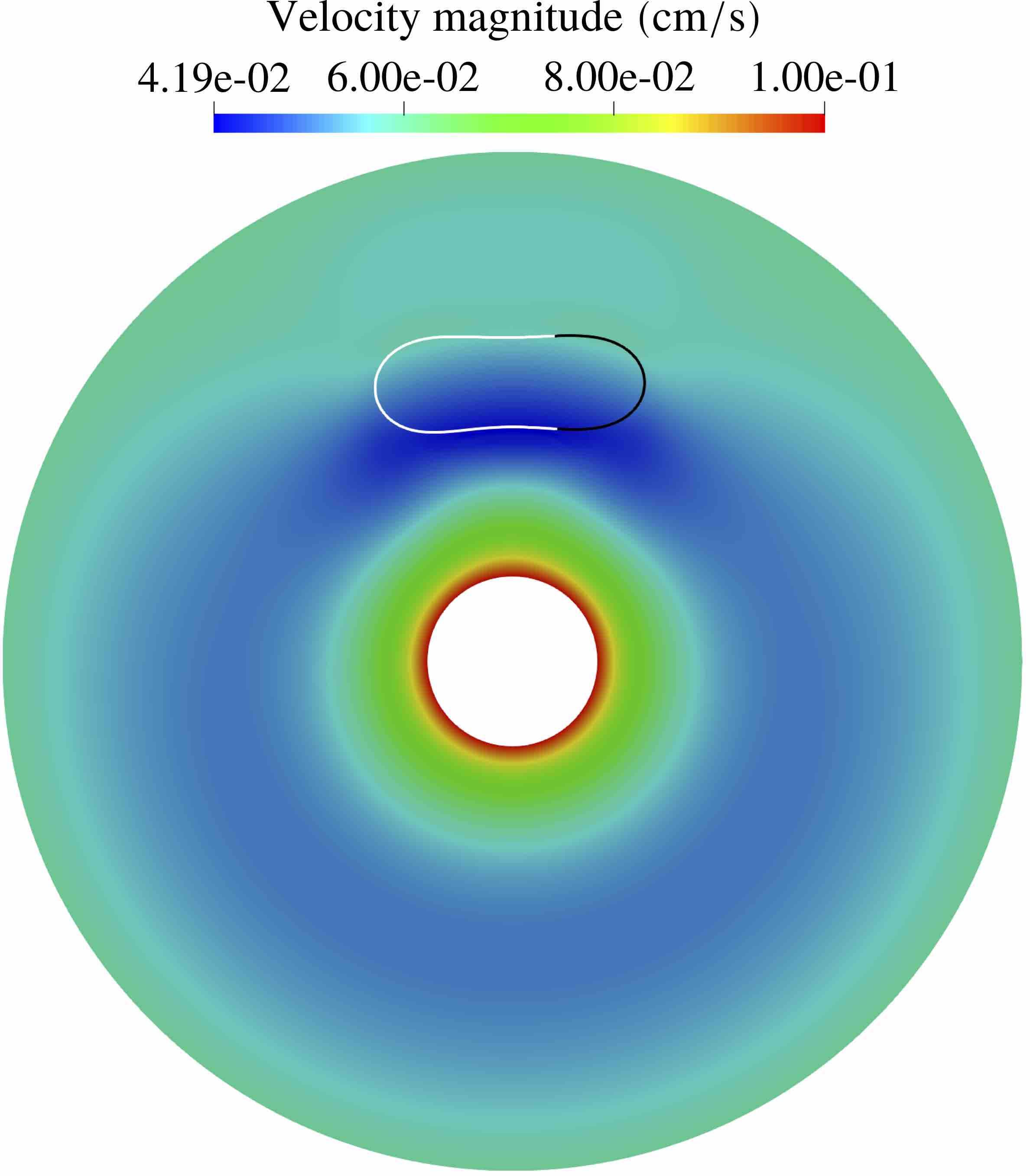}} \, \hspace*{+7mm}
\subfigure[$\Lambda = 100$, $t=2.994 \; \text{s}$]{\includegraphics[scale=0.06]{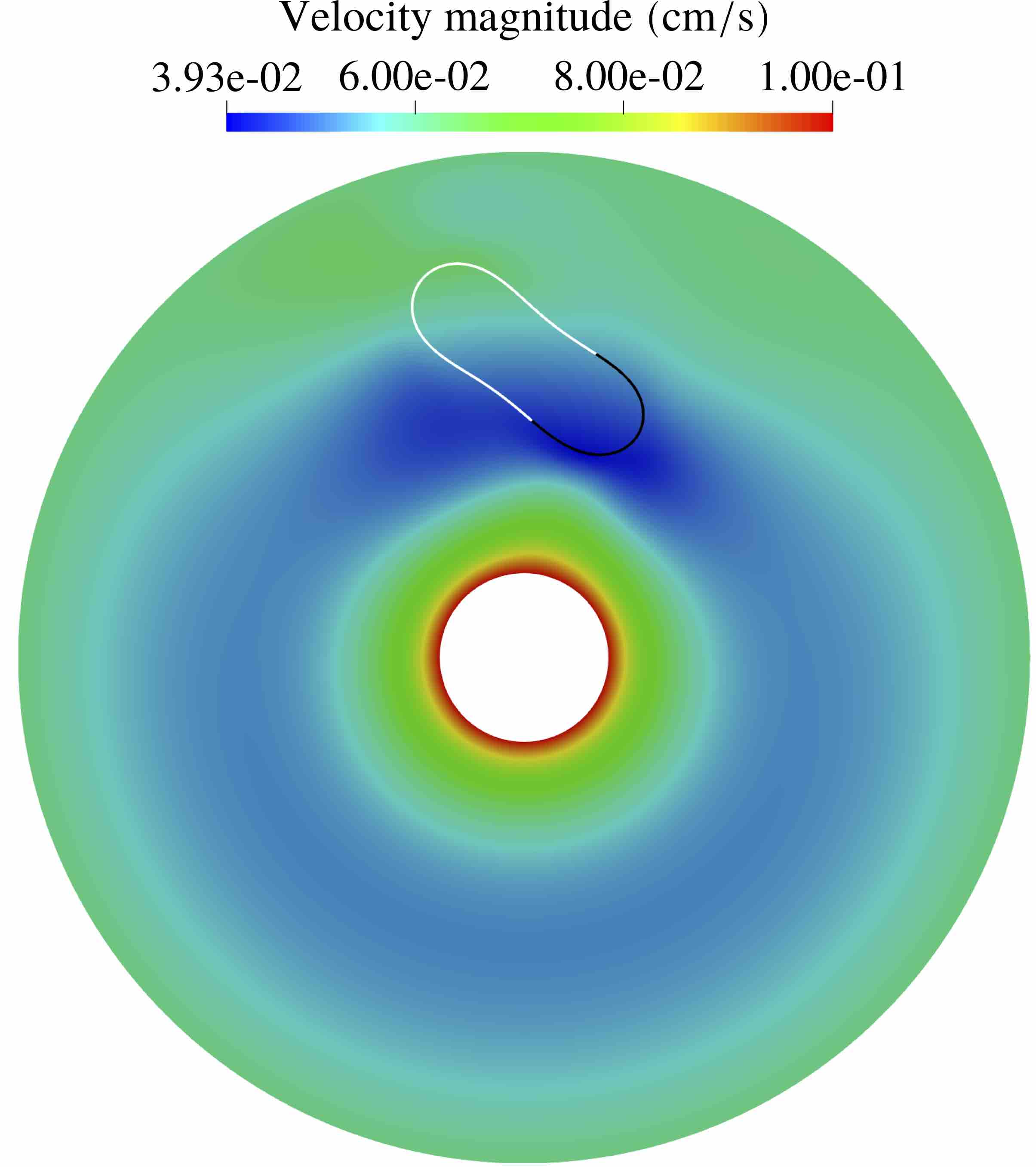}} \\
\caption{Two-dimensional vesicle dynamics in Taylor-Couette flow. (a)-(b) Vesicle with viscosity ratio 1.0 undergoes tank-treading motion. (c)-(d) Vesicle with viscosity ratio 100 undergoes tumbling motion. In all figures, some Lagrangian elements are represented in black color to know whether or not tank-treading motion is taking place.} \label{TCflow}
\end{figure}

% 18.91 0.528

\subsection{Vesicle dynamics in Taylor-Couette flow}

This example studies the dynamics of a vesicle in Taylor-Couette flow \cite{taylor1923viii}. As in Section 6.2, the viscosity contrast is varied over three orders of magnitude to study the vesicle dynamics.

The physical domain $\Omega$ is an annulus with inner radius $R_1 = 0.001 \;  \text{cm}$ and outer radius $R_2 = 0.006 \;  \text{cm}$. The vesicle is an ellipse whose semiaxes are $a = 1.6625 \text{e--}3 \;  \text{cm}$ and $b = 6.015 \text{e--}4 \;  \text{cm}$ ($R_0 = 0.001 \;  \text{cm}$), and its center is initially located at $(0.0 \; \text{cm},0.0035 \; \text{cm})$. Dirichlet boundary conditions are applied to obtain (if there was no vesicle) a Taylor-Couette flow with inner and outer angular velocities of $ 100 \;  \text{rad/s} $ and $ 10 \;  \text{rad/s} $, respectively. Both the fluid and the vesicle are initially at rest. The physical parameters defining this problem are the following: $\rho = 1.0  \; \textrm{g} / \textrm{cm}^3$, $\mu_o = 0.01 \; \textrm{g} / (\textrm{cm} \cdot \textrm{s})$, $\kappa = 2.0 \text{e--}10  \; \textrm{g} \cdot \textrm{cm}^2  /  \textrm{s}^2$, $C_I = 0.2 \; \textrm{g} / (\textrm{cm}^2 \cdot \textrm{s}^2)$, $\vec g_V = \vec 0$, and $\mu_i = 0.001, 0.002, 0.005, 0.01, 0.05, 0.08, 0.1, 0.2, 0.5,$ and $ 1.0 \; \textrm{g} / (\textrm{cm} \cdot \textrm{s})$. The dimensionless numbers of this problem are $\Delta = 0.7$, $\chi = 0.4$,  $C_a = 9.257$, $R_e = 0.0185$, and $\Lambda = 0.1$, 0.2, 0.5, 1, 2, 5, 8, 10, 20, 50, and 100, where the maximum shear rate (obtained in the inner circle) was used to compute $C_a$ and $R_e$.

The Eulerian mesh is composed of $80 \times 320$ elements with $k=2$, the Lagrangian mesh is composed of $256$ elements with $p=3$, and a time step $\Delta t = 5.\text{e--}5 \; \textrm{s}$ is used. For $\Lambda = 1$, the vesicle undergoes TT motion. The inclination angle is computed as the angle between the minor principal axis of inertia of the vesicle and the angular direction (flow direction in the Taylor-Couette flow). $e_{DIV}$ is smaller than 3.0e--10. The value of $e_{VC}$ at $t = 1.5 \; \text{s}$ (once the vesicle has already reached a steady inclination angle) is 3.8e--5. For the selected dilatation modulus, the relative perimeter change of the vesicle is less than 4.8e--4. Figs. \ref{TCflow} (a)-(b) show the velocity magnitude and the vesicle deformation for $\Lambda = 1.0$ at two different times. Some elements of the Lagragian mesh are represented in black color to show the TT motion of the vesicle. Figs. \ref{TCflow} (c)-(d) show the velocity magnitude and the vesicle deformation for $\Lambda = 100$ at two different times. For this viscosity contrast, the vesicle undergoes TU motion.

\begin{figure}[h!]
\centering
\subfigure[Inclination angle]{\includegraphics[scale=0.53]{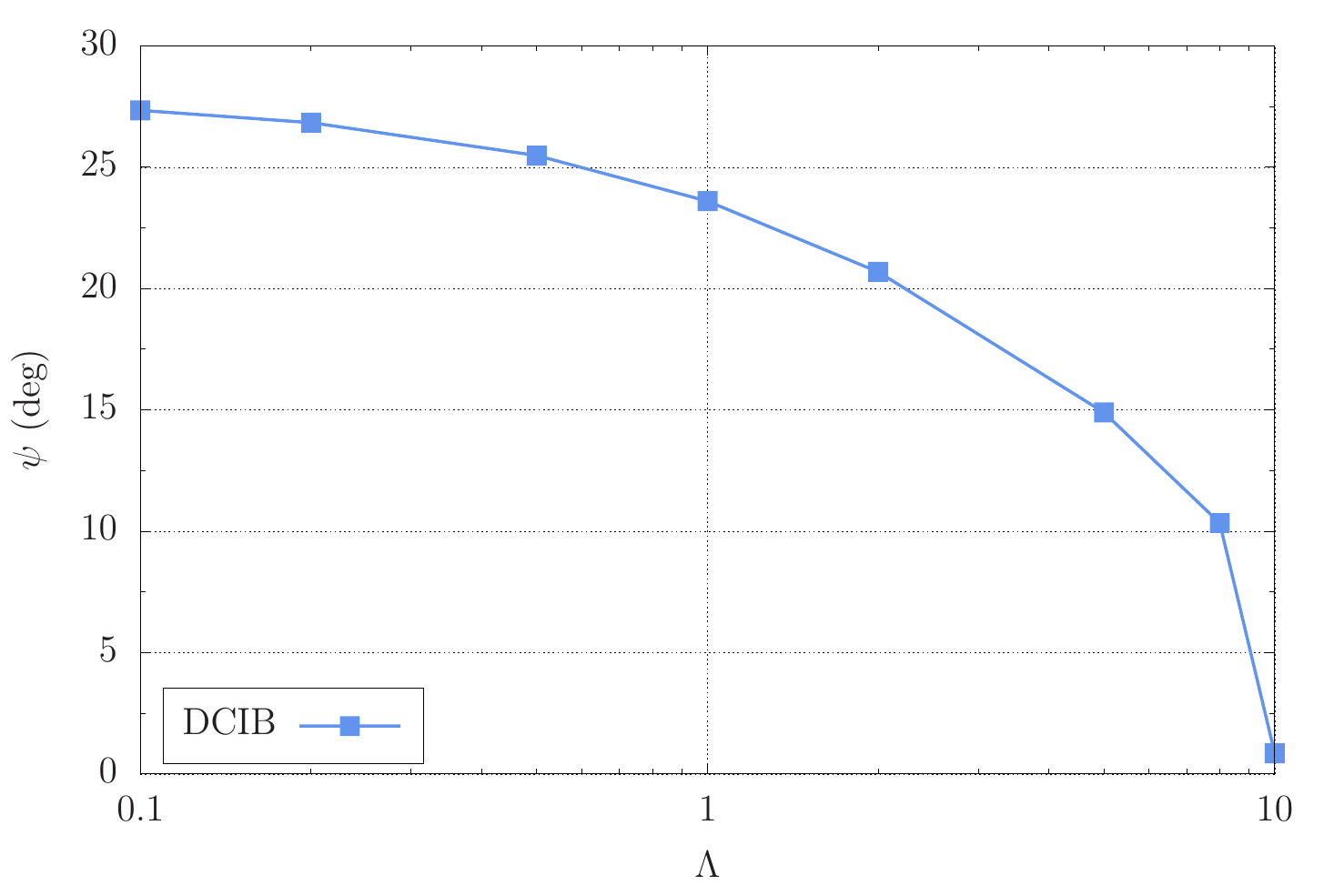}} \, \hspace*{-3mm}
\subfigure[Tumbling period]{\includegraphics[scale=0.53]{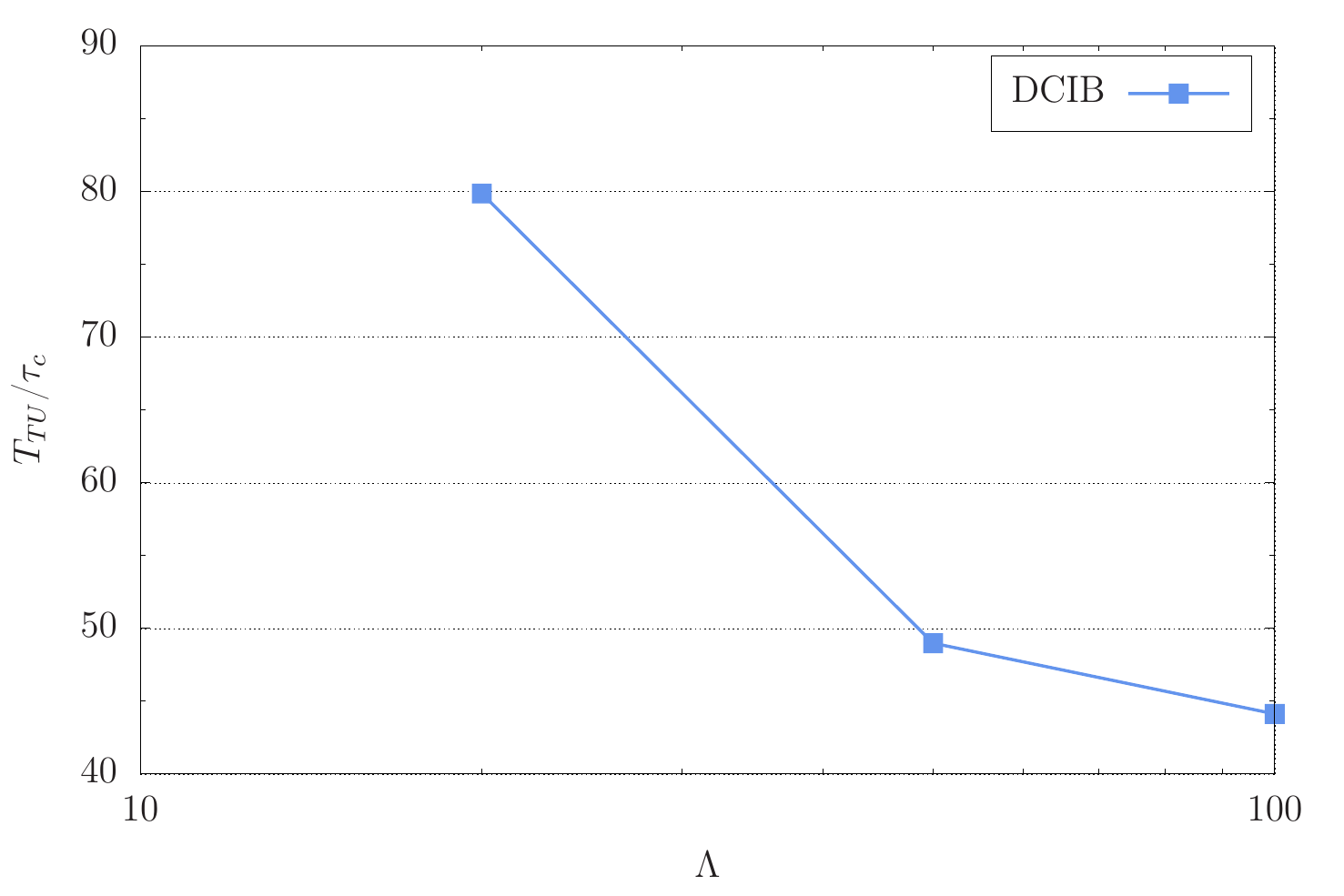}} \\
\caption{ Two-dimensional vesicle dynamics in Taylor-Couette flow. Inclination angle in the tank-treading regime. (b) Period in the tumbling regime.} \label{TCflowplots}
\end{figure}

As in Couette flow, a transition from TT motion to TU motion takes place as the viscosity contrast is increased. The inclination angle is measured in the TT regime and plotted in Fig. \ref{TCflowplots} (a). The period is measured in the TU regime and plotted in Fig. \ref{TCflowplots} (b). Here, we define the TU period as the time spent by the vesicle to undergo a complete turn in its rotation with respect to the angular direction (flow direction in the Taylor-Couette flow). In addition, the vesicle undergoes migration toward the inner circle in the TT regime, that is, the vesicle moves toward the region with higher flow line curvature. In a Taylor-Couette flow, the shear stress has its maximum at the inner circle and its minimum at the outer circle. Therefore, the vesicle is moving toward the region with higher shear stress. Note that in plane Poiseuille and Hagen-Poiseuille flows, a vesicle moves toward the region with lower shear stress instead \cite{abreu2014fluid}. This suggests that the dynamics of vesicles in flows with curvature (Taylor-Couette flow) is significantly different from flows with no curvature (plane Poiseuille and Hagen-Poiseuille flows). No significant radial migration is observed in the TU regime. The migration behavior in TT and TU regimes obtained in our simulations is consistent with the findings reported in \cite{ghigliotti2011vesicle}, which were obtained using the BIM and an unbounded flow that tries to mimic a Taylor-Couette flow.

\subsection{Capsule segregation in Hagen-Poiseuille flow}

% A heated polymeric sheet is incapable of sustaining any significant compressive stresses and its bending resistance is often neglected.

This example studies the segregation of two types of capsules with different size in a Hagen-Poiseuille flow.

\begin{figure}[h!]
\centering
\includegraphics[width=11cm]{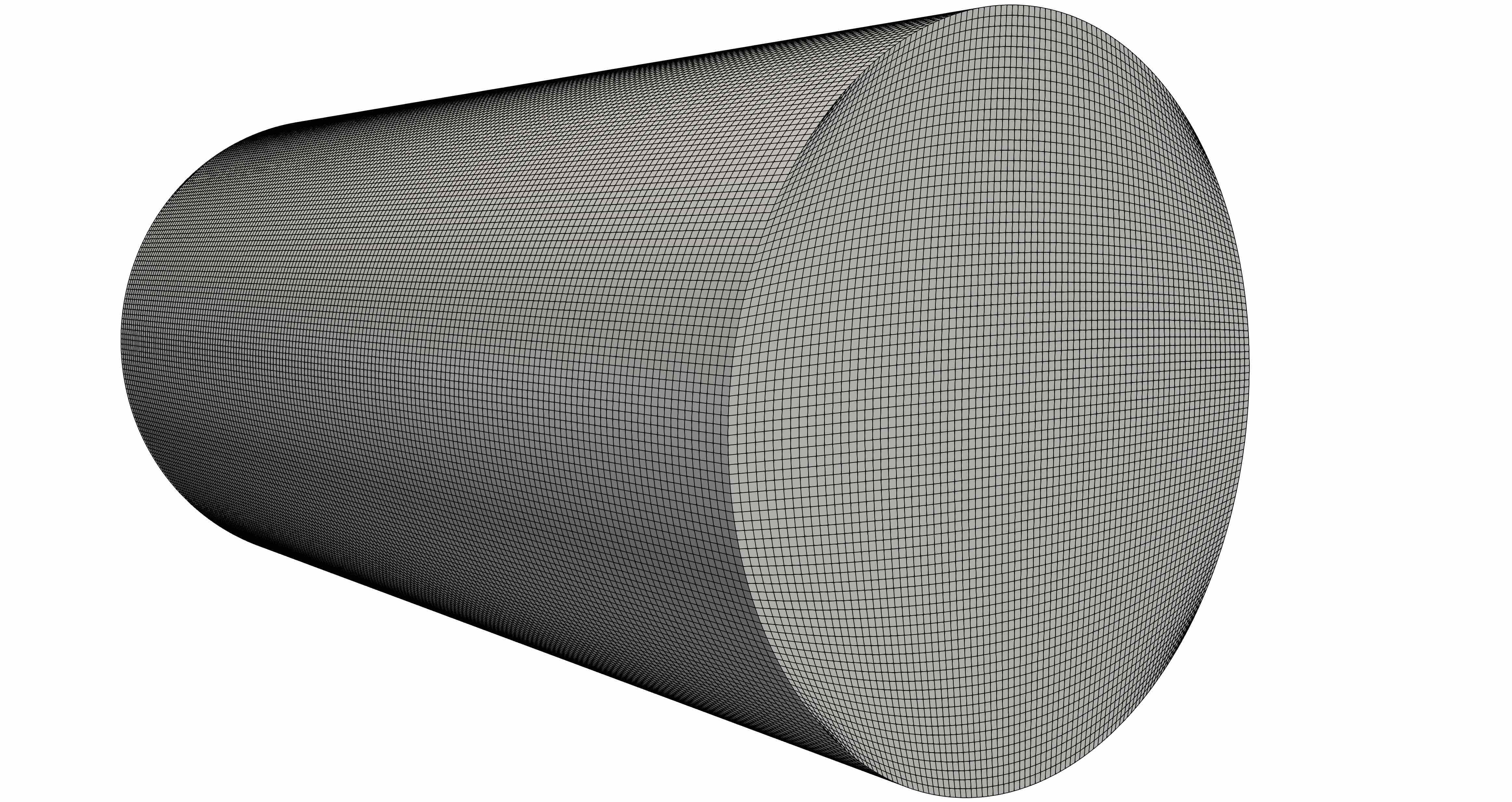}
\caption{Three-dimensional capsule segregation in Hagen-Poiseuille flow. The Eulerian mesh is composed of $80 \times 80 \times 160$ elements with $k=2$.} 
\label{Emesh}
\end{figure}

The physical domain $\Omega$ is a cylinder with radius $R = 0.005 \;  \text{cm}$ and length $L = 0.02 \;  \text{cm}$. A total of 36 spherical capsules modeled using the Skalak constitutive law are considered, 18 capsules with radius $R_s = 0.00065 \;  \text{cm}$ and 18 capsules with radius $R_l = 0.00091 \;  \text{cm}$. In complex simulations of capsules under flow in which the capsules are modeled with a membrane formulation, it is widespread to consider an isotropic tensile prestress \cite{lac2007hydrodynamic, kumar2014flow, maestre2019dynamics}. This tensile prestress is introduced to avoid significant compressive stresses, which would lead to unphysical buckling instabilities due to the lack of bending rigidity in the formulation. Note that this prestress can be introduced in a lab and can also naturally occur due to osmotic effects \cite{sherwood2003rates}. The isotropic tensile stress is obtained applying an inflation ratio $I = a/a_0 - 1$, where $a_0$ and $a$ are the radii of the capsule before and after the prestress is applied, respectively. Here, we apply an inflation ratio $I = 0.2$ to both types of capsules. Homogeneous Dirichlet boundary conditions for the velocity are applied at the cylinder wall. Periodic boundary conditions are applied in the flow direction. Both the fluid and the capsules are initially at rest. The physical parameters defining this problem are the following: $\rho = 1.0  \; \textrm{g} / \textrm{cm}^3$, $\mu_o = \mu_i = 0.01 \; \textrm{g} / (\textrm{cm} \cdot \textrm{s})$, $G_s = C_I = 0.012  \text{e--}3 \; \textrm{g} / (\textrm{cm}^2 \cdot \textrm{s}^2)$, and $\vec g_V = (0.0 \; \textrm{g} / ( \textrm{cm}^2 \cdot \textrm{s}^2 ),0.0 \; \textrm{g} / ( \textrm{cm}^2 \cdot \textrm{s}^2 ),800.0 \; \textrm{g} / ( \textrm{cm}^2 \cdot \textrm{s}^2 ))$. For the selected value of $\vec g_V$, a Hagen-Poiseuille flow with a wall shear rate $\dot{\gamma} = 200 \; \text{s}^{-1}$ would be obtained if there were no capsules. The dimensionless numbers of this problem are $C_a = 0.182$, $\chi = 0.2184$, $\Lambda = 1.0$, and $R_e = 0.02385$, where the radius of the larger capsules has been used to compute the dimensionless numbers.

\begin{figure}[h!]
\centering
\subfigure[$t = 0.000 \; \textrm{s}$]{\includegraphics[scale=0.06]{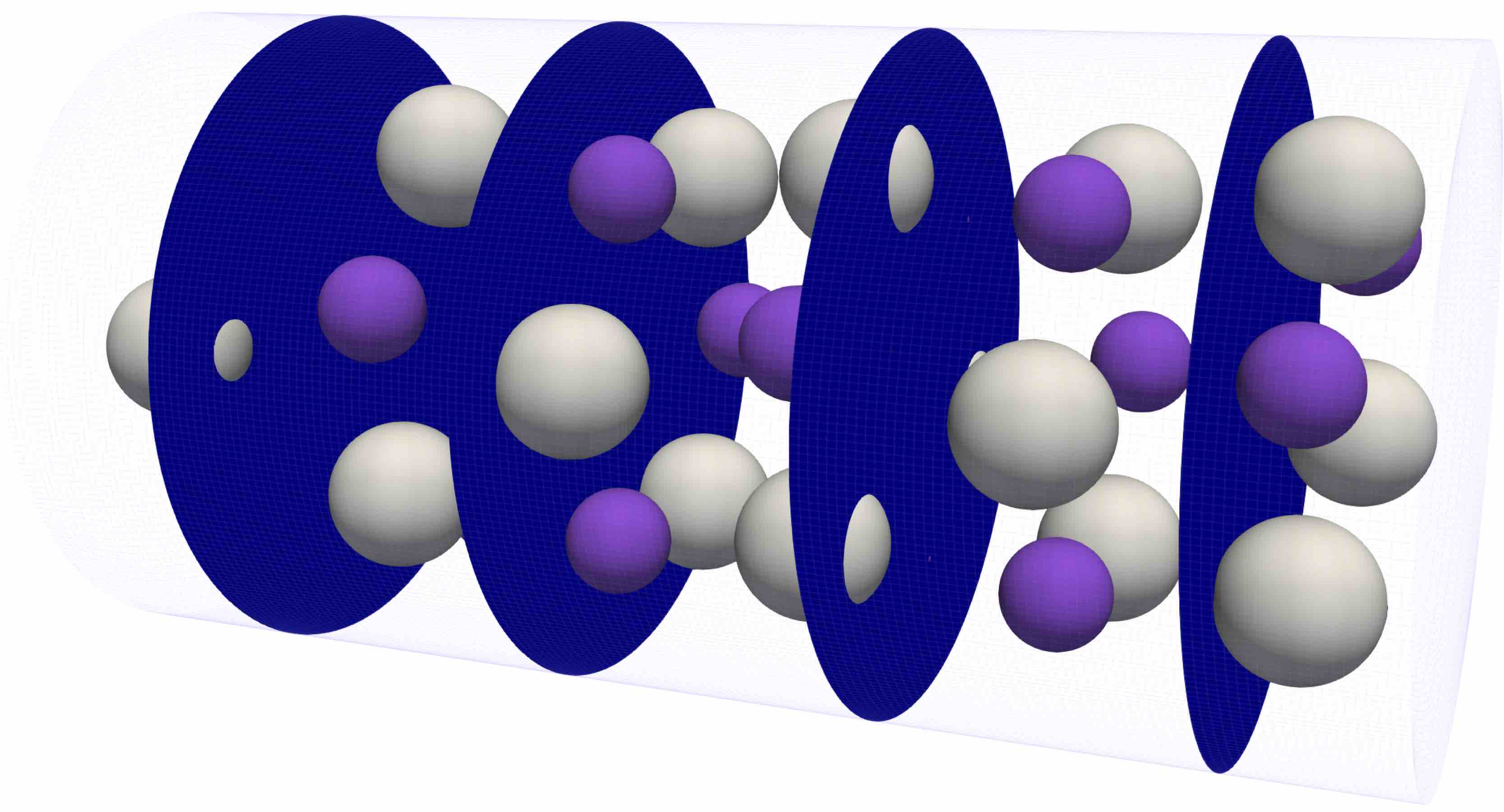}} 
\subfigure[$t = 0.875 \; \textrm{s}$]{\includegraphics[scale=0.06]{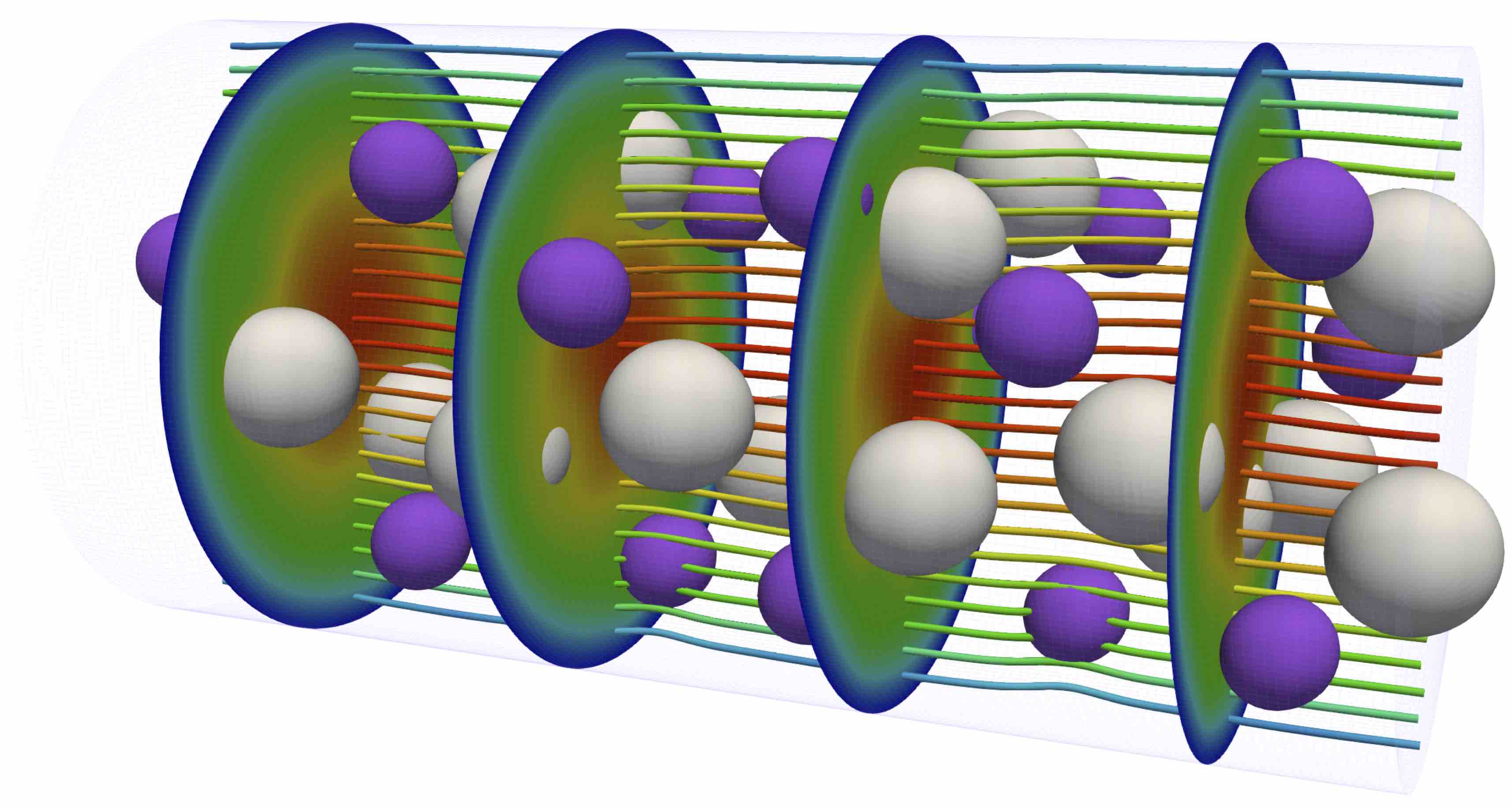}} \\
\subfigure[$t = 1.375 \; \textrm{s}$]{\includegraphics[scale=0.06]{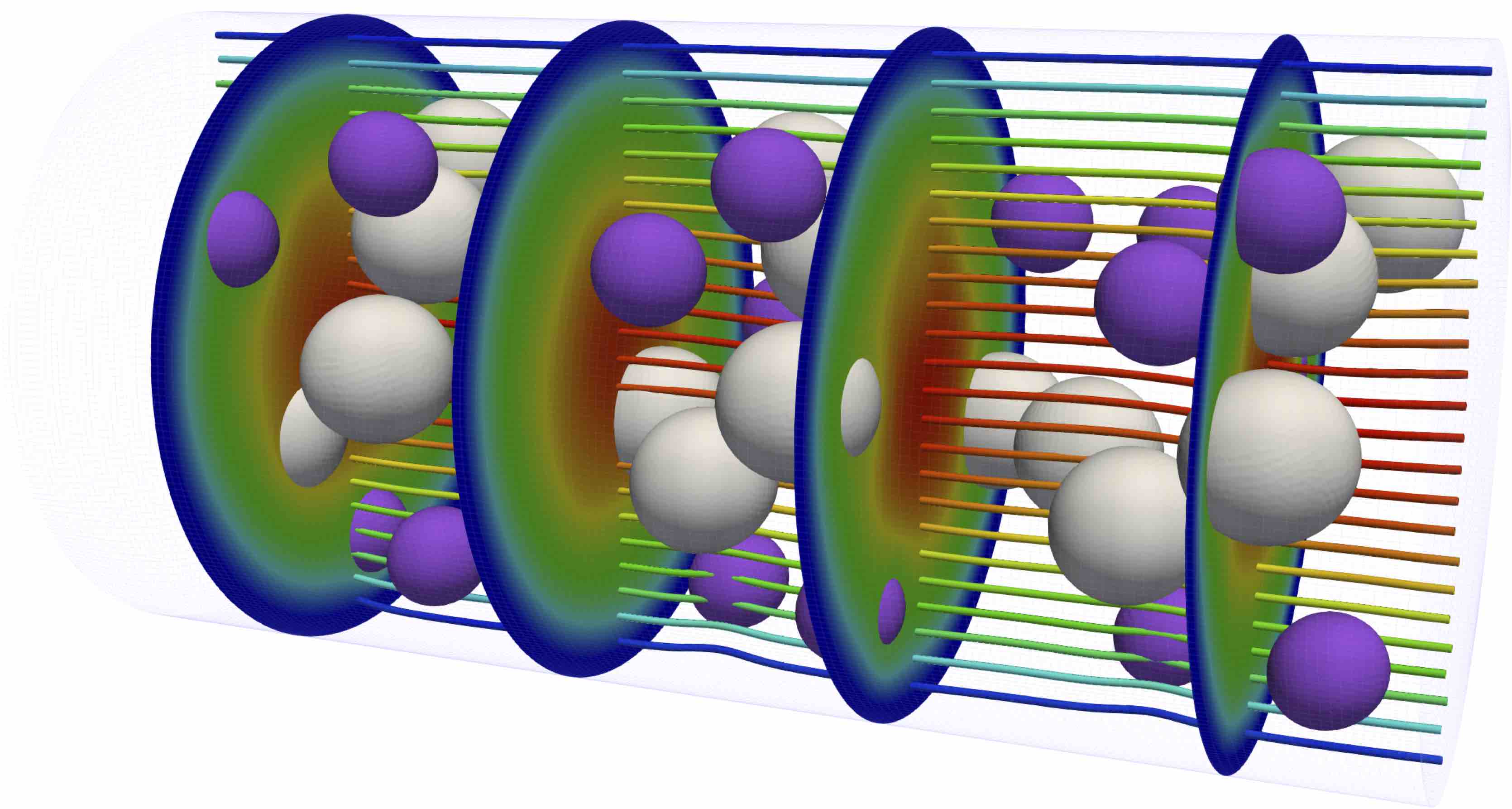}} 
\subfigure[$t = 2.100 \; \textrm{s}$]{\includegraphics[scale=0.06]{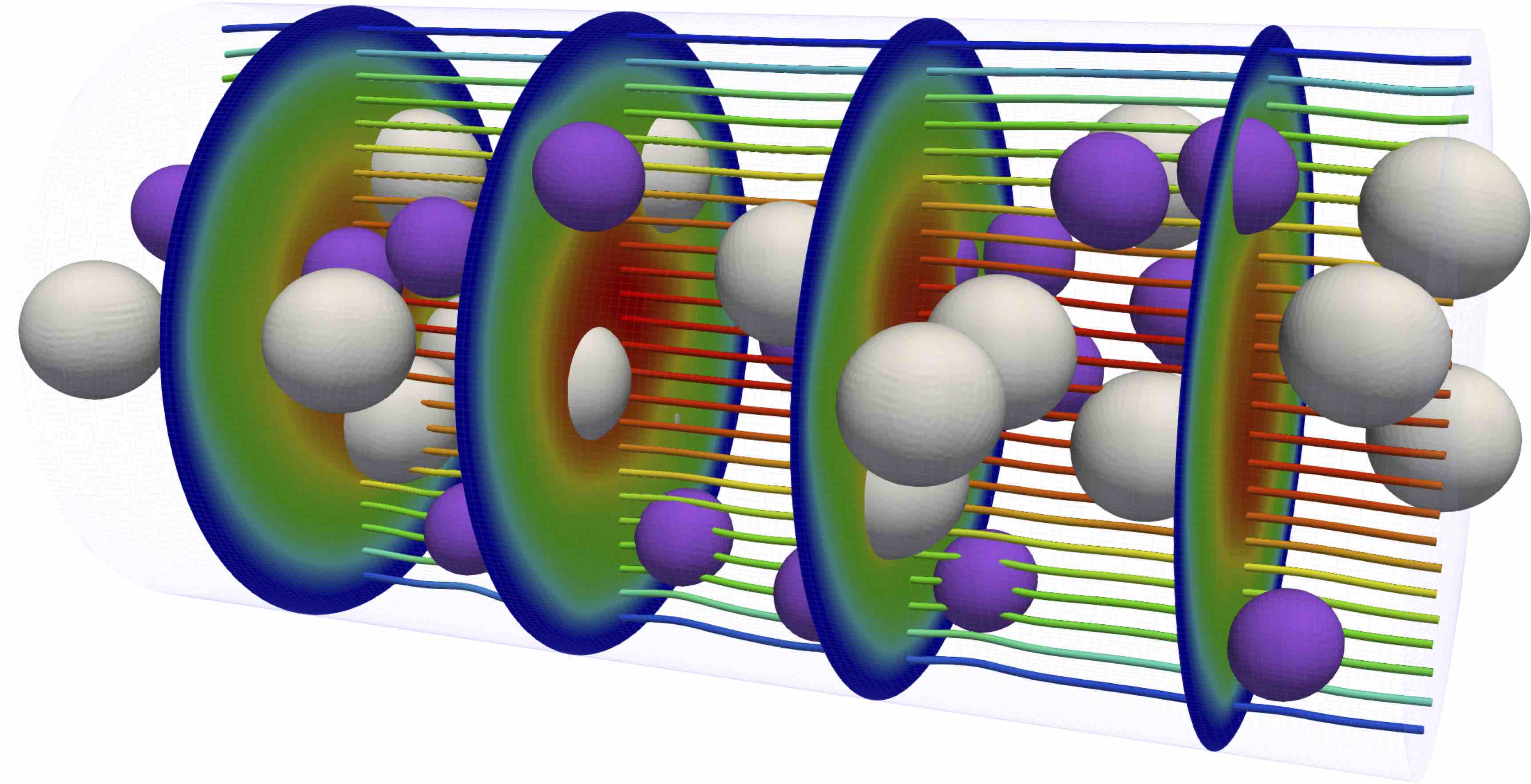}} \\
\caption{ Three-dimensional capsule segregation in Hagen-Poiseuille flow. The velocity magnitude and the two types of capsules are plotted at different times. The large and small capsules are colored in white and purple, respectively. The velocity magnitude is represented in 4 cutting planes and 25 streamlines. In all pictures, the velocity magnitude varies linearly from $0.0 \;\textrm{cm} / \textrm{s}$ (blue color) to  $0.41 \;\textrm{cm} / \textrm{s}$ (red color). The edges of the boundary elements of the Eulerian mesh are included in a see-through view.} \label{HPflow}
\end{figure}

The Eulerian mesh is composed of $80 \times 80 \times 160$ elements with $k=2$, each Lagrangian mesh is composed of 4,776 elements with $p=3$, and a time step $\Delta t = 2.5 \text{e--}4 \; \textrm{s}$ is used. The Eulerian mesh is plotted in Fig. \ref{Emesh}. $e_{DIV}$ is lower than 8.9e--11. For any capsule, the value of $e_{VC}$ at $t = 3.25 \; \text{s}$ (once the smaller capsules have traveled more than 1350 times their radius along the flow direction) is smaller than 9.0e--5. Figs. \ref{HPflow} (a)-(d) show the velocity magnitude and the two types of capsules at four different times. According to the mathematical model of the IB and FD methods, the capsules are not supposed to contact each other or overlap along the simulation due to the zero divergence of velocity and the no-slip and no-penetration conditions at the interface. However, due to discretization errors imposing the incompressibility constraint and the kinematic conditions at the interface, solids are found to unphysically overlap in most IB and FD methods \cite{borazjani2013fluid, Kamensky2015, kamensky2017immersogeometric, liao2015simulations, glowinski2001fictitious, saadat2018immersed}. This phenomenon is sometimes referred to as ``numerical sticking'' \cite{saadat2018immersed}. To prevent numerical sticking, terms based on collision theory or contact theory are often added in IB and FD methods \cite{borazjani2013fluid, Kamensky2015, kamensky2017immersogeometric, liao2015simulations, glowinski2001fictitious, saadat2018immersed}. As in \cite{casquero2018non}, we do not add any additional term to prevent overlap along the simulation and the distance among capsules is always at least equal to the element size of the Eulerian mesh. In other words, the DCIB method imposes the incompressibility constraint and the kinematic equation with enough accuracy to prevent numerical sticking.

For each type of capsule, the distance between the cylinder axis and the center of mass of the capsules is measured. This distance is denoted by $d_s$ and $d_l$ for the smaller and larger capsules, respectively. The quantities $d_s$ and $d_l$ are used to measure which type of capsule tends to travel closer to the cylinder wall. The initial position of the capsules, shown in Fig. \ref{HPflow} (a), is such that $d_s = d_l = 0.003 \;  \text{cm}$. Once the smaller capsules have traveled more than 1350 times their radius along the flow direction, the time averaged values of $d_s$ and $d_l$ are 3.108e--3 cm and 2.381e--3 cm, respectively. Therefore, the smaller capsules tend to travel closer to the cylinder wall. In \cite{kumar2014flow}, the segregation of spherical capsules in a Couette flow is studied using the boundary integral method. The authors found that the smaller capsules tend to travel closer to the parallel walls, which is analogous to the behavior found in this work in a Hagen-Poiseuille flow.

%For each type of capsule, the distance between the cylinder axis and the center of mass of the capsules is measured. This distance is denoted by $d_s$ and $d_l$ for the smaller and larger capsules, respectively. $d_s$ and $d_l$ are used to measure which type of capsule tends to travel closer to the cylinder wall. The initial position of the capsules, shown in Fig. \ref{HPflow} (a), is such that $d_s = d_l = 0.003 \;  \text{cm}$. Fig. \ref{HPflowplot} plots the time evolution of $d_s$ and $d_l$, which shows that the smaller capsules tend to travel closer to the cylinder wall. The time averaged values of $d_s$ and $d_l$ are XXX and XXX, respectively. In 
%\cite{kumar2014flow}, the segregation of spherical capsules in a Couette flow is studied using the boundary integral method. The authors found that the smaller capsules tend to travel closer to the parallel walls, which is analogous to the behavior found in this work in a Hagen-Poiseuille flow.

\section{Conclusions}

The DCIB method results in significantly improved volume conservation of the fluid inside closed co-dimension one solids in comparison with conventional IB methods \cite{griffith2012volume, griffith2012immersed, doddi2008lateral}. The DCIB method also outperforms previous IB methods tailored to tackle the issue of poor conservation of the inner fluid \cite{peskin1993improved, bao2017immersed}. More importantly, the DCIB method is as flexible and efficient as conventional IB methods, which is not the case of the other IB methods tailored to improve volume conservation (e.g., neither Dirichlet nor Neumann boundary conditions can be imposed in the DFIB method \cite{bao2017immersed}; only periodic boundary conditions can be imposed). All the gains in volume conservation of the inner fluid in the DCIB method are due to: (1) the negligible incompressibility errors at the Eulerian level produced by divergence-conforming B-splines, (2) a fully-implicit second-order accurate time discretization of the kinematic equation that computes the Lagrangian displacement from the Eulerian velocity, and (3) the higher inter-element continuity of divergence-conforming B-splines to avoid quadrature errors becoming the dominant discretization error of the kinematic equation that computes the Lagrangian displacement from the Eulerian velocity.

Two-dimensional vesicles and three-dimensional capsules are discretized by directly evaluating formulae of differential geometry for which $C^2$-continuous B-splines with periodic knot vectors and $C^1$-continuous analysis-suitable T-splines are used, respectively. Good agreement is found for the dynamics of vesicles and capsules in Couette flow with respect to simulations based on the boundary integral method and other IB methods. Leveraging the geometric flexibility of divergence-conforming B-splines, the dynamics of a vesicle in Taylor-Couette flow and the segregation of capsules with different sizes in Hagen-Poiseuille flow are studied. Our simulations reveal that (1) the tank-treading and tumbling motions that vesicles undergo in Couette flow prevail in Taylor-Couette flow and (2) the smaller capsules tend to travel closer to the wall and the larger capsules tend to travel closer to the cylinder axis.

\section*{Acknowledgements}

H. Casquero and Y.J. Zhang were supported in part by the PECASE Award N00014-16-1-2254 and NSF grant CBET-1804929. T.J.R. Hughes were partially supported by the Office of Naval Research, USA (Grant Nos. N00014-17-1-2119 and N00014-13-1-0500). C. Bona-Casas was supported by the Spanish Ministry of Economy and Competitiveness (MINECO/AEI/FEDER, UE) through project DPI2017-86610-P. This work used the Extreme Science and Engineering Discovery Environment (XSEDE), which is supported by National Science Foundation grant number OCI-1053575.  Specifically, it used the Bridges system, which is supported by NSF award number ACI-1445606, at the Pittsburgh Supercomputing Center (PSC). This work also used the computer resources at MareNostrum and the technical support provided by Barcelona Supercomputing Center (RES-FI-2018-3-0020, RES-IM-2019-2-0004).

% We also acknowledge the open source scientific libraries PETSc, PetIGA, PetIGA-MF, and their developers.

% and D. Toshniwal

%\section*{References}




\bibliographystyle{elsarticle-num}
\bibliography{./Bibliography}

\begin{thebibliography}{100}
\expandafter\ifx\csname url\endcsname\relax
  \def\url#1{\texttt{#1}}\fi
\expandafter\ifx\csname urlprefix\endcsname\relax\def\urlprefix{URL }\fi
\expandafter\ifx\csname href\endcsname\relax
  \def\href#1#2{#2} \def\path#1{#1}\fi

\bibitem{casquero2018non}
H.~Casquero, Y.~J. Zhang, C.~Bona-Casas, L.~Dalcin, H.~Gomez, Non-body-fitted
  fluid--structure interaction: {D}ivergence-conforming {B}-splines,
  fully-implicit dynamics, and variational formulation, Journal of
  Computational Physics 374 (2018) 625--653.

\bibitem{Peskin1972}
C.~Peskin, Flow patterns around heart valves: A numerical method, Journal of
  Computational Physics 10 (1972) 252--271.

\bibitem{Peskin1977}
C.~Peskin, Numerical analysis of blood flow in the heart, Journal of
  Computational Physics 25~(3) (1977) 220--252.

\bibitem{Peskin2002}
C.~Peskin, The immersed boundary method, Acta Numerica 11 (2002) 479--517.

\bibitem{boffi2008hyper}
D.~Boffi, L.~Gastaldi, L.~Heltai, C.~S. Peskin, On the hyper-elastic
  formulation of the immersed boundary method, Computer Methods in Applied
  Mechanics and Engineering 197~(25) (2008) 2210--2231.

\bibitem{boffi2003finite}
D.~Boffi, L.~Gastaldi, A finite element approach for the immersed boundary
  method, Computers \& Structures 81~(8) (2003) 491--501.

\bibitem{liu2006immersed}
W.~K. Liu, Y.~Liu, D.~Farrell, L.~Zhang, X.~S. Wang, Y.~Fukui, N.~Patankar,
  Y.~Zhang, C.~Bajaj, J.~Lee, et~al., Immersed finite element method and its
  applications to biological systems, Computer Methods in Applied Mechanics and
  Engineering 195~(13-16) (2006) 1722--1749.

\bibitem{saadat2018immersed}
A.~Saadat, C.~J. Guido, G.~Iaccarino, E.~S. Shaqfeh, Immersed-finite-element
  method for deformable particle suspensions in viscous and viscoelastic media,
  Physical Review E 98~(6) (2018) 063316.

\bibitem{mittal2005immersed}
R.~Mittal, G.~Iaccarino, Immersed boundary methods, Annual Review of Fluid
  Mechanics 37 (2005) 239--261.

\bibitem{glowinski1999distributed}
R.~Glowinski, T.-W. Pan, T.~I. Hesla, D.~D. Joseph, A distributed {L}agrange
  multiplier/fictitious domain method for particulate flows, International
  Journal of Multiphase Flow 25~(5) (1999) 755--794.

\bibitem{glowinski2001fictitious}
R.~Glowinski, T.~Pan, T.~Hesla, D.~Joseph, J.~Periaux, A fictitious domain
  approach to the direct numerical simulation of incompressible viscous flow
  past moving rigid bodies: application to particulate flow, Journal of
  Computational Physics 169~(2) (2001) 363--426.

\bibitem{baaijens2001fictitious}
F.~P. Baaijens, A fictitious domain/mortar element method for fluid-structure
  interaction, International Journal for Numerical Methods in Fluids 35~(7)
  (2001) 743--761.

\bibitem{van2004combined}
R.~Van~Loon, P.~D. Anderson, J.~De~Hart, F.~P. Baaijens, A combined fictitious
  domain/adaptive meshing method for fluid--structure interaction in heart
  valves, International Journal for Numerical Methods in Fluids 46~(5) (2004)
  533--544.

\bibitem{yu2005dlm}
Z.~Yu, A {DLM}/{FD} method for fluid/flexible-body interactions, Journal of
  Computational Physics 207~(1) (2005) 1--27.

\bibitem{peskin1993improved}
C.~S. Peskin, B.~F. Printz, Improved volume conservation in the computation of
  flows with immersed elastic boundaries, Journal of Computational Physics
  105~(1) (1993) 33--46.

\bibitem{griffith2012volume}
B.~E. Griffith, On the volume conservation of the immersed boundary method,
  Communications in Computational Physics 12~(2) (2012) 401--432.

\bibitem{strychalski2016intracellular}
W.~Strychalski, R.~D. Guy, Intracellular pressure dynamics in blebbing cells,
  Biophysical Journal 110~(5) (2016) 1168--1179.

\bibitem{boilevin2019numerical}
L.~Boilevin-Kayl, M.~A. Fern{\'a}ndez, J.-F. Gerbeau, Numerical methods for
  immersed {FSI} with thin-walled structures, Computers \& Fluids 179 (2019)
  744--763.

\bibitem{boilevin2019loosely}
L.~Boilevin-Kayl, M.~A. Fern{\'a}ndez, J.-F. Gerbeau, A loosely coupled scheme
  for fictitious domain approximations of fluid-structure interaction problems
  with immersed thin-walled structures, SIAM Journal on Scientific Computing
  41~(2) (2019) B351--B374.

\bibitem{bao2017immersed}
Y.~Bao, A.~Donev, B.~E. Griffith, D.~M. McQueen, C.~S. Peskin, An immersed
  boundary method with divergence-free velocity interpolation and force
  spreading, Journal of Computational Physics 347 (2017) 183--206.

\bibitem{mendez2014unstructured}
S.~Mendez, E.~Gibaud, F.~Nicoud, An unstructured solver for simulations of
  deformable particles in flows at arbitrary {R}eynolds numbers, Journal of
  Computational Physics 256 (2014) 465--483.

\bibitem{galvin2012stabilizing}
K.~J. Galvin, A.~Linke, L.~G. Rebholz, N.~E. Wilson, Stabilizing poor mass
  conservation in incompressible flow problems with large irrotational forcing
  and application to thermal convection, Computer Methods in Applied Mechanics
  and Engineering 237 (2012) 166--176.

\bibitem{john2017divergence}
V.~John, A.~Linke, C.~Merdon, M.~Neilan, L.~G. Rebholz, On the divergence
  constraint in mixed finite element methods for incompressible flows, SIAM
  Review 59~(3) (2017) 492--544.

\bibitem{peng2010multiscale}
Z.~Peng, R.~J. Asaro, Q.~Zhu, Multiscale simulation of erythrocyte membranes,
  Physical Review E 81~(3) (2010) 031904.

\bibitem{yazdani2012three}
A.~Yazdani, P.~Bagchi, Three-dimensional numerical simulation of vesicle
  dynamics using a front-tracking method, Physical Review E 85~(5) (2012)
  056308.

\bibitem{shen2017interaction}
Z.~Shen, A.~Farutin, M.~Thi{\'e}baud, C.~Misbah, Interaction and rheology of
  vesicle suspensions in confined shear flow, Physical Review Fluids 2~(10)
  (2017) 103101.

\bibitem{li2012volume}
Y.~Li, E.~Jung, W.~Lee, H.~G. Lee, J.~Kim, Volume preserving immersed boundary
  methods for two-phase fluid flows, International Journal for Numerical
  Methods in Fluids 69~(4) (2012) 842--858.

\bibitem{alauzet2016nitsche}
F.~Alauzet, B.~Fabr{\`e}ges, M.~A. Fern{\'a}ndez, M.~Landajuela, Nitsche-{XFEM}
  for the coupling of an incompressible fluid with immersed thin-walled
  structures, Computer Methods in Applied Mechanics and Engineering 301 (2016)
  300--335.

\bibitem{1003.000}
T.~J.~R. Hughes, J.~A. Cottrell, Y.~Bazilevs, Isogeometric analysis {CAD},
  finite elements, {NURBS}, exact geometry and mesh refinement, Computacional
  Methods in Applied Mechanics and Engineering 194 (2005) 4135--4195.

\bibitem{1002.000}
J.~A. Cottrell, T.~J.~R. Hughes, Y.~Bazilevs, Isogeometric Analysis Toward
  Integration of {CAD} and {FEA}, Wiley, 2009.

\bibitem{ruberg2014fixed}
T.~R{\"u}berg, F.~Cirak, A fixed-grid {B}-spline finite element technique for
  fluid--structure interaction, International Journal for Numerical Methods in
  Fluids 74~(9) (2014) 623--660.

\bibitem{Casquero2015}
H.~Casquero, C.~Bona-Casas, H.~Gomez, A {NURBS}-based immersed methodology for
  fluid-structure interaction, Computer Methods in Applied Mechanics and
  Engineering 284 (2015) 943--970.

\bibitem{Kamensky2015}
D.~Kamensky, M.-C. Hsu, D.~Schillinger, J.~A. Evans, A.~Aggarwal, Y.~Bazilevs,
  M.~S. Sacks, T.~J.~R. Hughes, An immersogeometric variational framework for
  fluid-structure interaction: {A}pplication to bioprosthetic heart valves,
  Computer Methods in Applied Mechanics and Engineering 284 (2015) 1005--1053.

\bibitem{Casquero2016a}
H.~Casquero, L.~Liu, C.~Bona-Casas, Y.~Zhang, H.~Gomez, A hybrid
  variational-collocation immersed method for fluid-structure interaction using
  unstructured {T}-splines, International Journal for Numerical Methods in
  Engineering 105~(11) (2016) 855--880.

\bibitem{Hsu2015}
M.-C. Hsu, D.~Kamensky, F.~Xu, J.~Kiendl, C.~Wang, M.~Wu, J.~Mineroff,
  A.~Reali, Y.~Bazilevs, M.~Sacks, Dynamic and fluid-structure interaction
  simulations of bioprosthetic heart valves using parametric design with
  {T}-splines and {F}ung-type material models, Computational Mechanics 55~(6)
  (2015) 1211--1225.

\bibitem{kamensky2017projection}
D.~Kamensky, J.~A. Evans, M.-C. Hsu, Y.~Bazilevs, Projection-based
  stabilization of interface lagrange multipliers in immersogeometric
  fluid–thin structure interaction analysis, with application to heart valve
  modeling, Computers \& Mathematics with Applications 74~(9) (2017)
  2068--2088.

\bibitem{bazilevs2017new}
Y.~Bazilevs, G.~Moutsanidis, J.~Bueno, K.~Kamran, D.~Kamensky, M.~C. Hillman,
  H.~Gomez, J.~Chen, A new formulation for air-blast fluid--structure
  interaction using an immersed approach: part {II}--coupling of {IGA} and
  meshfree discretizations, Computational Mechanics 60~(1) (2017) 101--116.

\bibitem{kadapa2016fictitious}
C.~Kadapa, W.~Dettmer, D.~Peri{\'c}, A fictitious domain/distributed lagrange
  multiplier based fluid--structure interaction scheme with hierarchical
  {B}-spline grids, Computer Methods in Applied Mechanics and Engineering 301
  (2016) 1--27.

\bibitem{kadapa2018stabilised}
C.~Kadapa, W.~Dettmer, D.~Peri{\'c}, A stabilised immersed framework on
  hierarchical {B}-spline grids for fluid-flexible structure interaction with
  solid--solid contact, Computer Methods in Applied Mechanics and Engineering
  335 (2018) 472--489.

\bibitem{heltai2017natural}
L.~Heltai, J.~Kiendl, A.~DeSimone, A.~Reali, A natural framework for
  isogeometric fluid--structure interaction based on {BEM}--shell coupling,
  Computer Methods in Applied Mechanics and Engineering 316 (2017) 522--546.

\bibitem{maestre20173d}
J.~Maestre, J.~Pallares, I.~Cuesta, M.~A. Scott, A 3{D} isogeometric {BE}--{FE}
  analysis with dynamic remeshing for the simulation of a deformable particle
  in shear flows, Computer Methods in Applied Mechanics and Engineering 326
  (2017) 70--101.

\bibitem{moutsanidis2018hyperbolic}
G.~Moutsanidis, D.~Kamensky, J.~Chen, Y.~Bazilevs, Hyperbolic phase field
  modeling of brittle fracture: {P}art {II}--immersed {IGA}--{RKPM} coupling
  for air-blast--structure interaction, Journal of the Mechanics and Physics of
  Solids 121 (2018) 114--132.

\bibitem{casquero2017nurbs}
H.~Casquero, C.~Bona-Casas, H.~Gomez, {NURBS}-based numerical proxies for red
  blood cells and circulating tumor cells in microscale blood flow, Computer
  Methods in Applied Mechanics and Engineering 316 (2017) 646--667.

\bibitem{scott1985norm}
L.~R. Scott, M.~Vogelius, Norm estimates for a maximal right inverse of the
  divergence operator in spaces of piecewise polynomials, ESAIM: Mathematical
  Modelling and Numerical Analysis 19~(1) (1985) 111--143.

\bibitem{zhang2011divergence}
S.~Zhang, Divergence-free finite elements on tetrahedral grids for $k \geq 6$,
  Mathematics of Computation 80~(274) (2011) 669--695.

\bibitem{neilan2015discrete}
M.~Neilan, Discrete and conforming smooth de {R}ham complexes in three
  dimensions, Mathematics of Computation 84~(295) (2015) 2059--2081.

\bibitem{falk2013stokes}
R.~S. Falk, M.~Neilan, Stokes complexes and the construction of stable finite
  elements with pointwise mass conservation, SIAM Journal on Numerical Analysis
  51~(2) (2013) 1308--1326.

\bibitem{guzman2017scott}
J.~Guzman, R.~Scott, The {S}cott-{V}ogelius finite elements revisited, arXiv
  preprint arXiv:1705.00020.

\bibitem{raviart1977mixed}
P.-A. Raviart, J.-M. Thomas, A mixed finite element method for 2-nd order
  elliptic problems, in: Mathematical Aspects of Finite Element Methods,
  Springer, 1977, pp. 292--315.

\bibitem{buffa2011isogeometricA}
A.~Buffa, C.~De~Falco, G.~Sangalli, Isogeometric analysis: stable elements for
  the 2{D} {S}tokes equation, International Journal for Numerical Methods in
  Fluids 65~(11-12) (2011) 1407--1422.

\bibitem{buffa2011isogeometricB}
A.~Buffa, J.~Rivas, G.~Sangalli, R.~V{\'a}zquez, Isogeometric discrete
  differential forms in three dimensions, SIAM Journal on Numerical Analysis
  49~(2) (2011) 818--844.

\bibitem{evans2013isogeometric}
J.~A. Evans, T.~J.~R. Hughes, Isogeometric divergence-conforming {B}-splines
  for the {D}arcy--{S}tokes--{B}rinkman equations, Mathematical Models and
  Methods in Applied Sciences 23~(04) (2013) 671--741.

\bibitem{evans2013isogeometric2}
J.~A. Evans, T.~J.~R. Hughes, Isogeometric divergence-conforming {B}-splines
  for the steady {N}avier--{S}tokes equations, Mathematical Models and Methods
  in Applied Sciences 23~(08) (2013) 1421--1478.

\bibitem{john}
J.~A. Evans, T.~J.~R. Hughes, Isogeometric divergence-conforming {B}-splines
  for the unsteady {N}avier--{S}tokes equations, Journal of Computational
  Physics 241 (2013) 141--167.

\bibitem{boffi2011finite}
D.~Boffi, N.~Cavallini, L.~Gastaldi, Finite element approach to immersed
  boundary method with different fluid and solid densities, Mathematical Models
  and Methods in Applied Sciences 21~(12) (2011) 2523--2550.

\bibitem{hesch2012continuum}
C.~Hesch, A.~Gil, A.~A. Carreno, J.~Bonet, On continuum immersed strategies for
  fluid--structure interaction, Computer Methods in Applied Mechanics and
  Engineering 247 (2012) 51--64.

\bibitem{guzman2013conforming}
J.~Guzm{\'a}n, M.~Neilan, Conforming and divergence-free stokes elements in
  three dimensions, IMA Journal of Numerical Analysis 34~(4) (2013) 1489--1508.

\bibitem{kamensky2017immersogeometric}
D.~Kamensky, M.-C. Hsu, Y.~Yu, J.~A. Evans, M.~S. Sacks, T.~J.~R. Hughes,
  Immersogeometric cardiovascular fluid--structure interaction analysis with
  divergence-conforming {B}-splines, Computer Methods in Applied Mechanics and
  Engineering 314 (2017) 408--472.

\bibitem{lim2019biomedical}
F.~Lim, Biomedical applications of microencapsulation, CRC press, 2019.

\bibitem{li2005liposome}
S.~Li, J.~Nickels, A.~F. Palmer, Liposome-encapsulated actin--hemoglobin
  ({LEAcHb}) artificial blood substitutes, Biomaterials 26~(17) (2005)
  3759--3769.

\bibitem{noireaux2004vesicle}
V.~Noireaux, A.~Libchaber, A vesicle bioreactor as a step toward an artificial
  cell assembly, Proceedings of the National Academy of Sciences 101~(51)
  (2004) 17669--17674.

\bibitem{andaloussi2013extracellular}
S.~E. Andaloussi, I.~M{\"a}ger, X.~O. Breakefield, M.~J. Wood, Extracellular
  vesicles: biology and emerging therapeutic opportunities, Nature Reviews Drug
  Discovery 12~(5) (2013) 347.

\bibitem{yanez2015biological}
M.~Y{\'a}{\~n}ez-M{\'o}, P.~R.-M. Siljander, Z.~Andreu, A.~Bedina~Zavec, F.~E.
  Borr{\`a}s, E.~I. Buzas, K.~Buzas, E.~Casal, F.~Cappello, J.~Carvalho,
  et~al., Biological properties of extracellular vesicles and their
  physiological functions, Journal of Extracellular Vesicles 4~(1) (2015)
  27066.

\bibitem{freund2014numerical}
J.~B. Freund, Numerical simulation of flowing blood cells, Annual review of
  fluid mechanics 46 (2014) 67--95.

\bibitem{lanotte2016red}
L.~Lanotte, J.~Mauer, S.~Mendez, D.~A. Fedosov, J.-M. Fromental, V.~Claveria,
  F.~Nicoud, G.~Gompper, M.~Abkarian, Red cells’ dynamic morphologies govern
  blood shear thinning under microcirculatory flow conditions, Proceedings of
  the National Academy of Sciences 113~(47) (2016) 13289--13294.

\bibitem{mauer2018flow}
J.~Mauer, S.~Mendez, L.~Lanotte, F.~Nicoud, M.~Abkarian, G.~Gompper, D.~A.
  Fedosov, Flow-induced transitions of red blood cell shapes under shear,
  Physical Review Letters 121~(11) (2018) 118103.

\bibitem{barthes2016motion}
D.~Barthes-Biesel, Motion and deformation of elastic capsules and vesicles in
  flow, Annual Review of fluid mechanics 48 (2016) 25--52.

\bibitem{abreu2014fluid}
D.~Abreu, M.~Levant, V.~Steinberg, U.~Seifert, Fluid vesicles in flow, Advances
  in Colloid and Interface Science 208 (2014) 129--141.

\bibitem{tsubota2006particle}
K.-i. Tsubota, S.~Wada, T.~Yamaguchi, Particle method for computer simulation
  of red blood cell motion in blood flow, Computer Methods and Programs in
  Biomedicine 83~(2) (2006) 139--146.

\bibitem{tsubota2010effect}
K.-i. Tsubota, S.~Wada, Effect of the natural state of an elastic cellular
  membrane on tank-treading and tumbling motions of a single red blood cell,
  Physical Review E 81~(1) (2010) 011910.

\bibitem{boedec20113d}
G.~Boedec, M.~Leonetti, M.~Jaeger, 3{D} vesicle dynamics simulations with a
  linearly triangulated surface, Journal of Computational Physics 230~(4)
  (2011) 1020--1034.

\bibitem{biben2011three}
T.~Biben, A.~Farutin, C.~Misbah, Three-dimensional vesicles under shear flow:
  Numerical study of dynamics and phase diagram, Physical Review E 83~(3)
  (2011) 031921.

\bibitem{zhao2011dynamics}
H.~Zhao, A.~P. Spann, E.~S. Shaqfeh, The dynamics of a vesicle in a wall-bound
  shear flow, Physics of Fluids 23~(12) (2011) 121901.

\bibitem{charrier1989free}
J.~Charrier, S.~Shrivastava, R.~Wu, Free and constrained inflation of elastic
  membranes in relation to thermoforming—non-axisymmetric problems, The
  Journal of Strain Analysis for Engineering Design 24~(2) (1989) 55--74.

\bibitem{shrivastava1993large}
S.~Shrivastava, J.~Tang, Large deformation finite element analysis of
  non-linear viscoelastic membranes with reference to thermoforming, The
  Journal of Strain Analysis for Engineering Design 28~(1) (1993) 31--51.

\bibitem{eggleton1998large}
C.~D. Eggleton, A.~S. Popel, Large deformation of red blood cell ghosts in a
  simple shear flow, Physics of Fluids 10~(8) (1998) 1834--1845.

\bibitem{bhalla2013unified}
A.~P.~S. Bhalla, R.~Bale, B.~E. Griffith, N.~A. Patankar, A unified
  mathematical framework and an adaptive numerical method for fluid--structure
  interaction with rigid, deforming, and elastic bodies, Journal of
  Computational Physics 250 (2013) 446--476.

\bibitem{Li2012}
X.~Li, J.~Zheng, T.~W. Sederberg, T.~J.~R. Hughes, M.~A. Scott, On linear
  independence of {T}-spline blending functions, Computer Aided Geometric
  Design 29 (2012) 63--76.

\bibitem{wei2017truncated}
X.~Wei, Y.~Zhang, L.~Liu, T.~J.~R. Hughes, Truncated {T}-splines:
  {F}undamentals and methods, Computer Methods in Applied Mechanics and
  Engineering 316 (2017) 349--372.

\bibitem{casquero2017arbitrary}
H.~Casquero, L.~Liu, Y.~Zhang, A.~Reali, J.~Kiendl, H.~Gomez, Arbitrary-degree
  {T}-splines for isogeometric analysis of fully nonlinear {K}irchhoff--{L}ove
  shells, Computer-Aided Design 82 (2017) 140--153.

\bibitem{toshniwal2017smooth}
D.~Toshniwal, H.~Speleers, T.~J.~R. Hughes, Smooth cubic spline spaces on
  unstructured quadrilateral meshes with particular emphasis on extraordinary
  points: {G}eometric design and isogeometric analysis considerations, Computer
  Methods in Applied Mechanics and Engineering 327 (2017) 411--458.

\bibitem{casquero2019}
H.~Casquero, X.~Wei, D.~Toshniwal, A.~Li, T.~J.~R. Hughes, J.~Kiendl, Y.~J.
  Zhang, Seamless integration of design and {K}irchhoff-{L}ove shell analysis
  using analysis-suitable unstructured {T}-splines, Computer Methods in Applied
  Mechanics and Engineering, under review.

\bibitem{thiebaud2013rheology}
M.~Thi{\'e}baud, C.~Misbah, Rheology of a vesicle suspension with finite
  concentration: {A} numerical study, Physical Review E 88~(6) (2013) 062707.

\bibitem{doddi2008lateral}
S.~K. Doddi, P.~Bagchi, Lateral migration of a capsule in a plane poiseuille
  flow in a channel, International Journal of Multiphase Flow 34~(10) (2008)
  966--986.

\bibitem{tryggvason2001front}
G.~Tryggvason, B.~Bunner, A.~Esmaeeli, D.~Juric, N.~Al-Rawahi, W.~Tauber,
  J.~Han, S.~Nas, Y.-J. Jan, A front-tracking method for the computations of
  multiphase flow, Journal of Computational Physics 169~(2) (2001) 708--759.

\bibitem{kaoui2008lateral}
B.~Kaoui, G.~Ristow, I.~Cantat, C.~Misbah, W.~Zimmermann, Lateral migration of
  a two-dimensional vesicle in unbounded {P}oiseuille flow, Physical Review E
  77~(2) (2008) 021903.

\bibitem{skalak1973strain}
R.~Skalak, A.~Tozeren, R.~Zarda, S.~Chien, Strain energy function of red blood
  cell membranes, Biophysical Journal 13~(3) (1973) 245--264.

\bibitem{lac2004spherical}
E.~Lac, D.~Barthes-Biesel, N.~Pelekasis, J.~Tsamopoulos, Spherical capsules in
  three-dimensional unbounded stokes flows: effect of the membrane constitutive
  law and onset of buckling, Journal of Fluid Mechanics 516 (2004) 303--334.

\bibitem{barthes2002effect}
D.~Barthes-Biesel, A.~Diaz, E.~Dhenin, Effect of constitutive laws for
  two-dimensional membranes on flow-induced capsule deformation, Journal of
  Fluid Mechanics 460 (2002) 211--222.

\bibitem{piegl2012nurbs}
L.~Piegl, W.~Tiller, The NURBS book, Springer Science \& Business Media, 2012.

\bibitem{liu2013isogeometric}
J.~Liu, L.~Ded{\`e}, J.~A. Evans, M.~J. Borden, T.~J.~R. Hughes, Isogeometric
  analysis of the advective {C}ahn--{H}illiard equation: spinodal decomposition
  under shear flow, Journal of Computational Physics 242 (2013) 321--350.

\bibitem{Dalcin2016}
L.~Dalcin, N.~Collier, P.~Vignal, A.~Cortes, V.~Calo, Pet{IGA}: A framework for
  high-performance isogeometric analysis, Computer Methods in Applied Mechanics
  and Engineering 308 (2016) 151 -- 181.

\bibitem{Jansen2000}
K.~Jansen, C.~Whiting, G.~Hulbert, Generalized-$\alpha$ method for integrating
  the filtered {N}avier-{S}tokes equations with a stabilized finite element
  method, Computer Methods in Applied Mechanics and Engineering 190 (2000)
  305--319.

\bibitem{Bazilevs2012}
Y.~Bazilevs, K.~Takizawa, T.~E. Tezduyar, Computational fluid-structure
  interaction: methods and applications, John Wiley \& Sons, 2012.

\bibitem{sarmiento2017petiga}
A.~F. Sarmiento, A.~M. C{\^o}rtes, D.~Garcia, L.~Dalcin, N.~Collier, V.~M.
  Calo, Pet{IGA}-{MF}: a multi-field high-performance toolbox for
  structure-preserving {B}-splines spaces, Journal of Computational Science 18
  (2017) 117--131.

\bibitem{cortes2017scalable2}
A.~C{\^o}rtes, L.~Dalcin, A.~Sarmiento, N.~Collier, V.~Calo, A scalable
  block-preconditioning strategy for divergence-conforming {B}-spline
  discretizations of the {S}tokes problem, Computer Methods in Applied
  Mechanics and Engineering 316 (2017) 839--858.

\bibitem{espath2016energy}
L.~Espath, A.~Sarmiento, P.~Vignal, B.~Varga, A.~Cortes, L.~Dalcin, V.~Calo,
  Energy exchange analysis in droplet dynamics via the
  {N}avier--{S}tokes--{C}ahn--{H}illiard model, Journal of Fluid Mechanics 797
  (2016) 389--430.

\bibitem{petsc-web-page}
S.~Balay, S.~Abhyankar, M.~F. Adams, J.~Brown, P.~Brune, K.~Buschelman,
  L.~Dalcin, V.~Eijkhout, W.~D. Gropp, D.~Kaushik, M.~G. Knepley, D.~A. May,
  L.~C. McInnes, K.~Rupp, B.~F. Smith, S.~Zampini, H.~Zhang, H.~Zhang, {PETS}c
  {W}eb page, \url{http://www.mcs.anl.gov/petsc} (2017).

\bibitem{Brune2015}
P.~R. Brune, M.~G. Knepley, B.~F. Smith, X.~Tu, Composing scalable nonlinear
  algebraic solvers, SIAM Review 57~(4) (2015) 535--565.

\bibitem{saad1993flexible}
Y.~Saad, A flexible inner-outer preconditioned {GMRES} algorithm, SIAM Journal
  on Scientific Computing 14~(2) (1993) 461--469.

\bibitem{Saad1986}
Y.~Saad, M.~H. Schultz, {GMRES}: {A} generalized minimal residual algorithm for
  solving nonsymmetric linear systems, SIAM Journal on Scientific and
  Statistical Computing 7 (1986) 856--869.

\bibitem{cortez2004resonance}
R.~Cortez, C.~Peskin, J.~Stockie, D.~Varela, Parametric resonance in immersed
  elastic boundaries, SIAM Journal on Applied Mathematics 65~(2) (2004)
  494--520.

\bibitem{cortez2004resonance2}
W.~Ko, J.~Stockie, Correction to "parametric resonance in immersed elastic
  boundaries", arXiv:1207.4744.

\bibitem{griffith2012immersed}
B.~E. Griffith, Immersed boundary model of aortic heart valve dynamics with
  physiological driving and loading conditions, International Journal for
  Numerical Methods in Biomedical Engineering 28~(3) (2012) 317--345.

\bibitem{yu2018error}
Y.~Yu, D.~Kamensky, M.-C. Hsu, X.~Y. Lu, Y.~Bazilevs, T.~J. Hughes, Error
  estimates for projection-based dynamic augmented lagrangian boundary
  condition enforcement, with application to fluid--structure interaction,
  Mathematical Models and Methods in Applied Sciences 28~(12) (2018)
  2457--2509.

\bibitem{lai2001remark}
M.-C. Lai, Z.~Li, A remark on jump conditions for the three-dimensional
  {N}avier-{S}tokes equations involving an immersed moving membrane, Applied
  mathematics letters 14~(2) (2001) 149--154.

\bibitem{li2008front}
X.~Li, K.~Sarkar, Front tracking simulation of deformation and buckling
  instability of a liquid capsule enclosed by an elastic membrane, Journal of
  Computational Physics 227~(10) (2008) 4998--5018.

\bibitem{ramanujan1998deformation}
S.~Ramanujan, C.~Pozrikidis, Deformation of liquid capsules enclosed by elastic
  membranes in simple shear flow: large deformations and the effect of fluid
  viscosities, Journal of Fluid Mechanics 361 (1998) 117--143.

\bibitem{taylor1923viii}
G.~I. Taylor, {VIII.} stability of a viscous liquid contained between two
  rotating cylinders, Philosophical Transactions of the Royal Society of
  London. Series A, Containing Papers of a Mathematical or Physical Character
  223~(605-615) (1923) 289--343.

\bibitem{ghigliotti2011vesicle}
G.~Ghigliotti, A.~Rahimian, G.~Biros, C.~Misbah, Vesicle migration and spatial
  organization driven by flow line curvature, Physical Review Letters 106~(2)
  (2011) 028101.

\bibitem{lac2007hydrodynamic}
E.~Lac, A.~Morel, D.~Barth{\`e}s-Biesel, Hydrodynamic interaction between two
  identical capsules in simple shear flow, Journal of Fluid Mechanics 573
  (2007) 149--169.

\bibitem{kumar2014flow}
A.~Kumar, R.~G.~H. Rivera, M.~D. Graham, Flow-induced segregation in confined
  multicomponent suspensions: effects of particle size and rigidity, Journal of
  Fluid Mechanics 738 (2014) 423--462.

\bibitem{maestre2019dynamics}
J.~Maestre, J.~Pallares, I.~Cuesta, M.~A. Scott, Dynamics of a capsule flowing
  in a tube under pulsatile flow, Journal of the Mechanical Behavior of
  Biomedical Materials 90 (2019) 441--450.

\bibitem{sherwood2003rates}
J.~Sherwood, F.~Risso, F.~Coll{\'e}-Paillot, F.~Edwards-L{\'e}vy, M.-C.
  L{\'e}vy, Rates of transport through a capsule membrane to attain {D}onnan
  equilibrium, Journal of Colloid and Interface Science 263~(1) (2003)
  202--212.

\bibitem{borazjani2013fluid}
I.~Borazjani, Fluid--structure interaction, immersed boundary-finite element
  method simulations of bio-prosthetic heart valves, Computer Methods in
  Applied Mechanics and Engineering 257 (2013) 103--116.

\bibitem{liao2015simulations}
C.-C. Liao, W.-W. Hsiao, T.-Y. Lin, C.-A. Lin, Simulations of two
  sedimenting-interacting spheres with different sizes and initial
  configurations using immersed boundary method, Computational Mechanics 55~(6)
  (2015) 1191--1200.

\end{thebibliography}


\end{document}